\documentclass[aps,prb,twocolumn,amsmath,amssymb,superscriptaddress,floatfix,eqsecnum,showpacs,nofootinbib]{revtex4-1}
\usepackage{amsmath}
\usepackage{graphicx}
\usepackage{latexsym}
\usepackage{amsmath,amssymb}
\usepackage{bm} 

\DeclareMathOperator{\sgn}{sgn}

\newcommand{\ord}[1]{\bm{\mathit{O}}\left(#1\right)}

\newcommand{\Hi}{H^{(i)}}
\newcommand{\Hf}{H^{(f)}}
\newcommand{\Hic}{\bar{H}^{(i)}}
\newcommand{\Hfc}{\bar{H}^{(f)}}

\newcommand{\bsub}{\begin{subequations}}
\newcommand{\esub}{\end{subequations}}

\newcommand{\gnd}{{| 0 \rangle}}
\newcommand{\gndc}{{\langle 0 |}}
\newcommand{\gndmu}{| \bar{0} \rangle}
\newcommand{\gndmuc}{\langle \bar{0} |}

\newcommand{\tp}{t'}

\newcommand{\rf}{\mathcal{R}}

\newcommand{\nr}{n_{\mathcal{R}}}
\newcommand{\np}{n_{+}}
\newcommand{\nm}{n_{-}}
\newcommand{\npm}{n_{\pm}}

\newcommand{\ket}[1]{\left|{#1}\right\rangle}

\newcommand{\ts}[1]{{\textstyle{#1}}}
\newcommand{\sss}[1]{{\scriptstyle{#1}}}

\usepackage{color}

\begin{document}

\title{
	Quantum quench spectroscopy of a Luttinger liquid:
	Ultrarelativistic density wave dynamics due to
	fractionalization in an XXZ chain
}

\author{Matthew S. Foster}
\email{psiborf@rci.rutgers.edu}
\affiliation{Center for Materials Theory, Department of Physics and Astronomy,
Rutgers University, Piscataway, NJ 08854, USA}

\author{Timothy C. Berkelbach}
\email{tcb2112@columbia.edu}
\affiliation{Department of Chemistry, Columbia University, New 
York, New York 10027, USA}

\author{David R. Reichman}
\affiliation{Department of Chemistry, Columbia University, New 
York, New York 10027, USA}

\author{Emil A. Yuzbashyan}
\affiliation{Center for Materials Theory, Department of Physics and Astronomy,
Rutgers University, Piscataway, NJ 08854, USA}

\date{\today}

\begin{abstract}
We compute the dynamics of localized excitations produced by a quantum 
quench in the spin 1/2 XXZ chain. Using numerics combining the density matrix
renormalization group and exact time evolution, as well as analytical arguments, 
we show that fractionalization due to interactions in the pre-quench state 
gives rise to ``ultrarelativistic'' density waves that travel at the maximum 
band velocity. The system is initially prepared in the ground state of the chain 
within the gapless XY phase, which admits a Luttinger liquid (LL) description at 
low energies and long wavelengths. The Hamiltonian is then suddenly quenched to 
a band insulator, after which the chain evolves unitarily. Through the gapped 
dispersion of the insulator spectrum, the post-quench dynamics serve 
as a ``velocity microscope,'' revealing initial state particle correlations via 
space time density propagation. We show that the ultrarelativistic wave production 
is tied to the particular way in which fractionalization evades Pauli-blocking in 
the zero-temperature initial LL state. 
\end{abstract}


\maketitle


\section{Introduction}

In the labyrinth of one-dimensional (1D) quantum many-body physics, the Luttinger liquid (LL) 
lurks around nearly every corner. It emerges as the low energy field theory 
description of interacting Bose gases, gapless quantum spin chains, fermion lattice models 
(Hubbard, etc.), electrons in metallic carbon nanotubes, and chiral quantum Hall 
edge states.\cite{BosonizationRev2,BosonizationRev3,Cazalilla04}
Luttinger liquid physics is universal: it reduces the complexities
of myriad microscopic models to the hydrodynamics of free bosons.

Despite its apparent simplicity, the LL description of interacting fermions 
exhibits a number of rather peculiar properties, due to the advent of 
quasiparticle \emph{fractionalization}. The elementary excitations of a LL
are collective density waves that carry fractional (electric or number) 
charge, relative to the ``bare'' fermionic constituents; injecting
a bare fermion into a LL causes it to ``break up'' into many pieces.
This collectivization of the dynamics due to fractionalization
leads to a host of predicted anomalies, 
including the low bias suppression of the tunneling density of states,
and perfect insulating behavior at zero temperature due to the
presence of even a single impurity.\cite{FisherGlazman96}
For spinful fermions, fractionalization induces spin-charge 
separation.\cite{BosonizationRev1,BosonizationRev3} 
Interestingly enough, zero temperature dc transport in a clean
quantum wire through ideal Fermi liquid leads shows no signature 
of fractionalization; the conductance is quantized to
$e^2/h$ per channel, irrespective of the interactions.\cite{LLDCTransport}

\begin{figure}[b!]
   \includegraphics[scale=0.28]{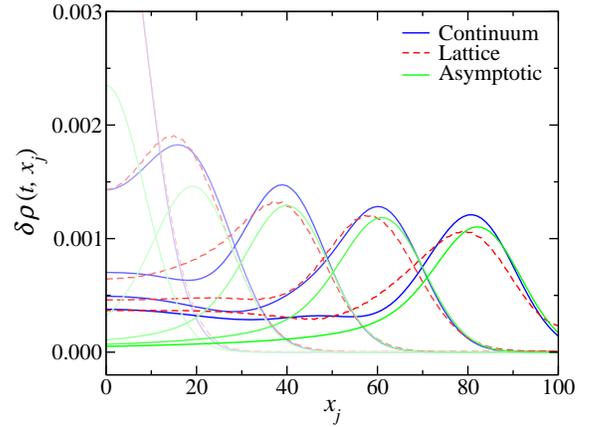}
   \caption{
	Ultrarelativistic wave generated from an initial density bump, 
	following an interacting quench.
	The Luttinger liquid ground state of an interacting XXZ chain 
	is time-evolved according to a non-interacting, band insulator Hamiltonian. 
	The density $\delta \rho(t,x_j)$ due to the inhomogeneity is plotted
	at time slices
	$t=$ 0, 12, 24, 36, and 48
	after the quench; fainter (bolder) traces depict earlier (later) times.
	The evolution is symmetric about $x_j=0$.
	In this figure, red dashed lines were obtained from a combination of DMRG and
	exact time evolution for the XXZ chain, while blue solid lines are
	the prediction of continuum sine-Gordon field theory.
	The curves marked ``asymptotic'' are the analytical result for the 
	``regularized supersoliton,'' obtained in Eq.~(\ref{SupersolitonRegEq}).
	The initial coupling strength is $\gamma = -0.872$, corresponding to $\sigma = 0.7$.
	The initial bump has width $\Delta = 12$ and weight $Q=0.10$;
	the mass gap is $M = 1/8$. 
	The continuum data obtains from numerical integration of Eq.~(\ref{ContEOM})
	with $\alpha = 0.75$ and $\zeta = 1$.
	}
   \label{fig:int_0.7_12}
\end{figure}

In this paper, we describe a ``transport'' effect that directly
exhibits fractionalization in a LL, observed in the dynamics
of a density fluctuation following a sudden quantum quench.
In a quantum quench, a system is prepared in an eigenstate 
of some initial Hamiltonian. In our case, we take the ground state 
of an XXZ chain with a non-uniform density profile,
which possesses a low-energy LL description.
At the time of the quench, by external means a sudden deformation
is affected upon the Hamiltonian, which subsequently drives
the unitary post-quench dynamics.
Here, the post-quench spectrum consists of non-interacting
fermions, with a Hamiltonian that possesses a gapped,
band insulator ground state (the XX chain in the presence
of a sublattice staggered external field). 
We show that the relative fractionalization of the pre-quench
system (due to interparticle interactions) leads to the production
of ``ultrarelativistic'' density waves after the quench.
These waves travel
at the maximum band velocity, and exhibit a 
particular shape set by the interaction strength. 
The propagating density waves are ``elementary excitations'' of the post-quench
non-equilibrium state; they occur because the fractionalized 
density inhomogeneity of the initial LL
``injects'' high momentum excitations into the post-quench band insulator. 
By contrast, under the same conditions a quench from the ground state of 
the non-interacting Fermi gas (the XX point of the XXZ chain) into the band 
insulator yields only dispersive density dynamics,
a consequence of Pauli-blocking.
Our setup can be viewed as a ``quench spectroscopy'' of fractionalization in a 
Luttinger liquid.

In the last decade, rapid experimental progress\cite{BlochDalibardZwerger08} 
in ultracold atoms and optical lattice gases has transported far-from-equilibrium 
many body physics fully into the quantum realm.  
In these systems, the quantum quench has emerged as a primary tool with
which to investigate dynamics. 
Quenches have been performed in boson\cite{Bloch02,Weiss06,Stamper-Kurn06,Weiler08}
and fermion\cite{Ott04,Sommer11,Schneider10} systems, with and without optical lattices,
in one, two, and three dimensions. An ultracold gas can be very well isolated from its
environment, and provides 
an unprecedented degree of control in terms of realizing
model systems and manipulating their parameters.\cite{BlochDalibardZwerger08}
Theoretical work has focused primarily on 
thermalization,\cite{Rigol07,Kollath07,Rigol09,MoeckelKehrein08,SabioKehrein10,Manmana07,
Barmettler09}, quantum critical scaling and defect 
production,\cite{KibbleZurek,Polkovnikov05,PolkovnikovGritsev08,Polkovnikov09,Cincio07}
and correlation functions in spatially homogeneous 
systems.\cite{Gritsev07,Polkovnikov07,IucciCazalilla10,Uhrig09,BarouchMcCoyDresden70,HastingsLevitov08,
Manmana09,CardyCalabrese06,Rossini09,Mathey10}
Prior art on Luttinger liquid, sine-Gordon, and XXZ chain quenches includes 
that of Refs.~\onlinecite{Gritsev07,Polkovnikov07,IucciCazalilla10,Uhrig09,SabioKehrein10,
Manmana07,HastingsLevitov08,Manmana09,Barmettler09,MosselCaux10,
LancasterMitra10,LancasterGullMitra10}.
Wavepackets have been previously employed in the study of excitations induced by a 
\emph{local} quench,\cite{Schmitteckert04,Kollath05,Langer09} in which the Hamiltonian 
deformation is restricted to a spatial subregion of the larger system. 
The characterization of spatially inhomogeneous dynamics following a \emph{global} parameter 
quench (as studied here) is a more recent 
development.\cite{MosselCaux10,LancasterMitra10,LancasterGullMitra10,supersoliton,Cai11}

Many of the previous schemes proposed or executed in the 
theoretical\cite{Rigol07,Kollath07,Heidrich-Meisner09,Rigol09,Kajala11} 
and 
experimental\cite{Bloch02,Weiss06,Sommer11,Schneider10}
literature can be termed ``hard quenches.''
In these works, large changes in a parameter value 
or trap geometry lead to the excitation of novel high energy 
states\cite{Kollath07,Heidrich-Meisner09,Bloch02,Weiss06} whose physics has 
little to do with the low-energy sectors of \emph{either} the
initial or final Hamiltonians 
(for an interesting exception, see Ref.~\onlinecite{Heidrich-Meisner08}).
Our goal in this paper is different: 
we use a ``soft quench'' (defined below) as a low-energy probe of the initial state.


\subsection{Overview}

\subsubsection{XXZ quench protocol; velocity microscope \label{SoftQuenchSec}}

\begin{figure}
   \includegraphics[scale=0.25]{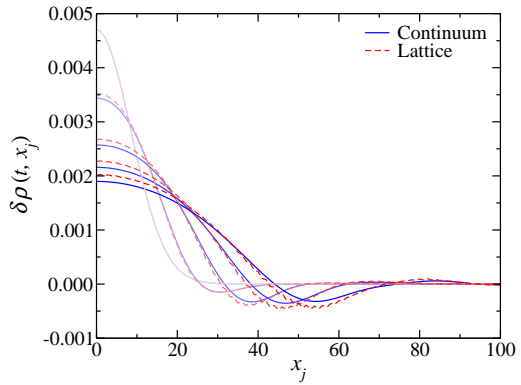}
   \caption{
	Dispersive decay of a density bump following a non-interacting
	($\gamma = \sigma = 0$), non-relativistic ($M \Delta = 3/2$) quench.
	This is the same as Fig.~\ref{fig:int_0.7_12}, but for 
	a quench from a non-interacting Fermi gas to a band insulator.
	Parameters $\Delta$, $Q$, and $M$ are as in Fig.~\ref{fig:int_0.7_12}.
	Red dashed lines are the results of exact diagonalization of the lattice 
	Hamiltonian, while blue solid lines are the continuum predictions. 
	}
   \label{fig:nonint_12}
\end{figure}

We study the dynamics following a quantum quench in the 1D spin 
$1/2$ XXZ chain. Working in the equivalent spinless (or spin-polarized) fermion
representation, we investigate the time evolution of the particle
density induced by a non-uniform initial state. 
Other works treating XXZ and sine-Gordon quenches subject to 
initial state inhomogeneity include Refs.~\onlinecite{MosselCaux10,LancasterMitra10,LancasterGullMitra10};
see Sec.~\ref{Extensions} for a discussion. 

We consider a system initially prepared in the ground state of the XXZ chain in its gapless XY phase, 
subject to an
external field.
The 
field 
induces
a localized ``bump'' in the density profile of 
the otherwise spatially homogeneous system.
This state is further characterized by the 
spin anisotropy $\gamma$ of the pre-quench 
$\hat{S}_i^z \, \hat{S}_{i+1}^z$ coupling, i.e.\ the four fermion interaction strength. 
The gapless XY phase of the XXZ chain admits a low energy Luttinger liquid (LL) 
description.\cite{BosonizationRev2,BosonizationRev3} 
At time $t = 0$, the system Hamiltonian is deformed discontinuously: 
$\gamma$ is set to zero, while a sublattice-staggered external 
field is simultaneously applied along the 
length of the chain, opening up a gap in the spectrum. 
In the fermion language, the ground state of the post-quench (``final'') Hamiltonian is a non-interacting band 
insulator with a doubled unit cell. 

The lattice quench with $\gamma = 0$ in the initial XY state is special, 
because both the initial and final Hamiltonians are non-interacting in the fermion language.
We 
dub this
the ``non-interacting'' quench; the exact solution can be written 
for the time evolution of the 
density expectation value. By contrast, for $\gamma \neq 0$ 
(``interacting'' quench) the initial Hamiltonian is interacting in the fermion language and not soluble by 
elementary means. Although the XXZ chain is integrable, 
the non-uniform density profile makes difficult the application
of the Bethe ansatz method.
Instead, in this paper we use the 
Density Matrix Renormalization Group (DMRG) to numerically compute correlation functions of the initial 
ground state. 
For both the non-interacting and interacting quenches, the dynamics generated by the non-interacting band 
insulator Hamiltonian are determined exactly. This allows us to avoid the use of more computationally intensive, 
time-dependent DMRG calculations. We exploit this advantage to analyze larger system sizes than previous 
numerical quench studies of the XXZ chain.\cite{Manmana07,Manmana09} 

The idea behind this setup is to use the quench
into a gapped, dispersive phase as a ``velocity microscope'' on the initial
correlated LL state. The non-uniform initial density profile creates additional
excitations \emph{on top of} the homogeneous bath induced by the
global parameter quench, leading to real space dynamics that can in principle be 
directly observed. 
Particles composing space time density fluctuations are excited with a broad
range of momenta; these are velocity-resolved by the dispersive post-quench
spectrum.
By contrast, time evolution with a generic gapless post-quench Hamiltonian 
in 1D (such as that governing a continuum conformal field theory) 
produces only pure left- and right-moving ``ultrarelativistic'' waves, regardless
of the structure of the initial state.\cite{CardyCalabrese06,IucciCazalilla10} 

Throughout this work we make the crucial assumption of a ``soft quench,''
defined as follows.
The magnitude of the gap in the post-quench Hamiltonian is specified by a 
dimensionless parameter $M a$, where $1/M$ gives the ``Compton wavelength''
for the low energy, massive excitations of the band insulator, and $a$ denotes 
the lattice spacing. In addition, we assume a Gaussian density inhomogeneity in 
the initial state of width $\Delta$. 
The assumption of a ``soft'' quench requires that 
\begin{equation}\label{SoftQuench}
	a \ll \frac{1}{M} \lesssim \Delta,
\end{equation}
i.e.\ that the low-energy Compton wavelength dwarfs the lattice
spacing, while the width of the initial state inhomogeneity exceeds
the Compton wavelength.
The first assumption guarantees that the gap opens in the low-energy
sector of the band Hamiltonian. The second $M \Delta \gtrsim 1$ 
assures that any excitation of large-momentum particles post-quench 
arises from the correlated character of the LL, and not the excessive 
``squeezing'' of the initial density bump.

Despite the requirement in Eq.~(\ref{SoftQuench}),
we will consider quenches with ``intermediate'' to ``large'' 
values of the initial XXZ interaction strength $\gamma$,
approaching the ferromagnetic transition at $\gamma = -1$.
It is far from obvious that a change from $|\gamma| \lesssim 1$
to $\gamma = 0$ preserves the notion of a ``soft quench''
as articulated above. Because the low-energy description
throughout the gapless phase is a LL, it is nevertheless
the case.

\begin{figure}
   \includegraphics[scale=0.725]{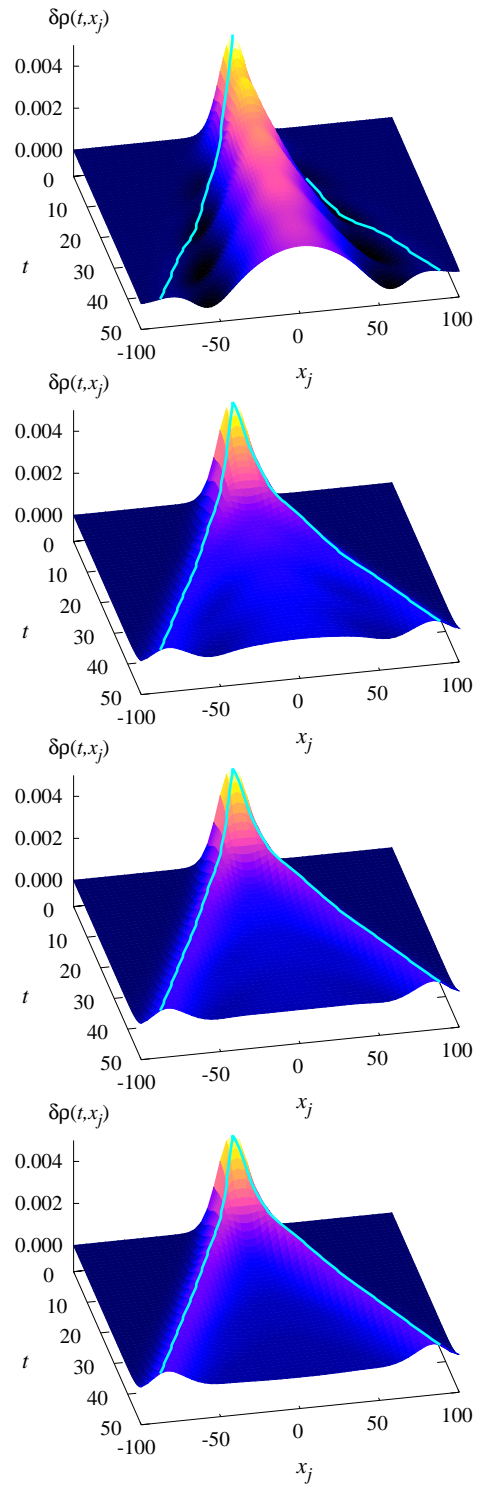}
   \caption{
	Post-quench evolution of a density bump: dependence on initial state
	interparticle interaction strength.
	Each subpanel exhibits a three-dimensional view of a lattice quench 
	(obtained by DMRG + exact time evolution)
	into the gapped band insulator; 
	$Q$, $\Delta$, and $M$ are the same as in 
	Figs.~\ref{fig:int_0.7_12} and \ref{fig:nonint_12}.
	The four frames depict quenches with increasing initial state interactions, 
	$\sigma=0$ (non-interacting), 0.4, 0.7, and 1.0 (top to bottom).
	The cyan line demarks the maximum band propagation velocity (``speed of light''), 
	$v_{\rm max}(M=1/8)\approx1.77$.
	}
   \label{fig:3d_gapped_12}
\end{figure}

\subsubsection{Sine-Gordon and ``Supersolitons''}

The XXZ quench can be interpreted as a lattice version of the continuum sine-Gordon field
theory analyzed previously in Ref.~\onlinecite{supersoliton}. In that work, spatiotemporal
dynamics were computed in a quench across a quantum critical point. 
In Ref.~\onlinecite{supersoliton}, a LL ground state subject to an inhomogeneous density modulation was 
time-evolved according to a translationally invariant, post-quench Hamiltonian favoring a gapped Mott ground state. 
The Mott Hamiltonian generating the dynamics was chosen to reside at the Luther-Emery\cite{LutherEmery74} 
point of the sine-Gordon model, where the excitation spectrum consists entirely of non-interacting, massive Dirac 
fermions.\cite{Rajaraman,BosonizationRev2,BosonizationRev3} 
In the XXZ chain quench studied here, the final state band insulator carriers play the role of the 
non-interacting Dirac fermions that compose the sine-Gordon spectrum at the Luther-Emery point;
the Mott gap of the sine-Gordon model is here substituted by the band gap. 

A localized density inhomogeneity in the sine-Gordon quench 
launches ultrarelativistic, non-dispersing traveling waves, dubbed ``supersolitons''
in Ref.~\onlinecite{supersoliton}. The supersoliton exhibits a rigid   
shape, propagates at the ``speed of light'' $v_F$ (the Fermi velocity), and 
possesses an amplitude that grows in time as $t^{\sigma/2}$. The exponent $\sigma \geq 0$
characterizes the fractionalization of the initial LL state relative to the final 
Mott insulator. For the case $\sigma = 0$ (non-interacting quench), there is no fractionalization and no supersoliton;
the density dynamics of such a quench with $M \Delta \gtrsim 1$ show only dispersive broadening.

In this work, we demonstrate that the supersoliton arises in the sine-Gordon quench
for $\sigma > 0$ due to the particular way in which LL fractionalization 
evades Pauli-blocking. This is made explicit through a calculation of the local 
phase space (Wigner) distribution in the pre-quench LL.
The result is a power-law occupation of momentum states in the post-quench 
insulator that translates into a singular peak at $v_F$ in the corresponding
(local) velocity distribution. 
Because velocity is conserved by the post-quench Hamiltonian, the spectral
weight associated to the singularity is translated at $v_F$.
By contrast, a non-interacting quench with 
$\sigma = 0$ and $M \Delta \gg 1$ excites only small velocities
$v \lesssim v_F/M \Delta$. 

A key point is that it is the long-distance behavior of correlations
in the initial state that permits the evasion of Pauli-blocking
in the fractionalized case. Although lattice details can and do 
modify the ultraviolet behavior of correlations in the XXZ chain
considered here, the fundamental distinction between non-interacting
and interacting quenches remains a robust feature of the soft quench
satisfying Eq.~(\ref{SoftQuench}).

\subsubsection{Preview of numerical results}

We defer a detailed discussion of our XXZ chain quench results 
to the main text; the impatient reader may consult Sec.~\ref{Conc} for a summary. 
Instead, we exhibit a few graphs that demonstrate the
qualitative difference between the 
interacting and non-interacting
quenches. 
Fig.~\ref{fig:int_0.7_12} shows the ``ultrarelativistic'' density wave launched
in an interacting quench satisfying the constraint in Eq.~(\ref{SoftQuench}).
(We set the lattice spacing $a = 1$). By contrast, Fig.~\ref{fig:nonint_12} depicts 
a non-interacting quench; in this case, only dispersive broadening
of the initial density inhomogeneity is seen.
The parameters in these two figures are
the same, except for the interaction strength, 
quantified by a parameter $\sigma(\gamma)$. For a non-interacting quench 
one has $\sigma(\gamma = 0) = 0$; otherwise $\sigma > 0$ and increases 
monotonically with $|\gamma|$. 
The evolution of an XXZ chain quench as a function of the interaction
strength $\sigma$ is depicted as a 3D plot sequence 
in Fig.~\ref{fig:3d_gapped_12}. 

The blue continuous curves in Figs.~\ref{fig:int_0.7_12} and \ref{fig:nonint_12} 
are obtained using 
an ultraviolet-regularized version of the sine-Gordon quench
studied in Ref.~\onlinecite{supersoliton}. The regularization models
the effects of neglected lattice scale details in a very crude 
way.
The regularization cuts off the amplification of the supersoliton
predicted for the pure sine-Gordon quench; it also leads to 
a modification of its interaction-dependent shape. Because of the
close agreement between the field theory and lattice results,
we interpret the ultrarelativistic density wave appearing
in the interacting quench (Figs.~\ref{fig:int_0.7_12} and 
\ref{fig:3d_gapped_12}) as a ``regularized'' supersoliton.

We emphasize that the quench dynamics described in this paper are fully quantum 
coherent; the absence of interparticle scattering in the post-quench band insulator 
prevents dephasing or thermalization. The ``fractionalized'' density dynamics reflect 
the many-body entanglement of the initial gapless state. Future work incorporating 
integrability-preserving interactions post-quench could prove particularly interesting, 
as discussed in the Conclusion.


\subsection{Outline}

The organization of this paper is as follows.
In Sec.~\ref{Q Setup}, we define the pre- and post-quench XXZ
Hamiltonians and set up the dynamics to be computed. 
In Sec.~\ref{Sine-Gordon}, we provide a comprehensive analysis 
linking the XXZ chain quench studied here to the corresponding 
version in the continuum, low-energy sine-Gordon field theory.
We begin in Sec.~\ref{WavepacketDyn} with a quick review of single particle 
relativistic wavepacket mechanics, where we emphasize the distinction 
between ``relativistic'' and ``non-relativistic'' wavepacket propagation.
In Sec.~\ref{SG Def}, we describe the solution to the pure sine-Gordon
quench. We identify the supersoliton, discussed previously in 
Ref.~\onlinecite{supersoliton}.
The global and local (Wigner) distributions induced in the lattice
and continuum quenches are discussed in Sec.~\ref{DistFuncs},
wherein the origin of the supersoliton is revealed. 
In Sec.~\ref{SG Reg}, the ultraviolet modifications
of the sine-Gordon theory necessary to model the lattice quench
are articulated, and relevant time scales are defined.

Numerical results obtained for the time evolution of the
XXZ chain quench are presented and discussed in Sec.~\ref{Results}. 
Results for the non-interacting and interacting quenches are
exhibited and compared to the regularized sine-Gordon theory. 
We summarize our conclusions in Sec.~\ref{Conc}, and finish
with a discussion of open questions. 
The asymptotic analysis method used to obtain key analytical
results is explicated in Appendix~\ref{APP--AA}.
Appendix~\ref{APP--FracSec} recapitulates
the notion of fractionalization in a Luttinger liquid. 
In Appendix~\ref{APP--DistInhomog}, we derive the local (Wigner) velocity distributions
induced by the initial state inhomogeneity, in the interacting
and non-interacting continuum quenches.

\section{
Quench setup
\label{Q Setup}
}


\subsection{Lattice model \label{Lat model}}

In a (sudden) quantum quench, one prepares the system in an eigenstate 
of an initial Hamiltonian $\Hi$, and subsequently time evolves under a 
different final Hamiltonian, $\Hf$.
We 
consider the XXZ spin $1/2$ Heisenberg chain,
\begin{align}
	H 
	=& 
	-2 J 
	\sum_{i} \left(\hat{S}^x_i \hat{S}^x_{i+1} + \hat{S}^y_i \hat{S}^y_{i+1}
	-
	\gamma 
	\hat{S}^z_i \hat{S}^z_{i+1}
	\right)
	\nonumber\\
	&-
	\sum_i
	\mu_{i} \hat{S}^z_i.
\end{align}
Via the Jordan-Wigner transformation, the spin chain is equivalent to
a model of spinless (or spin-polarized) fermions whose Hamiltonian is 
given by
\begin{equation}\label{H_ferm_gen}
	H = -J \sum_i c^\dagger_i c_{i+1} + {\rm H.c.} + 2 J \gamma \sum_i \delta n_i \delta n_{i+1} 
	- \sum_i \mu_i \delta n_i,
\end{equation}
where $J$ denotes the nearest-neighbor hopping amplitude, $2 J \gamma$ is a nearest-neighbor 
density-density interaction strength, and $\mu_i$ represents a site-dependent chemical potential. 
In Eq.~(\ref{H_ferm_gen}), $c_i$ and $c_j^\dagger$ satisfy
$c_i c_j^\dagger + c_j^\dagger c_i = \delta_{i j}$, and $\delta n_i \equiv c_i^\dagger c_i - 1/2$.
We will quench from a ground state in the gapless XY phase of this Hamiltonian 
(labeled by the dimensionless interaction strength $\gamma$)
to a non-interacting, band insulator state.
The latter is induced via the application of a unit cell doubling, sublattice-staggered chemical potential.  
Furthermore, by applying a localized Gaussian chemical potential, we will induce a density inhomogeneity 
into the initial state whose dynamics will reveal the effects of our quantum quench.

\begin{figure}
   \includegraphics[width = 0.3\textwidth]{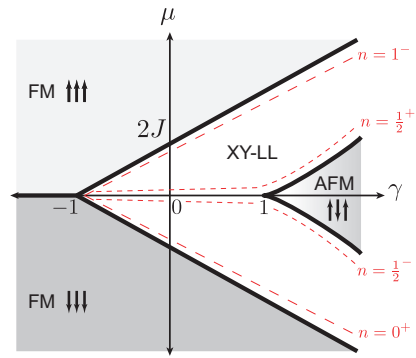}
   \caption{\label{XXZPhaseDiag}
	$T = 0$ 
	ground state
	phase diagram for the XXZ chain.\cite{Sutherland}
	Red dashed lines correspond to constant fermion density contours.
	}
\end{figure}

The zero temperature phase diagram for the XXZ chain in Eq.~(\ref{H_ferm_gen}) with $\mu_i = \mu$ (const.)
is sketched in Fig.~\ref{XXZPhaseDiag}. At zero chemical potential, the chain is in its gapless, power-law correlated
XY phase for $-1 < \gamma \leq 1$. For $\gamma > 1$ ($\gamma < -1$), the spin chain assumes
long-range Ising antiferromagnetic (ferromagnetic) order and the spectrum gaps out. 
We note that the thickness of the chemical potential window over which power-law XY order occurs
(at fermion densities between 0 and 1 per site) narrows to zero upon approaching the
ferromagnetic transition at $\gamma = -1$.

For the quantum quench studied here,
the initial and final lattice Hamiltonians are given by 
\begin{subequations}\label{ham}
\begin{align}
	\Hi 
	=& 
	-J
	\left[
	\sum_i c^\dagger_i c_{i+1} + {\rm H.c.} 
	- 2 \gamma 
	\sum_i \delta n_i \delta n_{i+1} 
	\right]
	\nonumber\\
	&
	- \sum_i \mu^{(0)}_i \delta n_i, 
	\label{HiDef}	
	\\
	\Hf 
	=& 
	-J 
	\left[
	\sum_i c^\dagger_i c_{i+1} + {\rm H.c.} - 
	2 M a
	\sum_i (-1)^i \delta n_i 
	\right].
	\label{HfDef}
\end{align}
\end{subequations}
We assume periodic boundary conditions in a chain of $N = L/a$ sites,
with $a$ the lattice spacing, so that $c_{N+1} = c_{1}$. 
We always take $N$ to be an even integer. 
The initial Hamiltonian $\Hi$ is tuned to reside in its gapless XY
phase, so that $-1 < \gamma \leq 1$.

In Eq.~(\ref{HiDef}) above,
\begin{equation}\label{gaussian}
	\mu^{(0)}_i 
	= 
	\frac{Q \sqrt{\pi}}{\Delta} \frac{u(\gamma)}{K(\gamma)}
	e^{-x_i^2/\Delta^2}
\end{equation}
is the localized chemical potential used to introduce a particle 
density inhomogeneity near the center of the chain; we have introduced the 
spatial coordinate $x_i = (i - N/2)a$ such that $x_{N/2} = 0$ and 
$x_{N/2 + 1} = a$ straddle the chain center. The parameters $Q$ and $\Delta$ 
set the ``strength'' and width of the Gaussian potential, respectively.  
Two additional parameters that enter into Eq.~(\ref{gaussian}) are the
sound velocity $u$ and the Luttinger parameter $K$. These coefficients
completely determine the character of the low-energy field theory description
of the XXZ chain in its critical XY phase, in equilibrium.
In the absence of an external chemical potential, 
$u$ and $K$ can be obtained from the Bethe ansatz, yielding\cite{BosonizationRev2}
\begin{subequations}\label{u,K Bethe ansatz}
\begin{align}
	u(\gamma) 
	&= 
	J a
	\frac{\pi\sqrt{1-\gamma^2}}{\arccos(\gamma)}, 
	\label{u Bethe ansatz}
	\\
	K(\gamma) & = \frac{\pi}{2 [\pi - \arccos(\gamma)]},
	\label{K Bethe ansatz}
\end{align}
\end{subequations}
such that $u(0) = 2 J a \equiv v_{F}$ and $K(0) = 1$. Here, $v_F$ denotes 
the band Fermi velocity at half-filling with $M = 0$. 
We have included the ratio $u/K$
in the definition of the 
local potential so as to keep the initial
induced density inhomogeneity approximately constant with
varying interaction strength $\gamma$.  

The spectrum of $\Hf$ [Eq.~(\ref{HfDef})] is
\begin{equation}\label{spec}
	E_k = 
	\pm 2 J \sqrt{\cos^2(k a) + \left(M a\right)^2}. 
\end{equation}
The staggered potential, which doubles the unit cell, introduces
a bandgap in the spectrum at $k_F = \pi/2a$ with magnitude 
	$E_g = 4 J M a$.

Based on the 
analysis of the continuum sine-Gordon quench in 
Ref.~\onlinecite{supersoliton},
we expect the post-quench
system response to be governed by the dynamical exponent,
\begin{align}\label{sigma}
	\sigma(\gamma) 
	&\equiv 
	\frac{1}{2}\left[K(\gamma) + \frac{1}{K(\gamma)}\right]-1 
	\nonumber\\
	&= 
	\frac{2[\arcsin(\gamma)]^2}{\pi^2+2\pi\arcsin(\gamma)}.
\end{align}
Such an interaction-dependent exponent
characterizes
the (critical) power-law behavior exhibited by correlation functions in gapless 
1D quantum systems
that possess a low-energy LL description.
At $\gamma = 0$ (non-interacting quench), $\sigma$ assumes 
its minimum value of zero. At $\gamma = 1$, on the precipice of the instability to Ising antiferromagnetism,
$\sigma = 1/4$. By contrast, $\sigma$ diverges upon approaching $\gamma = -1$
from above. 

In what follows, we set $J=1$ and $a=1$, thereby measuring energies in units of 
the transfer integral $J$ and distances in units of the lattice spacing $a$.
Our observable of interest will be the time-evolved density at each site of the lattice, i.e.
\begin{equation}
	\rho(t,x_i) \equiv \langle \delta n_i(t) \rangle
	=
	\gndc e^{i\Hf t} c^\dagger_i c_i e^{-i \Hf t} \gnd -1/2,
\end{equation}
where $\gnd$ is the ground state of the initial Hamiltonian $\Hi$.


\subsection{Dynamics; non-interacting quench \label{Nonint Exact sol}}

For both the interacting ($\gamma \neq 0$) and non-interacting ($\gamma = 0$) quenches, 
the dynamics are obtained by solving the Heisenberg equation of motion for 
the annihilation operator $c_i(t)$ at site $i$ using $\Hf$ [Eq.~(\ref{HfDef})].
The result is
\begin{align}\label{LatDyn}
	c_i(t) 
	=& 
	\sum_{j=1}^N 
	c_j(0) 
	\begin{aligned}[t]
	&\left\{ 
	G^{(1)}(t,i-j)
	\right.
	\\
	&\quad 
	+ \left[ (-1)^i + (-1)^j \right] 
	G^{(2)}(t,i-j)
	\\
	&\quad 
	\left.
	+ (-1)^{i+j} 
	G^{(3)}(t,i-j) 
	\right\} 
	\end{aligned}
	\nonumber \\
	\equiv& \sum_{j=1}^N \mathcal{G}_{ij}(t) c_j(0), 
\end{align} 
where 
$c_j(0)$ 
denotes the Schr\"odinger picture operator, and
\begin{equation}
	G^{(a)}(t,j)
	= \frac{1}{N}\sum_{n_k = 1}^{N/2} 
	\exp\left(i \frac{2\pi n_k \, j}{N} \right) 
	\tilde{G}^{(a)}\left(t,\frac{2\pi n_k}{N}\right),
\end{equation}
with
\begin{subequations}\label{GreensFT}
\begin{align}
	\tilde{G}^{(1)}(t,k) 
	& = \cos(E_k t) - i \frac{
	\epsilon_k
	}{E_k} \sin(E_k t), \\
	\tilde{G}^{(2)}(t,k) 
	& = - i \frac{2 M}{E_k} \sin(E_k t), \\
	\tilde{G}^{(3)}(t,k) 
	& = \cos(E_k t) + i \frac{
	\epsilon_k
	}{E_k} \sin(E_k t).
\end{align}
\end{subequations}
In 
Eq.~(\ref{GreensFT}), 
$\epsilon_k = -2\cos(k)$ 
and $E_k$ was defined by Eq.~(\ref{spec}).
The post-quench dynamics of the number density
are subsequently given by
\begin{equation}\label{dynamics}
	\rho(t,x_i)
	= \sum_{j,j^\prime = 1}^{N} 
	\left[ 
	\mathcal{G}^*_{ij}(t)
	\mathcal{G}_{ij^\prime}(t)
	\;
	\mathcal{C}(x_j,x_{j^\prime})
	\right], 
\end{equation}
where all information about the initial state is encoded in the static
correlation function (single-particle density matrix)
\begin{equation}\label{ICCorr}
	\mathcal{C}(x_j,x_{j^\prime}) \equiv 
	\gndc c_j^\dagger(0) c_{j^\prime}(0) \gnd;
\end{equation} 
$\gnd$ denotes the ground state of $\Hi$.

For the special case of the non-interacting quench, 
$\mathcal{C}(x_j,x_{j^\prime})$
is obtained by diagonalizing an $N \times N$ matrix. We denote the single particle Hamiltonian 
implied
in Eq.~(\ref{HiDef}) with $\gamma = 0$ by $\hat{h}$. In this case, the correlator is given by
\begin{equation}\label{NonIntCorr}
	\mathcal{C}(x_j,x_{j^\prime})
	=
	\left[
	\hat{U}
	\hat{P}(-\hat{h}_D)
	\hat{U}^\dagger
	\right]_{j,j^\prime},
\end{equation}
where $\hat{U}^\dagger \hat{h} \hat{U} = \hat{h}_{D}$ diagonalizes the single
particle Hamiltonian, and $\hat{P}(-\hat{h}_D)$ projects onto the (filled)
negative energy states of the diagonalized $\hat{h}_{D}$.
Combining Eqs.~(\ref{NonIntCorr}) and (\ref{dynamics}) gives the formal
solution to the non-interacting quench. In practice, because of the inhomogeneity,
we compute the single-particle matrix in Eq.~(\ref{NonIntCorr}) numerically.


\section{Continuum vs.\ lattice \label{Sine-Gordon}}


\subsection{Warm-up: Relativistic wavepacket dynamics of a single massive Dirac particle \label{WavepacketDyn}}

Before turning to the continuum sine-Gordon quench, we pause to consider a toy problem:
the time evolution of a Gaussian wavepacket for a single, massive Dirac fermion in 1D.
This material is standard, but we include it to emphasize several important points 
regarding disparate regimes of relativistic wave propagation, and to clarify the similarities and differences
between single particle wave packet mechanics and the many particle quantum quench problem studied in this paper.

The post-quench, band insulator Hamiltonian in Eq.~(\ref{HfDef}) exhibits a gap 
$E_g = 2 v_F M$, 
centered at $k = k_F = \pi/2$. Linearizing and truncating the band 
structure to modes near $k_F$, one obtains 
\begin{equation}\label{HfLowNRG}
	\Hfc = \int d x \, \psi^\dagger \hat{h} \psi,
\end{equation}
where the single particle Hamiltonian is given by 
\begin{equation}\label{Diracham}
	\hat{h} = v_{F} \left[-i \hat{\sigma}^3 \frac{d}{d x} + M \hat{\sigma}^2\right],
\end{equation}
and the 2-component Dirac spinor $\psi(x)$ has the Fourier transform
\begin{equation} \label{DiracDef}
	\psi(k)
	\equiv
	\begin{bmatrix}
	\psi_1(k)\\
	\psi_2(k)
	\end{bmatrix}
	=
	\begin{bmatrix}
	e^{-i \pi/4} c(k + k_F)\\
	e^{i \pi/4} c(k - k_F)
	\end{bmatrix},
\end{equation}
with $0 \leq |k|\leq \Lambda \ll k_F$ ($\Lambda$ is a momentum cutoff).
The components $\psi_1$ and $\psi_2$ denote right- and left-movers in the massless limit.
In Eq.~(\ref{Diracham}), we have introduced a set of Pauli matrices $\{\hat{\sigma}^{1,2,3}\}$
acting in the pseudospin space of $\psi$.\cite{Ref--Pauli matrix}
The Fermi velocity $v_F = 2$; below we absorb it into the primed time,
\begin{equation}\label{PrimeTimeDef}
	t' \equiv v_F t.
\end{equation}
In this section we take the system size $L \rightarrow \infty$.

We assume a Gaussian initial wavefunction for a particle in its
rest frame,
\begin{equation}\label{DiracInitial}
	\Psi_0(x) 
	= 
	\frac{1}{\left(\pi \Delta^2 \right)^{1/4}}
	e^{-x^2 / 2 \Delta^2} 
	\begin{bmatrix}
	\Psi_{0,1}\\
	\Psi_{0,2}
	\end{bmatrix},
\end{equation}
with $|\Psi_{0,1}|^2 = |\Psi_{0,2}|^2 = 1/2$. In this equation and
the ones that follow, $\Psi$ denotes a single particle wavefunction;
its time evolution is determined by $\hat{h}$ in Eq.~(\ref{Diracham})
via the Schr\"odinger equation. 
It is useful to write the solution at times $t \geq 0$ in two different ways.
One way is
\begin{align}\label{DiracMom}
	\Psi(t,x) 
	&\equiv
	\Psi_{+}(t,x) + \Psi_{-}(t,x),
\end{align}
where the components $\Psi_{\mu = \pm}(t,x)$ are defined via
\begin{align}
	\Psi_{\mu}(t,x)
	=
	\frac{\sqrt{\Delta} M}{(4 \pi)^{3/4}}
	\int_{-\infty}^{\infty} 
	&
	d z \,
	e^{
	-\left(\frac{M \Delta}{2}\right)^2 \left[\cosh(2 z) - 1 \right]
	}
	\nonumber\\
	& 
	\times
	e^{
	- i \mu A(t',x) 
	\cosh 
	\left[
	z - \mu
	z_0(t',x)
	\right]
	}
	\nonumber\\
	&
	\times
	\left(\Psi_{0,1} 
	- \mu i  
	e^{- \mu z} \Psi_{0,2} \right)
	\begin{bmatrix}
	e^{\mu z}\\
	\mu i
	\end{bmatrix}.
	\label{DiracMomComp}
\end{align}
An alternative representation for $\Psi(t,x)$ is given by
\begin{align}
	\Psi(t,x)
	=&
	\frac{1}{\left(\pi \Delta^2 \right)^{1/4}}
	\left\{
	\!
	\begin{bmatrix}
	\Psi_{0,1} \, e^{-(x-t')^2/2 \Delta^2} \\ 
	\Psi_{0,2} \, e^{-(x+t')^2/2 \Delta^2}
	\end{bmatrix}
	\right.
	\nonumber\\
	&
	\left.
	+
	\int_{-t}^{t}
	d y
	\,
	e^{-\frac{(x - y)^2}{2 \Delta^2}}
	\!
	\begin{bmatrix}
	\sss{\bar{G}^{(1)}(t',y)} & \sss{\bar{G}^{(2)}(t',y)} \\
	\sss{-\bar{G}^{(2)}(t',y)} & \sss{\bar{G}^{(3)}(t',y)}
	\end{bmatrix}
	\!\!
	\begin{bmatrix}
	\Psi_{0,1} \\
	\Psi_{0,2}
	\end{bmatrix}
	\!
	\right\}\!,
	\label{DiracPos}
\end{align}
where
\begin{subequations}\label{ContGFs}
\begin{align}
	\bar{G}^{(1)}(t,y) = & -\frac{M}{2}\left[\frac{t + y}{\sqrt{t^2-y^2}}\right] J_1\left[A(t,y)\right],\\
	\bar{G}^{(2)}(t,y) = & -\frac{M}{2} J_0\left[A(t,y)\right], \\
	\bar{G}^{(3)}(t,y) = & -\frac{M}{2}\left[\frac{t - y}{\sqrt{t^2-y^2}}\right] J_1\left[A(t,y)\right],
\end{align}
\end{subequations}
denote the $M$-dependent components of the Green's functions [from the continuum limit of Eq.~(\ref{GreensFT})].
In Eq.~(\ref{DiracMomComp}), 
$\tanh(z_0) = x/t'$, while $A(t',y) = M \sqrt{t'^2 - y^2}$.
Eqs.~(\ref{DiracMom}) and (\ref{DiracMomComp}) follow 
from the momentum eigenstate expansion for the
time evolution operator, 
while Eq.~(\ref{DiracPos}) obtains from the real space propagation amplitude.
In Eq.~(\ref{ContGFs}),
the symbols $J_{\{0,1\}}$ denote Bessel functions of the first kind.

A basic consequence of relativistic quantum field theory is that a 
single particle cannot be confined to a region smaller than its Compton 
wavelength $1/M$. Localization to smaller scales induces particle energies 
in excess of the mass gap; in a many-particle theory, this
typically leads to pair production out of the vacuum.

In single particle relativistic wave mechanics, one instead finds qualitatively different
behavior for initial confinements $\Delta \gg 1/M$ (``non-relativistic'')
and $\Delta \ll 1/M$ (``relativistic''). We consider first the non-relativistic case. 
For $M \Delta \gg 1$, the argument of the exponential in Eq.~(\ref{DiracMomComp}) 
can be expanded to quadratic order in $z$. In this approximation, 
one obtains 
\begin{align}
	|\Psi&(t,x)|^2 
	=
\nonumber\\
	&\frac{1}{4 \sqrt{\pi} \Delta \delta(t')}
	e^{- x^2 / \Delta^2 \delta^2(t')}
	\nonumber\\
	&\times
	\left\{
	\begin{aligned}
	2& 
	- 
	\frac{2}{\delta(t')} \cos\left[2 M t' - \phi_0(t',x)\right]
	\\
	+& \frac{2 t'}{M\Delta^2 \delta(t')} \sin\left[2 M t' - \phi_0(t',x)\right]
	\\
	+&e^{1/\left[M \Delta \delta(t')\right]^2 + 2 t' x/M^2 \Delta^4 \delta^2(t')}
	\\&
	\times
	\left[
	\begin{aligned}
	&1 + \frac{1}{\delta(t')} \cos\left[ 2 M t' - \phi_{+}(t',x) \right]
	\\
	&- \frac{t'}{M \Delta^2 \delta(t')} \sin\left[ 2 M t' - \phi_{+}(t',x) \right]
	\end{aligned}
	\right]	
	\\
	+&e^{1/\left[M \Delta \delta(t')\right]^2 - 2 t' x/M^2 \Delta^4 \delta^2(t')}
	\\&
	\times
	\left[
	\begin{aligned}
	&1 + \frac{1}{\delta(t')} \cos\left[ 2 M t' - \phi_{-}(t',x) \right]
	\\
	&- \frac{t'}{M\Delta^2 \delta(t')} \sin\left[ 2 M t' - \phi_{-}(t',x) \right]
	\end{aligned}
	\right]
	\end{aligned}
	\right\},
	\label{DiracNonRelProp}
\end{align}	
where the scale factor $\delta(t) = \sqrt{1 + t^2 / M^2 \Delta^4}$;
$\phi_{\{0,+,-\}}(t,x)$ denote some phase factors.\cite{Ref--Phi Plus Minus}
For the initial spinor components in Eq.~(\ref{DiracInitial}), we have made
the choice $\Psi_{0,\{1,2\}} = \exp( \mp i \pi /4)/\sqrt{2}$, 
so that $\Psi_{0}(x)$ is invariant under time-reversal and parity operations.\cite{Ref--Dirac Parity TRI}

To the lowest order in $1/M\Delta$, Eq.~(\ref{DiracNonRelProp}) reduces to the 
usual non-relativistic formula
\[
	|\Psi(t,x)|^2 
	=
	\frac{1}{\sqrt{\pi} \Delta \delta(t')}
	e^{- x^2 / \Delta^2 \delta^2(t')}.
\]
At smaller values of $1/M\Delta$, the oscillatory character of Eq.~(\ref{DiracNonRelProp}) becomes
important, and one observes the ``Zitterbewegung'' phenomenon: the evolving probability
density exhibits an undulatory envelope that beats at the ``interband'' frequency $\omega = 1 / 2 M$. 
These oscillations occur because the eigenstate synthesis of the initial Gaussian [Eq.~(\ref{DiracInitial})]
requires larger contributions from negative energy states as the width $\Delta$ is narrowed.

\begin{figure}[b]
   \includegraphics[width = 0.3\textwidth]{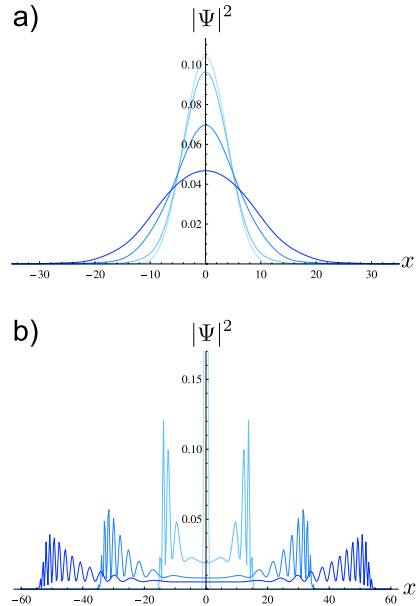}
   \caption{\label{DiracPlot}
	Examples of single particle, massive Dirac equation
	wave packet propagation, obtained via numerical integration
	of 
	Eqs.~(\ref{DiracMom}) and (\ref{DiracMomComp}).
	The case {\bf (a)} corresponds to
	a ``non-relativistic'' initial condition, the Gaussian in 
	Eq.~(\ref{DiracInitial}) with 
	$\Delta = 5/M$. 
	The
	``relativistic'' case is illustrated in {\bf (b)}, with 
	an initial 
	$\Delta = 0.5/M$. 
	Here $M \equiv 1$, and
	data is shown at times $t = 0$, $15$, $35$, and $55$; 
	fainter (bolder) traces depict earlier (later) times.
	}
\end{figure}

In the ultrarelativistic limit $M = 0$, Eq.~(\ref{DiracPos}) 
implies that
\[
	|\Psi(t,x)|^2
	=
	\frac{1}{2\sqrt{\pi}\Delta}
	\left[
	e^{-(x-t')^2/\Delta^2}+e^{-(x+t')^2/\Delta^2}
	\right].
\]
By contrast, when $0 < M \Delta \ll 1$, the propagation is relativistic
but dispersive. 
In the long time limit, the second term on the right-hand side of Eq.~(\ref{DiracPos})
is dominated by the diagonal Green's functions. Close to the right lightcone edge 
$|x - t| \lesssim \Delta$, for $t' \gg t'_{\mathsf{disp}}$ one obtains
\begin{align}\label{SingPartRelDisp}
	\Psi(t,x)
	\sim
	&\frac{1}{\left(\pi \Delta^2 \right)^{1/4}}
	e^{-(x-t')^2 / 2\Delta^2}
	f\left(\frac{t'}{t'_{\mathsf{disp}}},\frac{x - t'}{\Delta}\right)
	\begin{bmatrix}
	\Psi_{0,1} \\
	0
	\end{bmatrix}
	\nonumber\\
	&+
	\text{dispersive background},
\end{align}
where
\[
	t'_{\mathsf{disp}}
	=
	1/2 M^2 \Delta.
\]
The function
\[
	f(\alpha,\beta)
	= 
	\int_{0}^{\infty} d w \left[1 - e^{-\beta w^2/\alpha - w^4/2 \alpha^2} \right] J_{1}(w)
\]
\emph{vanishes} in the limit $\alpha \rightarrow \infty$.

In the XXZ chain quantum quench studied in this paper, it will prove
essential to distinguish relativistic vs.\ non-relativistic initial 
conditions using the width $\Delta$ of the Gaussian chemical potential 
inhomogeneity in Eq.~(\ref{gaussian}) and the bandgap parameter $M$ in the
post-quench Hamiltonian $H_f$ [Eq.~(\ref{HfDef})].  
Examples of single particle non-relativistic and relativistic propagation are shown
in Fig.~\ref{DiracPlot}.


\subsection{Sine-Gordon quench and ``supersolitons'' \label{SG Def}}

We now consider the continuum limit of the XXZ quench defined by $\Hi$ and $\Hf$ [Eqs.~(\ref{HiDef}) 
and (\ref{HfDef}). This problem was previously analyzed in Ref.~\onlinecite{supersoliton}.
In this section, we provide the 
solution to the sine-Gordon quench and a brief recapitulation of the results found 
in Ref.~\onlinecite{supersoliton}. 
In Sec.~\ref{SG Reg}, we consider the modification of these results due to the presence of irrelevant 
operators (i.e., lattice-scale details left out of the renormalizable continuum field theory).

The massive Dirac continuum limit for the final state Hamiltonian $\Hfc$ was 
derived in the last section, Eqs.~(\ref{HfLowNRG})--(\ref{DiracDef}). 
Since this Hamiltonian is non-interacting, 
we can construct a formal solution to the quench dynamics by solving
the Heisenberg equations of motion for the Dirac spinor $\psi(t,x)$
[c.f.\ Eq.~(\ref{LatDyn})].
The result is 
\begin{subequations}\label{LEEvolve}
\begin{align}
	\psi(t,x)
	=&
	\int_{-t'}^{t'}
	d y
	\,
	\hat{\mathcal{J}}(t',y)
	\,
	\psi(0,x - y),
\end{align}
where
\begin{align}\label{JDef}
	\hat{\mathcal{J}}(t,y)
	&\equiv
	\hat{\mathcal{J}}_0(t,y)
	+
	\hat{\mathcal{J}}_M(t,y),
	\\
	\hat{\mathcal{J}}_0(t,y)
	&=
	\begin{bmatrix}
	\delta(y - t) & 0 \\
	0 & \delta(y + t)
	\end{bmatrix},
	\label{J0Def}
	\\
	\hat{\mathcal{J}}_M(t,y)
	&=
	\begin{bmatrix}
	\bar{G}^{(1)}(t,y) & \bar{G}^{(2)}(t,y) \\
	-\bar{G}^{(2)}(t,y) & \bar{G}^{(3)}(t,y)
	\end{bmatrix}.
	\label{JMDef}
\end{align}
\end{subequations}
Eq.~(\ref{LEEvolve}) is identical to the propagation amplitude 
transcribed in the last section [Eq.~(\ref{DiracPos})], after replacing the single
particle wave function $\Psi_{0}(x)$ with the Schr\"odinger picture field operator
$\psi(0,x)$. 
In these equations, $t' = v_F t$ [Eq.~(\ref{PrimeTimeDef})], while 
the Green's functions $\bar{G}^{(1,2,3)}(t,y)$ 
were defined by Eq.~(\ref{ContGFs}).

The post-quench fermion density is given by
\begin{align}\label{DensityEvolveSG}
	\rho(t,x)
	=
	\int_{-t'}^{t'}
	d y_1 
	\int_{-t'}^{t'}
	d y_2 
	&	
	\left[
	\hat{\mathcal{J}}^\dagger(t',y_1)
	\hat{\mathcal{J}}(t',y_2)
	\right]_r^{\phantom{r} s}
	\nonumber\\
	&
	\times\mathcal{C}^{r}_{\phantom{r} s}(x - y_1, x - y_2),
\end{align}
where all information about the initial state is encoded in the
correlation function 
\begin{equation}\label{ICCorrSG}
	\mathcal{C}^{r}_{\phantom{r} s}(x_1,x_2)
	\equiv
	\gndmuc
	\psi^{r \, \dagger}(0,x_1)
	\psi_{s}(0,x_2)
	\gndmu.
\end{equation}
In Eqs.~(\ref{DensityEvolveSG}) and (\ref{ICCorrSG}), the indices 
$r,s \in \{1,2\}$; 
repeated indices are summed. 

The pre-quench system is described by the ket $\gndmu$, which is
taken as the ground state of the Luttinger liquid Hamiltonian
\begin{align}\label{HiLowNRG}
	\Hic
	= 
	\int
	d x
	&
	\left[
	-
	v_F 
	\,
	\psi^\dagger 
	\left(
	i\,
	\hat{\sigma}^3
	\frac{d}{d x}
	\right)
	\psi
	\right.
	\nonumber\\
	&
	\left.
	\quad
	-
	\mu^{(0)}(x)
	: \psi^\dagger \psi :
	+
	2 \gamma \, v_F
	:
	\psi^\dagger \psi \; \psi^\dagger \psi 
	:
	\right].
\end{align}
Eq.~(\ref{HiLowNRG}) gives the continuum limit of $\Hi$ in Eq.~(\ref{HiDef}),
after discarding all irrelevant operators;
here, $\mu^{(0)}(x)$ represents the long 
wavelength, continuum approximation to the lattice potential $\mu_{i}^{(0)}$.\cite{Ref--Sublat Stag Chem Pot}
The symbol $:\ldots:$ denotes normal-ordering.

Using abelian bosonization rules,\cite{BosonizationRev1,BosonizationRev2,BosonizationRev3}
we rewrite Eq.~(\ref{HiLowNRG}) as
\begin{align}\label{HiLowNRGB}
	\Hic
	&=
	\int 
	d x
	\left[
	\frac{u K}{2}
	\left( 
	\frac{d \phi}{d x}
	\right)^2
	+
	\frac{u}{2 K}
	\left( 
	\frac{d \theta}{d x}
	\right)^2
	- \frac{\mu^{(0)}(x)}{\sqrt{\pi}} 
	\frac{d \theta}{d x}
	\right].
\end{align}	
In our conventions, the fermion current components are
bosonized as
\begin{equation}\label{Currents}
	\left\{J^0, J^1\right\} 
	\equiv
	\left\{\psi^\dagger \psi, \psi^\dagger \hat{\sigma}^3 \psi\right\} 
	=
	\left\{\frac{1}{\sqrt{\pi}}\frac{d \theta}{d x}, \frac{1}{\sqrt{\pi}} \frac{d \phi}{d x}\right\}, 
\end{equation}
and satisfy $[J^0(x), J^1(x')] = - (i/\pi) (d / d x) \delta(x - x')$.
The sound velocity $u$ and the Luttinger parameter $K$ in Eq.~(\ref{HiLowNRGB})
are given by
\begin{align}\label{LuttParamsSG}
	u = \frac{v_F}{K}, \; K = \frac{1}{\sqrt{1 + \frac{4 \gamma}{\pi}}}.
\end{align}
With $v_F = 2$, these agree with the Bethe ansatz results in Eq.~(\ref{u,K Bethe ansatz}) 
only to the first order in $\gamma$. 

When expressed in terms of the boson variables, the post-quench massive Dirac Hamiltonian
in Eq.~(\ref{HfLowNRG}) becomes the sine-Gordon model
\begin{align}\label{HfLowNRGB}
	\Hfc
	=
	v_F
	\int 
	d x
	&
	\left[
	\frac{1}{2}
	\left( 
	\frac{d \phi}{d x}
	\right)^2
	+
	\frac{1}{2}
	\left( 
	\frac{d \theta}{d x}
	\right)^2
	\right.
	\nonumber\\
	&
	\left.
	\quad
	+
	\frac{M}{\pi \alpha}
	\cos\left(\sqrt{4 \pi} \theta \right)
	\right].
\end{align}	
The variable $\alpha$ appearing in the prefactor of the cosine term carries 
units of length, and is formally introduced by 
the bosonization procedure.\cite{BosonizationRev1,BosonizationRev2} 

While $\Hic$ [Eq.~(\ref{HiLowNRGB})] assumes a non-interacting form when expressed in boson
variables, $\Hfc$ becomes the non-linear sine-Gordon theory. By contrast, 
$\Hfc$ [Eq.~(\ref{HfLowNRG})] is non-interacting in terms of the Fermi field $\psi$, 
while $\Hic$ [Eq.~(\ref{HiLowNRG})] incorporates
four fermion interactions. For a quench with $\gamma \neq 0$, there is no common language 
in which both $\Hic$ and $\Hfc$ can be simultaneously expressed as non-interacting Hamiltonians. 
We refer to this generic scenario as the ``interacting'' quench
in the sine-Gordon theory. We reserve the appellation ``non-interacting'' for the exceptional case 
with $\gamma = 0$, where both $\Hic$ and $\Hfc$ are bilinear in fermions.
The post-quench dynamics exhibited for each case are different, as discussed below.

Expressing the fermions in Eq.~(\ref{ICCorrSG}) as vertex operators in the
bosonic language,\cite{BosonizationRev2} the initial state correlation function components evaluate to
\begin{subequations}\label{BosCorrs}
\begin{align} 
	\mathcal{C}^{1}_{\phantom{i} 1}(x_1,x_2)
	=&
	\frac{i c_N \alpha^\sigma}{2\pi}
	\exp\left[
	i \frac{K}{u} \int_{x_1}^{x_2} d y \, \mu^{(0)}(y) 
	\right]
	\nonumber\\
	&\times
	\frac{\sgn(x_1 - x_2)}{|x_1 - x_2|^{\sigma + 1}},
	\label{BosCorrsRight}
	\\
	\mathcal{C}^{2}_{\phantom{i} 2}(x_1,x_2)
	=&
	\frac{-i c_N \alpha^\sigma}{2\pi}
	\exp\left[
	-i \frac{K}{u} \int_{x_1}^{x_2} d y \, \mu^{(0)}(y) 
	\right]
	\nonumber\\
	&\times
	\frac{\sgn(x_1 - x_2)}{|x_1 - x_2|^{\sigma + 1}},
	\\
	\mathcal{C}^{1}_{\phantom{i} 2}(x_1,x_2)
	=&
	\mathcal{C}^{2}_{\phantom{i} 1}(x_1,x_2)
	=
	0 \; (L \rightarrow \infty).
\end{align}
\end{subequations}
In these equations, the external chemical potential manifests in a gauge ``string''
due to the axial anomaly.\cite{QFT,BosonizationRev3}
The coefficient $\alpha$ was introduced in Eq.~(\ref{HfLowNRGB}); 
the parameter $c_N$ is a numerical normalization constant.\cite{Ref--cN} 
The off-diagonal components of 
$\mathcal{C}^{r}_{\phantom{r} s}(x_1,x_2)$ ($r \neq s$) 
vanish in the thermodynamic limit $L \rightarrow \infty$.

The essential character of the initial Luttinger liquid state is encoded in the 
dynamic exponent $\sigma$, defined as
\begin{equation}\label{SigmaDefSG}
	\sigma \equiv \frac{1}{2}\left(\frac{1}{K} + K\right) - 1.
\end{equation}
For the non-interacting quench, $K = 1$ and $\gamma = \sigma = 0$. By contrast,
any $K \neq 1$ ($\gamma \neq 0$) gives $\sigma > 0$.\cite{Ref--SigmaLE}
Eq.~(\ref{BosCorrs}) implies that $\sigma$ is twice the anomalous scaling dimension of $\psi$ 
in the initial LL 
ground state.

Using Eq.~(\ref{BosCorrs}), the integrals appearing in the final expression for the post-quench density expectation in 
Eq.~(\ref{DensityEvolveSG}) are ultraviolet convergent for $0 \leq \sigma < 1$. 
Over this range of initial conditions, we obtain a cutoff-independent prediction
for the post-quench evolution of the number density in the continuum sine Gordon
field theory. Similar expressions with identical convergence properties may be 
obtained for the kinetic and potential energy densities due to the inhomogeneous 
initial state chemical potential.\cite{supersoliton}

The general characteristics of the long-time density dynamics implied by
Eq.~(\ref{DensityEvolveSG}) and (\ref{BosCorrs}) for a generic initial state $\mu^{(0)}(x)$ were
discussed in Ref.~\onlinecite{supersoliton}. In this paper, we restrict our attention
to the waves induced by a localized, Gaussian initial inhomogeneity,
\begin{equation}\label{GaussianSG}
	\frac{K}{\mu}
	\mu^{(0)}(x) 
	= 
	\frac{Q \sqrt{\pi}}{\Delta}
	e^{-x^2/\Delta^2}.
\end{equation}
Combining Eqs.~(\ref{GaussianSG}), (\ref{BosCorrs}), and (\ref{DensityEvolveSG}) 
gives an exact integral expression for the post-quench density expectation evolution
after the sine-Gordon quench. In the long time limit 
$t \gg 1/ v_F M$, 
the requisite integrals 
yield to a systematic asymptotic 
analysis,
as explained in Appendix~\ref{APP--AA}.
One thereby obtains the exact leading asymptotic behavior
\begin{align}\label{supersolitonSG}
	\rho(t,x)
	=&
	\frac{Q}{2\sqrt{\pi}\Delta} e^{-(x - t')^2/\Delta^2}
	\nonumber\\
	&-
	\frac{Q}{2 \Delta}
	\frac{\Gamma(1 - \sigma)}{\Gamma\left(\frac{1 + \sigma}{2}\right)}
	\left[
	\frac{(M \alpha)^2 t'}{\sqrt{2} \Delta}
	\right]^{\sigma/2}
	F_{\sigma}\left(\frac{x - t'}{\Delta}\right)
	\nonumber\\
	&+
	\{x \rightarrow -x\},
\end{align}
where
\begin{equation}\label{FDef}
	F_\sigma(z) \equiv \exp(-z^2/2) D_{\sigma/2}(\sqrt{2} z),
\end{equation}
and $D_\nu(x)$ denotes the parabolic cylinder function.  
In Eq.~(\ref{supersolitonSG}), we have used the explicit expression
for the normalization constant $c_N$.\cite{Ref--cN} Regardless,
for $\sigma > 0$ (interacting quench) the prefactor of the second term 
in Eq.~(\ref{supersolitonSG}) is in some sense arbitrary,
due to the $\alpha$ factor. This ambiguity can be resolved if
a conventional normalization is adopted for the vertex function
correlators in Eq.~(\ref{BosCorrs}).\cite{Ref--CFT} 
The derivation of Eq.~(\ref{supersolitonSG}) is sketched in Appendix~\ref{APP--AA}.

\begin{figure}
\includegraphics[width=0.4\textwidth]{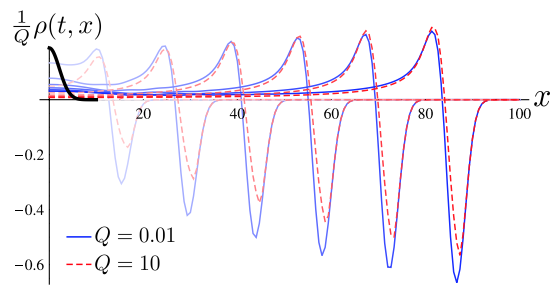}
\caption{
	The right-moving ``supersoliton'' obtained in the interacting sine-Gordon quench.
	The number density evolution after a Luttinger liquid to insulator quench is depicted 
	for a Gaussian initial density profile (heavy black line),  
	with $\sigma=0.7$, $\Delta = 3$, and $M = 15/16$, 
	obtained via numerical integration of the exact bosonization result 
	[Eqs.~(\ref{DensityEvolveSG}) 
	and (\ref{BosCorrs}), using Eq.~(\ref{GaussianSG})]. 
	Time series for two different $Q$ are plotted; 
	the densities are normalized relative to these. 
	The evolution is reflection symmetric about $x = 0$. 
\label{supersolitonSGFig}}
\end{figure}

For the interacting quench with $0 < \sigma < 1$, Eq.~(\ref{supersolitonSG}) 
describes the propagation of right and left moving ``supersolitons''
launched from the Gaussian initial condition, in the long time limit.
A right-moving supersoliton is depicted in Fig.~\ref{supersolitonSGFig}.
From the equation, it is evident that the supersoliton does not disperse.
In the long-time limit, the response to the initial chemical potential
(and thus the initial state density inhomogeneity) is linear, regardless
of the strength of $Q$ in Eq.~(\ref{GaussianSG}). The supersoliton
features an amplitude that grows in time as power law, with growth
exponent $\sigma/2$; subleading terms neglected in Eq.~(\ref{supersolitonSG})
decay for $\sigma > 0$.
The peculiar shape of the supersoliton implied by Eq.~(\ref{FDef}) obtains
because the quench kernel effectively takes a fractional derivative $(d / d x)^{\sigma/2}$
of the input profile.\cite{supersoliton} 
The total number fluctuation
induced by the inhomogeneity is conserved by the supersoliton, since the second term
in Eq.~(\ref{supersolitonSG}) integrates to zero over $x \in \mathbb{R}$.\cite{Ref--Causality}
Finally, we note that Eq.~(\ref{supersolitonSG}) holds for generic $M \Delta$: the supersoliton 
arises for both ``relativistic'' ($M \Delta \gg 1$) and ``non-relativistic'' 
($M \Delta \ll 1$) initial density profiles (c.f.\ Sec.~\ref{WavepacketDyn}).

\begin{figure}[b]
\includegraphics[width=0.4\textwidth]{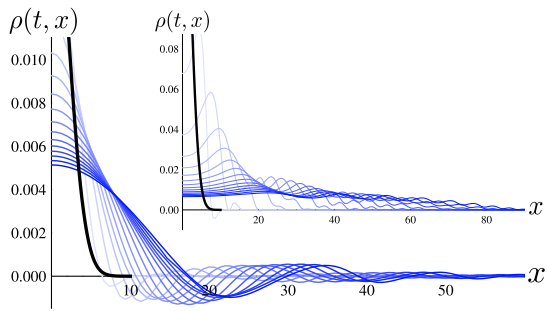}
\caption{
The number density evolution as in Fig.~\ref{supersolitonSGFig},
but for the non-interacting quench $K = 1$ ($\sigma = 0$). 
The initial bump (heavy black line) has area $Q = 0.1$ in the main figure and $Q=1$ in the inset;
in both cases $\Delta = 3$ and $M = 15/16$. The evolution is reflection symmetric about $x = 0$. Now there is no fractionalization of the 
initial LL quasiparticles with respect to the insulator and, consequently, the dynamics
are simply dispersive with no supersolitons or inhomogeneity growth.
\label{nonintqSGFig}}
\end{figure}

By contrast, the non-interacting quench with $\sigma = 0$ exhibits no amplification.
For $M \Delta \gg 1$ (``non-relativistic'' initial condition), one finds simple dispersive broadening,
qualitatively similar to the single particle wavepacket spreading in Fig.~\ref{DiracPlot}(a).
Examples of non-interacting quenches with 
non-relativistic initial conditions are shown in 
Fig.~\ref{nonintqSGFig}. For the non-interacting quench, the response is given 
entirely by terms neglected as subleading (for $\sigma > 0$) in Eq.~(\ref{supersolitonSG});
indeed, the right-hand side of this expression vanishes for $\sigma = 0$.

Non-interacting quenches with ``relativistic'' ($M \Delta \ll 1$)
initial conditions exhibit a different behavior, qualitatively similar to the single particle wavepacket 
evolution depicted in Fig.~\ref{DiracPlot}(b): the initial Gaussian density bump
blows apart into left and right-moving wave trains, with leading edges that 
rip
along lightcone. 
In this sense, the non-interacting quench with $M \Delta \ll 1$ behaves
similar to the supersoliton, which also propagates relativistically. 
At short timescales $t \lesssim 1/v_F M$, the interacting and non-interacting quenches in fact
exhibit qualitatively similar dynamics 
for relativistic initial conditions.
However, the non-interacting quench evolution shows no amplification in the long time limit, and the 
generated wave train exhibits no static, non-dispersing structure.
In the numerical results for the XXZ chain quench presented in Sec.~\ref{Results}, we will restrict our
attention to non-relativistic initial conditions, in order to avoid possibly confusing 
the supersoliton with the trivial (and essentially single particle) 
effect of squeezing the initial density wavepacket to a width narrower 
than the Compton wavelength.

Finally, we note that setting $M = 0$ in Eq.~(\ref{supersolitonSG}) gives the result appropriate
to a Luttinger liquid to Luttinger liquid quench--the initial density disturbance is merely
propagated along the light cone without dispersion, as expected for dynamics generated by
a critical state. This case was previously considered in Ref.~\onlinecite{LancasterMitra10}.

The physics of the interacting and non-interacting quenches are fundamentally distinguished by 
the advent of quasiparticle \emph{fractionalization} in the interacting case.
The interacting nature of the pre-quench initial ground state $\gndmu$ 
\emph{relative to the post-quench Hamiltonian} $\Hfc$ is implied by Eq.~(\ref{HiLowNRG}) with $\gamma \neq 0$, 
which expresses $\Hic$ in terms of the ``final state'' fermion $\psi$. The presence of interparticle 
interactions in $\Hic$ means the fermion $\psi$ is not a ``natural'' propagating degree of freedom 
in the initial state Luttinger liquid.\cite{Ref--PrimaryField} 
Equivalently, the ``quasiparticles'' of the LL carry a fraction $\sqrt{K}$ of the $\psi$ fermion
number charge; we say that the initial LL state is \emph{fractionalized} with respect
to the final band insulating state. This notion is made explicit in Appendix~\ref{APP--FracSec}.
Fractionalization due to the presence of interparticle interactions is a ubiquitous 
feature in 1D, responsible e.g.\ for spin-charge separation in quantum wires.\cite{BosonizationRev2,BosonizationRev3}

We interpret the supersoliton and the amplification effect that 
arises
in the sine-Gordon quench
for the interacting case as a \emph{spectroscopy} of the initial LL state.\cite{supersoliton} 
The key ingredients are quasiparticle fractionalization of the initial state relative to the
excitation spectrum of the post-quench Hamiltonian, and fact that the post-quench
Hamiltonian is gapped. 
In the next section and in Appendix \ref{APP--DistInhomog}, we show that fractionalization leads to an anomalous momentum
dependence in the Wigner distribution function of the excited post-quench quasiparticles, 
due to the inhomogeneity. The low-energy dispersion of the gapped final state characterized
by $M$ translates this into a divergent velocity distribution, giving rise to the 
supersoliton. By contrast, for a non-relativistic density profile with $M \Delta \gg 1$,
Pauli-blocking suppresses the excitation of large velocities in the non-interacting quench.
The distinction arises due to the long-distance behavior of correlations in the initial state,
and is not destroyed by lattice effects. At the same time, we will see in Sec.~\ref{SG Reg} 
that the advent of the lattice does modify the post-quench dynamics, but in way that can
be parametrically controlled by the system size.


\subsection{Quasiparticle distribution functions: continuum and lattice quenches \label{DistFuncs}}

We consider the post-quench distribution of \emph{final state} quasiparticles,
in the sine-Gordon and lattice quenches. 
Time evolution is generated by the final state Hamiltonian, 
which is translationally invariant and non-interacting in terms of band fermions
for both the lattice [$\Hf$, Eq.~(\ref{HfDef})] and continuum [$\Hfc$, Eq.~(\ref{HfLowNRG})] 
theories. 
The global momentum
distribution of excited quasiparticles induced by the quench
in each case constitutes a static quantity, which does not
encode information about the density inhomogeneity. 
The physics of the supersoliton resides in the local Wigner function,
which is discussed subsequently.

\subsubsection{Global distribution function}\label{GlobalDist}

We consider first the continuum sine-Gordon quench. The final state Hamiltonian
$\Hfc$ in Eq.~(\ref{HfLowNRG}) and (\ref{Diracham}) can be rewritten as
\begin{align}\label{HfLowNRGPH}
	\Hfc 
	= \int \frac{d k}{2 \pi} \, \varepsilon_k 
	\left[a_k^\dagger a_k + b_k^\dagger b_k\right],
\end{align}
where $a_k$ ($b_k$) annihilates a particle (hole) with momentum $k$,
and
\begin{equation}\label{DiracSpec}
	\varepsilon_k = 
	v_F 
	\sqrt{k^2 + M^2}.
\end{equation}
The particle and hole operators are related to $\psi$ via
\begin{align}\label{Psitoab}
	\begin{bmatrix}
	\psi_1(k)\\
	\psi_2(k)
	\end{bmatrix}
	=&
	\frac{a_k}{\sqrt{1 + s^2(k)}}
	\begin{bmatrix}
	1\\
	i s(k)
	\end{bmatrix}
	\nonumber\\
	&+
	\frac{b_{-k}^\dagger}{\sqrt{1 + s^2(-k)}}
	\begin{bmatrix}
	1\\
	- i s(-k)
	\end{bmatrix},
\end{align}
where 
\[
	s(k) \equiv
	\frac{\varepsilon_k - v_F k}{v_F M}.
\]
We define the occupation numbers
\begin{align}\label{DistDefSG}
	\begin{aligned}
	n_{+}(k) &\equiv \gndmuc a_k^\dagger a_k \gndmu,
	\\
	n_{-}(k) &\equiv \gndmuc b_k^\dagger b_k \gndmu,
	\end{aligned}
\end{align}
in which $\gndmu$ denotes the ground state of the pre-quench
Hamiltonian $\Hic$, Eqs.~(\ref{HiLowNRG}) and (\ref{HiLowNRGB}).

For the translationally invariant case $\mu^{(0)}(x) = 0$,
one can show that
\begin{equation}\label{OccNumSG}
	n_{+}(k)
	=
	n_{-}(k)
	=
	\frac{1}{2}
	+
	\frac{k}{\varepsilon_k}
	\mathcal{F}(k),
\end{equation}
where
\begin{equation}\label{FDefSG}	
	\mathcal{F}(k)
	=
	\gndmuc \psi^{1 \, \dagger}(k) \psi_1(k) \gndmu
	-\frac{1}{2}
\end{equation}
and 
$\mathcal{F}(-k) = - \mathcal{F}(k)$.
The form of Eqs.~(\ref{OccNumSG}) and (\ref{FDefSG}) follows from 
Eq.~(\ref{Psitoab}) and the imposition of the sum rule (canonical anticommutation
relations) upon correlators of the fermion components $\psi_{i}(k)$.

In the case of the non-interacting quench ($\sigma = 0$), one
finds 
$\mathcal{F}(k) = -(1/2) \sgn(k)$, 
so that
\begin{equation}\label{DistNI}
	n_{\pm}(k)
	=
	\frac{1}{2}
	\left(
	1 - \frac{|k|}{\sqrt{k^2 + M^2}}
	\right).
\end{equation}
This occupancy factor peaks to a value of one-half at $k = 0$, and
decays as $M^2/k^2$ for $|k| \gg M$. The density of particles or holes excited by the quench
is thus ultraviolet finite and equal to $M/2 \pi$. The associated kinetic energy density
is given by the difference of the pre- and post-quench Hamiltonian zero point energy densities, 
and is logarithmically divergent. 

The calculation for the interacting case is more subtle, 
due to an ultraviolet divergence.
We must compute
\begin{align}\label{FDefI}
	\gndmuc \psi^{1\,\dagger}(k) \psi_1(k) \gndmu
	&=
	\frac{c_N \alpha^\sigma}{2 \pi}
	\int_{-\infty}^{\infty}
	d x \, 
	\frac{e^{i k x}(\alpha + i x)}{\left( \alpha^2 + x^2 \right)^{1 + \sigma/2}},
\end{align}
which is the Fourier transform of the initial state correlator in Eq.~(\ref{BosCorrsRight})
with $\mu^{(0)} = 0$, retaining the soft 
cutoff $\alpha$.\cite{BosonizationRev2}
The prefactor $c_N$ appears explicitly in Ref.~\onlinecite{Ref--cN}. 
Performing an expansion in $k \alpha$ and extracting 
$\mathcal{F}(k)$, 
we finally obtain
\begin{align}\label{DistI}
	n_{\pm}(k)
	=&
	\frac{1}{2}
	\left[
	1
	-
	\frac{|k|}{\sqrt{k^2 + M^2}}
	\frac{\Gamma\left(\frac{1 - \sigma}{2}\right)}{\Gamma\left(\frac{1 + \sigma}{2}\right)}
	\left(
	\frac{|k| \alpha}{2}
	\right)^\sigma
	\right]
	\nonumber\\
	&+ \ord{|k|\alpha},
\end{align}
valid for $0 \leq \sigma < 1$.
Eq.~(\ref{DistI}) holds only for $|k| \alpha$ small, where the second term on the right-hand
side trails the first. At such wavevectors, $n_{\pm}$ is enhanced relative to the
non-interacting case in Eq.~(\ref{DistNI}), indicating that the interacting quench induces
a stronger excitation of the post-quench quasiparticles. Clearly Eq.~(\ref{DistI}) becomes
unphysical for sufficiently large $|k|$; 
the global distribution function cannot be uniquely
defined (i.e., its value will depend upon the regularization procedure) in the continuum,
interacting sine-Gordon quench. 

\begin{figure}[t]
   \includegraphics[width=0.48\textwidth]{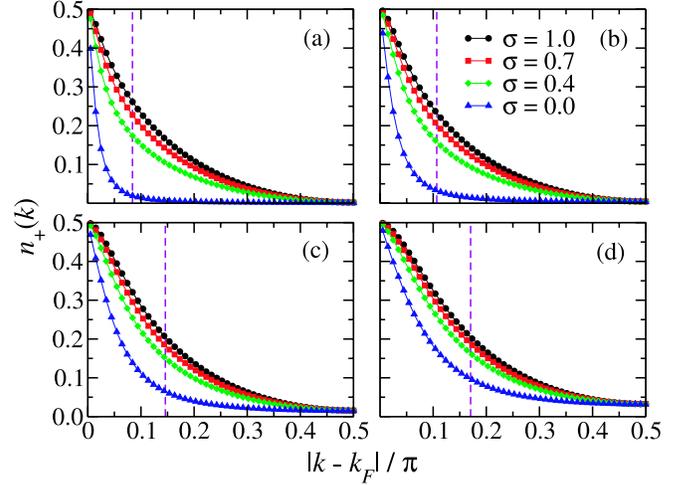}
   \caption{Occupancy of momentum states in the upper (conduction) band
	after the lattice XXZ quench (obtained by DMRG),
	for the four values of the dynamical exponent $\sigma$ given in the legend.
	The occupancies are plotted for four different choices of the band-gap parameter,
	$M=$ 3/40 (a), 1/8 (b), 1/4 (c), and 3/8 (d).
	In each subplot, the dashed vertical line marks the wavenumber 
	$k_{\rm max}(M)$
	at which the 
	final state Hamiltonian band group velocity is maximized, 
	$v(k_{\rm max};M) = \textrm{max}\left[ dE_k/d k \right] \equiv v_{\rm max}(M)$;
	see Sec.~\ref{VeloSec}, Eqs.~(\ref{vmax_k}) and (\ref{kstar}).
	}
   \label{fig:nk}
\end{figure}

In Fig.~\ref{fig:nk}, we exhibit $n_{+}(k)$ for the lattice quench in a finite
size system of 
202 sites, obtained via numerical density matrix renormalization group calculations
(see Sec.~\ref{Results} for details).
The occupancy is defined as in Eq.~(\ref{DistDefSG}), except that the continuum
state $\gndmu$ is replaced by $\gnd$,
the ground state of $\Hi$ in Eq.~(\ref{HiDef}); $a_k$ now denotes the lattice conduction band annihilation operator.

\subsubsection{Wigner function, fractionalization, and the origin of the supersoliton}\label{WDSec}

Eqs.~(\ref{DistNI}) and (\ref{DistI}) have been calculated for the case of the homogeneous quench.
In the infinite system size limit, these equations also
apply in the presence of an arbitrary initial state chemical potential $\mu^{(0)}(x)$ that vanishes
faster than $1/x$ as $|x| \rightarrow \infty$. The effects of the inhomogeneous 
density profile in the initial state can be extracted from the ``local'' (Wigner) distribution function.\cite{LP}

The main idea is conveyed by the ground state
Wigner function for the right-moving fermion $\psi_{1}(x) \equiv \rf(x)$
in the inhomogeneous Luttinger liquid, defined as
\begin{align}\label{WDDef}
	\nr(k;R)
	\equiv&
	\int
	d x_d \,
	e^{-i k x_d}
	\gndmuc
	\rf^\dagger\left(R - {\textstyle{\frac{x_d}{2}}}\right)
	\rf\left(R + {\textstyle{\frac{x_d}{2}}}\right)
	\gndmu
	\nonumber\\
	=&
	\int
	\frac{d q}{2\pi} \,
	e^{i q R}
	\gndmuc
	\tilde{\rf}^\dagger\left(k - {\textstyle{\frac{q}{2}}}\right)
	\tilde{\rf}\left(k + {\textstyle{\frac{q}{2}}}\right)
	\gndmu.
\end{align}
Here $\gndmu$ is the ground state of $\Hic$ in Eq.~(\ref{HiLowNRG}),
which has the density expectation $\gndmuc \rf^\dagger(x) \rf(x) \gndmu = \rho_0(x)/2$, where 
\begin{equation}\label{rho0Def}
	\rho_{0}(x) \equiv \frac{K}{\pi u} \mu^{(0)}(x).
\end{equation}
The global momentum distribution and the position space density 
expectation value can both be extracted from the Wigner function
(Appendix~\ref{APP--DistInhomog}). Although we employ it here
to gain intuition about the local momentum profile induced by 
$\rho_0(x)$, strictly speaking $n_{\mathcal{R}}(k;R)$ cannot be
interpreted as a probability distribution\cite{Gutzwiller} (it can take negative 
values), because momentum and position are canonically conjugate quantum
observables.

We let $\delta\nr(k;R)$ denote the linear response due
to $\rho_{0}$(x), after subtracting off the homogeneous
(global) background. 
Using the correlation function in Eq.~(\ref{BosCorrsRight}),
we find
\begin{align}
	\delta\nr&(k;R)
	\nonumber\\
	=&
	c_N \alpha^\sigma
	\begin{aligned}[t]
	\int&
	\frac{d q}{2\pi}
	\frac{\tilde{\rho_0}(q)}{q}
	\,e^{i q R}
	\\&
	\times
	\left[
	\begin{aligned}
	&
	\sgn\left(k + \frac{q}{2}\right)
	\mathcal{G}_{\sigma}\left(\left|k + \frac{q}{2}\right|;\zeta\right)
	\\&
	-
	\sgn\left(k - \frac{q}{2}\right)
	\mathcal{G}_{\sigma}\left(\left|k - \frac{q}{2}\right|;\zeta\right)
	\end{aligned}
	\right],
	\end{aligned}
	\label{WD}
\end{align}
where
\begin{align}\label{GDef}
	\mathcal{G}_\sigma(|p|;\zeta)
	\equiv
	\int_{0}^{\infty}
	\frac{d y}{y}
	\frac{1}{
	\left(y^2 + \zeta^2\right)^{\sigma/2}
	}
	\sin\left(|p| y \right).
\end{align}
The parameter $\zeta$ is an ultraviolet regularization length, introduced
here for later use in the context of the lattice quench defined
in Sec.~\ref{Lat model} (see Sec.~\ref{SG Reg}). The pure continuum 
theory has $\zeta = 0$, for which Eq.~(\ref{GDef}) is convergent
when $0 \leq \sigma < 1$.

For the special case of a non-interacting Fermi gas ($\sigma = 0$), 
$\mathcal{G}_{0}(|p|;\zeta) = \pi/2$, \emph{independent} of $|p|$, so that 
\begin{align}
	\delta\nr(k;R)
	=&
	\frac{\pi}{2}
	\begin{aligned}[t]
	\int&
	\frac{e^{i q R} \,d q}{2\pi}
	\frac{\tilde{\rho_0}(q)}{q}
	\\&
	\times
	\left[
	\sgn\left(k + \frac{q}{2}\right)
	-
	\sgn\left(k - \frac{q}{2}\right)
	\right].
	\end{aligned}
	\label{WDNI}
\end{align}
This expression vanishes for $|q| < 2 |k|$, 
i.e.\ unless the creation and annihilation operators in Eq.~(\ref{WDDef})
carry momentum with opposite signs. This is a simple consequence
of Pauli blocking. The result can be understood via perturbation theory: let 
$\gndmu \equiv \gndmu_0 + \ket{\delta}$; $\gndmu_{0}$ denotes the homogeneous
vacuum, while $\ket{\delta}$ gives the response to $\rho_0(x)$.
To first order in $\ket{\delta}$, Eq.~(\ref{WDDef}) vanishes unless the product
$\tilde{\rf}^\dagger\left(k - {\textstyle{\frac{q}{2}}}\right)
\tilde{\rf}\left(k + {\textstyle{\frac{q}{2}}}\right)$ creates
a particle-hole pair in $\gndmu_0$ (acting to the left or to the right);
Eq.~(\ref{WDNI}) obtains from the overlap between this state and $\ket{\delta}$.

We consider a $\rho_0(x)$ localized in position space, of characteristic width $\Delta$. 
For $|k| \gg 1/\Delta$, Eq.~(\ref{WDNI}) implies that the $k$-dependence of the Wigner function 
is slaved to follow that of $\tilde{\rho_0}(q)$, with $q \sim 2 k$. For a Gaussian inhomogeneity, this 
means a Gaussian fall off of the Wigner function in $k$. No matter how wide or narrow the initial 
packet is made, the large-$k$ asymptotic is always strongly suppressed.

By contrast, the situation for the interacting Luttinger liquid is quite different,
Eq.~(\ref{WD}) with $\sigma > 0$. Then, the kernel $\mathcal{G}_\sigma(|p|;\zeta)$ depends upon $|p|$, 
and allows a contribution to $\delta\nr(k;R)$ from $|q| < 2 |k|$, violating the Pauli blocking 
condition. For $|k| \gg 1/\Delta$, the dominant contribution 
comes from $|q| \ll 2 |k|$, and the $q$-integration gives $\delta\nr(k;R) \propto \rho(R)$.
The $k$-dependence comes entirely from $d \mathcal{G}_\sigma(|k|;\zeta)/ d |k|$, and is 
independent of the initial inhomogeneity profile:
the $R$- and $k$-dependencies of the Wigner function \emph{factorize}.
For $0 < \sigma < 1$ and $\zeta = 0$, the leading term in the $|k| \gg 1/\Delta$ 
limit of Eq.~(\ref{WD}) goes as
\begin{align}\label{WDInt}
	\delta\nr(k;R)
	\sim
	c(\sigma) \,
	\alpha^\sigma |k|^{\sigma - 1}
	\rho_0(R),
\end{align}
where the prefactor satisfies $c(0) = 0$. 

In the sine-Gordon quench, the initial ``momentum distribution'' implied by Eq.~(\ref{WD}) or
(\ref{WDInt}) translates into a ``velocity distribution'' through the massive post-quench dispersion
relation in
Eq.~(\ref{DiracSpec});
details are presented in Appendix~\ref{APP--DistInhomog}.
In the non-interacting case, for a non-relativistic
initial condition with $M \Delta \gg 1$, only small velocities $v \lesssim v_F/M \Delta$
are excited 
[Eq.~(\ref{VelocDistNI}) in Appendix~\ref{APP--DistInhomog}].
By contrast, an interacting quench with Eq.~(\ref{WDInt}) induces 
a non-integrable divergence in the velocity distribution at the ``speed of light''
$v = v_F$
[Eq.~(\ref{VelocDistI1})],
signaling the presence of the non-dispersing supersoliton
(recall that velocity is conserved by the post-quench Hamiltonian).
Thus, the supersoliton arises due to the particular way in which quasiparticle
fractionalization evades Pauli blocking in the initial LL ground state. 

To gain further insight,
consider a many-particle product state in a relativistic quantum theory,
e.g.\ $N$ particles are placed into $N$ plane wave states
with momenta $\{k_{i}\}$. In the thermodynamic limit $N \rightarrow \infty$,
the system is described by a continuous distribution function $n(k)$, a 
well-defined (classical) observable for a product state.
Suppose further that
the corresponding velocity distribution $n(v)$  
exhibits a  delta-function-like singularity at $v=v_F$. 
Then, a fraction of the density (determined by the weight of the singularity) 
at each point in space is translated at the speed of light. In particular, a 
fraction of any initial density inhomogeneity will propagate at $v=v_F$ without dispersion.

Even though this intuitive velocity distribution picture helps
reveal
the physical origin of the supersoliton, it is not entirely satisfactory. 
For example, a finite number of massive particles traveling at the speed of light 
implies an infinite kinetic energy.
By contrast, although it propagates ultrarelativistically,
the supersoliton carries finite total energy beyond that
induced by the homogeneous quench.\cite{supersoliton} The initial
(inhomogeneous) Luttinger liquid is very far from a product state
of the post-quench spectrum;
indeed, the appearance of the exponent $\sigma$ in 
Eqs.~(\ref{DistI}) and (\ref{WDInt}) indicates that quantum coherence
(entanglement) plays a dominant role in the fractionalized
density dynamics of the quench. 
Further,
the Wigner function $\delta n(k;R)$
is not really a phase space distribution function,\cite{Gutzwiller} 
and the non-integrable velocity singularity in Eq.~(\ref{VelocDistI1}),
Appendix~\ref{APP--DistInhomog} does not imply an extensive 
mass or energy flow. Instead, we interpret this divergence as signaling
the supersoliton, an emergent, collective excitation of the post-quench 
non-equilibrium state that travels with velocity $v_F$. 

Let us also note that
the Wigner 
distribution post-quench is not a static object; the spatially-varying 
momentum profile in Eq.~(\ref{WDInt}) implies that the shape of the 
density wave can evolve. In the unregularized sine-Gordon quench, this is 
the amplification effect exhibited in Fig.~\ref{supersolitonSGFig} and Eq.~(\ref{supersolitonSG}).

The ultraviolet effects induced by the presence of a lattice cannot alter the
fundamental distinction between interacting and non-interacting quenches, 
because the long-distance behavior of $\gndmuc \rf^\dagger(0) \rf(x) \gndmu$ determines 
the efficacy of Pauli blocking in Eq.~(\ref{WD}) through Eq.~(\ref{GDef}). 
In the next section, we will nevertheless see that the modification of the short-distance
structure of the theory [e.g., $\zeta > 0$ in Eq.~(\ref{GDef})] 
does influence the post-quench dynamics. 

Further details about the post-quench Wigner function can be found in 
Appendix~\ref{APP--DistInhomog}, where explicit formulae are given for
the associated velocity distributions in the non-interacting and interacting quenches,
incorporating the effects of ultraviolet regularization.


\subsection{Irrelevant operators and UV regularization \label{SG Reg}}

So far, we have focused primarily on the Luttinger liquid to band insulator
quench in the continuum sine-Gordon model.
For a Gaussian initial density bump, the leading asymptotic result for the long-time
limit [Eq.~(\ref{supersolitonSG})] predicts the emergence of the supersoliton:
a non-dispersive, relativistically-propagating density wave with an amplitude
that grows as $t^{\sigma/2}$. This result applies
to the integrable sine-Gordon field theory, in the absence of additional 
perturbations. 
We have considered only a 
particular case by assuming the
non-interacting post-quench 
Hamiltonian in Eq.~(\ref{HfLowNRG}). This corresponds to the special Luther-Emery (LE) point 
in the sine-Gordon phase diagram.\cite{BosonizationRev2,BosonizationRev3,supersoliton} 
Away from this point, $\Hfc$ would acquire a four-fermion interaction as in 
Eq.~(\ref{HiLowNRG}); 
bosonization links the sine-Gordon and massive Thirring models 
in the general case.\cite{QFT} 
We postpone a discussion of the effects of interparticle collisions in the post-quench evolution 
until the end of Sec.~\ref{Conc}.

The sine-Gordon field theory 
arises as the low-energy description of many 1D 
solid state
and cold atomic systems,\cite{BosonizationRev2,Cazalilla04,Cirac10}
including
the XXZ chain quench introduced in Sec.~\ref{Lat model} between $\Hi$ and $\Hf$ in Eq.~(\ref{ham}).
Details of the original ``microscopic'' formulation are expected to 
appear in the effective low-energy field theory as irrelevant operators.\cite{RG}
Irrelevant operators typically exert a negligible effect
upon low-energy, long-wavelength properties in a zero temperature field theory.
Finiteness of correlation functions (up to logarithmic divergences subsumed by renormalization) 
and insensitivity to irrelevant operators go hand-in-hand.\cite{RG,QFT}
  
By contrast, the influence of irrelevant operators upon the strong non-equilibrium dynamics 
generated by a sudden quench remains largely unexplored 
territory. The incorporation of generic perturbations destroys some special
properties that may be enjoyed by a given renormalizable theory, 
such as conformal invariance or, in the case of the 1D sine-Gordon model, integrability. 
On general grounds, a non-integrable many-body 
system prepared in an initial, non-thermal state is expected to thermalize (presumably 
due to quantum chaotic dynamics) in the long time limit.  

In this paper, we do not mount a broad assault
upon the important topics of 
integrability-breaking perturbations and thermalization. 
Even in equilibrium, the impact of irrelevant operators and integrability 
on correlation functions \emph{at non-zero temperature}  
remains a contentious 
issue.\cite{DBT1,DBT2,DBT3,DBT4,DBT5,DBTRev,DamleSachdev98,Fujimoto99,Altshuler06,Langer09}
Here, we limit our focus to the post-quench wave train dynamics exemplified by the supersoliton. 
In particular, we would like to understand how 
irrelevant operators, or equivalently, lattice scale details and the presence of 
a finite ultraviolet cutoff, modify or suppress the supersoliton. 
Our considerations in this section will be used to interpret the numerical 
results for the XXZ chain quench presented in Sec.~\ref{Results}.

\subsubsection{Irrelevant operators: some examples}

The XXZ chain quench introduced in Sec.~\ref{Lat model} takes the ground state 
$\gnd$
of the XY phase Hamiltonian $\Hi$ in Eq.~(\ref{HiDef}), and evolves this state 
forward in time using the gapped band insulator Hamiltonian $\Hf$ defined via Eq.~(\ref{HfDef}).
In the continuum field theory limit, lattice microscopics induce the addition of 
irrelevant operators to the sine-Gordon model Hamiltonians $\Hic$ and $\Hfc$
[Eqs.~(\ref{HiLowNRG}), (\ref{HiLowNRGB}), and (\ref{HfLowNRG}), (\ref{HfLowNRGB})]. 
We now enumerate a few examples.

The \emph{least} irrelevant operators $\{\mathcal{O}_i(x)\}$ invariant 
under continuum versions of all lattice symmetries 
(time-reversal, parity, 
lattice translational invariance) 
carry the scaling dimension $x_{i} = 4$ when added to the non-interacting 
Dirac Hamiltonian in either Eq.~(\ref{HiLowNRG}) (with $\gamma = 0$) or (\ref{HfLowNRG}).
We consider first the umklapp interaction operator\cite{BosonizationRev1}
\begin{align}\label{umklapp}
	\mathcal{O}_{u}(x) 
	&\equiv 2 
	\left[
	\left(\psi_1^\dagger \psi_2\right)^2
	+
	\left(\psi_2^\dagger \psi_1\right)^2
	\right]
	\nonumber\\
	&=
	-\frac{1}{\left(\pi\alpha\right)^2}
	\cos\left[2\sqrt{4 \pi} \theta \right].
\end{align}
This operator appears as a lattice-induced modification (via $\Hi$)
of the initial Luttinger liquid Hamiltonian $\Hic$, Eq.~(\ref{HiLowNRG})
or (\ref{HiLowNRGB}).\cite{BosonizationRev1} The dimension of the umklapp operator is $x_1 = 4 K$, so
that the associated coupling constant has dimension $y_1 = 2(1 - 2K)$, where $K$ denotes 
the Bethe ansatz Luttinger parameter in Eq.~(\ref{K Bethe ansatz}).\cite{RG} Thus the umklapp operator
has $y_1(K = 1) = -2$ at the free fermion point, while $y_1(K = 1/2) = 0$ at the
threshold of the Ising antiferromagnetic order [Eq.~(\ref{K Bethe ansatz}) 
with $\gamma \rightarrow 1$]. In our lattice quenches, we will focus
upon $\gamma < 0$, so that $K > 1$ and umklapps are strongly irrelevant.

As a second example, we consider the effect of band curvature (at half-filling), which
gives the operator
\begin{align}\label{curvature}
	\mathcal{O}_{3}(x)
	\equiv& 
	- \psi^\dagger
	\left(
	i \hat{\sigma}^3 \frac{d^3}{d x^3} 
	\right)
	\psi
	\nonumber\\
	=&
	-\frac{1}{4}
	\left[
	\left(\frac{d^2 \phi}{d x^2}\right)^2
	+
	\left(\frac{d^2 \theta}{d x^2}\right)^2
	\right]
	\nonumber\\
	&
	-
	\frac{\pi}{2^3}
	:
	\left[
	\left(
	\frac{d \phi}{d x} 
	+
	\frac{d \theta}{d x} 
	\right)^4
	+
	\left(
	\frac{d \phi}{d x} 
	-
	\frac{d \theta}{d x} 
	\right)^4
	\right]
	:
\end{align}
$\mathcal{O}_3$
arises as a modification of both $\Hic$ and $\Hfc$,
due to the cosine dispersion of the lattice model.

We note that while the band curvature operator in Eq.~(\ref{curvature}) 
is bilinear in terms of fermions, both the umklapp and band
curvature operators induce interparticle interactions in the
boson language. This complication makes it difficult to determine
the influence of either upon the interacting LL initial state $\gndmu$.

One can in principle treat non-bilinear irrelevant operators 
perturbatively, but several difficulties
arise in attempting to account for their effects.
First, the perturbation theory is badly ultraviolet divergent,
and depends upon the way in which these divergences are regularized.
A second, more serious (but intimately related) problem arises
because the effects of irrelevant operators become strong at
short distances. In the context of the quench, the goal is
to construct the initial ground state correlator in
Eq.~(\ref{ICCorrSG}), accounting for the effects of lattice scale
details. These details should translate into a modification of the
ideal Luttinger liquid correlation functions in Eq.~(\ref{BosCorrs})
at short distances. However, the effects of irrelevant operators
become strong in precisely this limit; the result is that perturbation
theory breaks down, and a systematic accounting is only possible
via an exact or approximate non-perturbative resummation. 

Nevertheless, we show that a regularized version of the continuum
sine-Gordon quench can be constructed which gives a reasonably
good match to our finite system size numerics presented in Sec.~\ref{Results}. 
To motivate the regularization scheme that we employ, we consider
the effect of a finite-ranged density-density interaction.\cite{BosonizationRev2} 
Instead of Eq.~(\ref{HiLowNRGB}), one has the Hamiltonian
\begin{align}\label{finite range int}
	\Hic
	=&
	\frac{1}{2}
	\int d x \,
	\left[
	\left(\frac{d \phi}{d x}\right)^2
	+
	\left(\frac{d \theta}{d x}\right)^2
	\right]
	\nonumber\\
	&+
	\frac{1}{2 \pi}
	\int d x \, d x' \,
	v(x - x')\;
	\frac{d \theta}{d x}(x) 
	\frac{d \theta}{d x'}(x')
\end{align}
where 
$v(x) \equiv (2 \gamma/\zeta) \exp(-|x|/\zeta)$ 
is a Yukawa-type potential
that integrates to $4 \gamma$, regardless of the range $\zeta$. The limit
$\zeta \rightarrow 0$ gives the purely local interaction implemented in 
Eq.~(\ref{HiLowNRGB}). 

Unlike the non-bilinear boson operators associated with 
umklapp and band curvature effects discussed above,
the finite-range interaction in Eq.~(\ref{finite range int}) can be
treated non-perturbatively. The result is a modification of the
Luttinger liquid correlator in Eq.~(\ref{BosCorrs});
in the homogeneous limit with
$\mu^{(0)}(x) = 0$, one obtains
\begin{align}\label{BosCorrsFR}
	\mathcal{C}^{1}_{\phantom{i} 1}(x,0)
	=&
	\frac{i c_N}{2\pi}
	\frac{\sgn(x)}{|x|}
	\mathfrak{C}(x),
\end{align}
where
\begin{align}\label{FRCorrDef}
	\mathfrak{C}(x)
	=&
	\exp\left\{
	\int_{0}^{\infty} \frac{d q}{q} \sigma(q) 
	\left[
	\cos\left( q x \right) 
	-1
	\right]
	\right\}
	\nonumber\\
	\sim&
	\left[
	\frac{\zeta^2}{\zeta^2 + \beta \, x^2}
	\right]^{\sigma(0)/2}.
\end{align}
In this equation,
\begin{subequations}\label{FRCorrParam}
\begin{align}
	\sigma(q)
	&\equiv
	\frac{1}{2}
	\left[
	K(q)
	+
	K^{-1}(q)
	\right]
	-1,
	\\
	K(q)
	&\equiv
	\left[
	1 
	+ 
	\frac{4 \gamma}{\pi}
	\frac{1}{1 + (\zeta q)^2}
	\right]^{-1/2}.
\end{align}
\end{subequations}
The variable $\beta$ in Eq.~(\ref{FRCorrDef}) is some numerical constant.
The effect of a finite interaction range $\zeta > 0$ is to reduce the short-range 
scaling behavior ($|x| \lesssim \zeta$) of the $\psi$ fermion LL correlation functions 
in Eq.~(\ref{BosCorrs}) to that of free fermions.

\subsubsection{Regularized sine-Gordon theory \label{SGRegTheory}}

A systematic approach to incorporating lattice scale details into
the sine-Gordon quench would require the inclusion of all irrelevant operators
with a given scaling dimension, say. This task is made difficult by the
interacting nature of most such operators. The problem is compounded by
the fact that the influence of all irrelevant operators becomes strong
in the ultraviolet, which is precisely the regime where lattice scale
effects are expected to manifest.\cite{Ref--RQFT}

\begin{figure}
\includegraphics[width=0.48\textwidth]{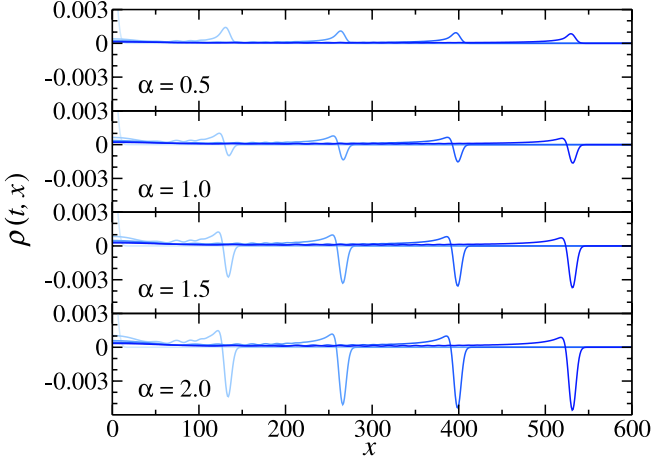}
\caption{
	The $\zeta$-regularized supersoliton obtained by numerically integrating
	Eq.~(\ref{ContEOM}). Here we have set $\zeta = 1$, $\Delta = 6$,
	and plotted data for four values of $\alpha$.
	The interaction exponent $\sigma = 0.7$.
	We have assigned $M = 3/2 \Delta$, so that the dynamics reside 
	within the ``non-relativistic'' transport regime, as discussed in Sec.~\ref{WavepacketDyn}.
	For all but the smallest value $\alpha$ depicted,
	the characteristic ``s'' shape of the supersoliton is identified at 
	sufficiently long times. In comparison to the pure sine-Gordon model result shown
	in Fig.~\ref{supersolitonSGFig}, the amplitude of the regularized supersoliton 
	saturates at times 
	$t \gtrsim t_\zeta$.
	Two inequivalent definitions for 
	$t_\zeta$
	are provided by Eqs.~(\ref{tcDef1}) and (\ref{tcDef2}).	
\label{SupersolitonReg}}
\end{figure}

In the following, we sidestep these difficulties with a phenomenological approach. 
Compare the lattice and continuum initial state correlation functions $\mathcal{C}(x_j,x_{j'})$ 
and $\mathcal{C}^{r}_{\phantom{r} s}(x_1,x_2)$ defined by Eqs.~(\ref{ICCorr}) and (\ref{ICCorrSG}).
While the long-distance behaviors of these functions should be compatible, the short-distance
behaviors clearly differ. The continuum LL correlation functions in Eq.~(\ref{BosCorrs}) exhibit
a power-law divergence as $x_1 \rightarrow x_2$ governed by twice the scaling dimension $(\sigma + 1)/2$; 
by contrast, the lattice correlator satisfies
$\mathcal{C}(x_j,x_j) =\gndc c^\dagger_j c_j \gnd = 1/2 + \ord{Q}$,
independent of $\sigma$ to lowest order.
[$|Q| \ll 1$ characterizes the small localized inhomogeneity induced by the chemical potential in 
Eq.~(\ref{gaussian}).] 
To capture the effects of the lattice, we must cut off the divergence at zero argument in the 
continuum bosonization approximation to the \emph{lattice} correlation function. We do this by incorporating 
a finite range $\zeta$ associated with the nearest-neighbor density-density interactions in $\Hi$ 
[Eq.~(\ref{HiDef})]. 
We obtain
\begin{align}\label{BosCorrsReg}
	\mathcal{C}(x_j,x_{j'})	
	=&
	\gndc c^\dagger(x_j) c(x_j') \gnd
	\nonumber\\
	\sim&
	\frac{c_N \alpha^\sigma}{\pi(x_j - x_j')}
	\left[
	\frac{1}{(x_j - x_j')^2 + \zeta^2}
	\right]^{\sigma/2}
	\nonumber\\
	&\times
	\sin
	\left[
	k_F (x_j - x_j')
	+
	\pi
	\int_{x_j'}^{x_j}
	d y
	\,
	\rho_{0}(y)
	\right],
\end{align}
where $\rho_0(y)$ denotes the initial density profile
[Eq.~(\ref{rho0Def})],
$c_N$ is the normalization 
constant from Eq.~(\ref{BosCorrs}), and $k_F = \pi/2$ is the Fermi wavevector 
at half-filling. In Eq.~(\ref{rho0Def}), $K$ and $u$ denote the 
Luttinger parameter and the sound velocity [for which we will employ the Bethe ansatz
results in Eq.~(\ref{u,K Bethe ansatz})].
To compare to the lattice quench, we use Eq.~(\ref{sigma}) for the exponent 
$\sigma(\gamma)$.

The correlator in Eq.~(\ref{BosCorrsReg}) depends upon two length scales $\alpha$
and $\zeta$ not defined in the lattice theory. 
While the pure sine-Gordon model results from the limit $\zeta \rightarrow 0$, 
the parameter $\alpha$ is always non-zero [c.f.~Eq.~(\ref{BosCorrs})]; its evaluation in the context
of the lattice model would require a Bethe ansatz calculation.\cite{BosonizationRev2}
In comparing to numerics, we will fix $\zeta = a = 1$ ($a$ denotes the lattice spacing), 
consistent with nearest-neighbor interactions, but we will treat $\alpha$ as a fitting
parameter. Our choices for $\zeta$ and $\alpha$ will not prescribe the value 
$\mathcal{C}(x_j,x_j) \equiv 1/2 + \ord{Q}$, except for the non-interacting quench $\sigma = 0$;
rather, we adjust $\alpha$ to fit the long-range part of the correlator to the lattice numerics, since
the regularized continuum approximation is still expected to behave the worst at short
distances.

Incorporating the same $\zeta$-regularization into the component
correlators in Eq.~(\ref{BosCorrs}) and using the result in 
Eq.~(\ref{DensityEvolveSG}), one can analyze the
``ultraviolet regularized'' version of the sine-Gordon quench
studied in the last section. 
The regularized post-quench density is expressed as the integral
\begin{align}\label{ContEOM}
	\rho&(t,x) 
	= 
	\frac{1}{2} 
	\left[
	\rho_0(x - t^\prime)+ 
	\rho_0(x + t^\prime)
	\right]
	\nonumber\\
	&+
	\frac{c_N \alpha^\sigma}{\pi}\int_{-t^\prime}^{t^\prime} d y \,
		\frac{
		\sin\left[\pi \int_{x-t^\prime}^{x-y} d z \, \rho_0(z) \right]
		}{
		(t^\prime - y) \left[\zeta^2 + (t^\prime - y)^2 \right]^{\sigma/2} 
		}
	\,	
	\bar{G}^{(1)}(t^\prime,y)
	\nonumber \\
	&
	-
	\frac{c_N \alpha^\sigma}{\pi}\int_{-t^\prime}^{t^\prime} d y \, 
		\frac{
		\sin\left[\pi \int_{x+t^\prime}^{x-y} d z \, \rho_0(z) \right]
		}{
		(t^\prime + y) \left[\zeta^2 + (t^\prime + y)^2 \right]^{\sigma/2}
		} 
	\, 
	\bar{G}^{(3)}(t^\prime,y)
	\nonumber \\
	& 
	+ \frac{c_N \alpha^\sigma}{2\pi} \int_{-t^\prime}^{t^\prime} d y_1 \int_{-t^\prime}^{t^\prime} d y_2 \,
		\frac{
		\sin\left[\pi \int_{x-y_1}^{x-y_2} d z \, \rho_0(z)\right]
		}{
		(y_1-y_2) \left[\zeta^2 + (y_1-y_2)^2 \right]^{\sigma/2}
		} 
	\,
	\nonumber\\
	&\phantom{\frac{c_N \alpha^\sigma}{2\pi} \int_{-t^\prime}^{t^\prime} d y_1}\times
	\left[
	\begin{aligned}
	&\bar{G}^{(1)}(t^\prime,y_1) \bar{G}^{(1)}(t^\prime,y_2) 
	\\
	&+
	\bar{G}^{(3)}(t^\prime,y_1) \bar{G}^{(3)}(t^\prime,y_2)
	\\
	&+
	2 \bar{G}^{(2)}(t^\prime,y_1) \bar{G}^{(2)}(t^\prime,y_2)
	\end{aligned}
	\right],
\end{align}
where $\bar{G}^{(1,2,3)}(t,y)$
denote the unregularized continuum Green's functions in Eq.~(\ref{ContGFs});
the primed time $t' \equiv v_F t$ [Eq.~(\ref{PrimeTimeDef})].

\begin{figure}[b]
\includegraphics[width=0.45\textwidth]{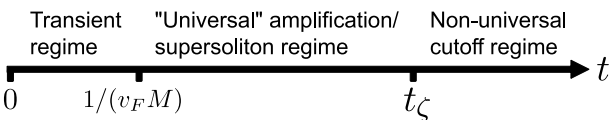}
\caption{
Dynamical regimes of the lattice quench between $\Hi$ and $\Hf$
in Eq.~(\ref{ham}), based on considerations of the regularized 
sine-Gordon model, as discussed in the text. The quench occurs
at $t = 0$. The cutoff time $t_\zeta$ 
can be defined by either Eq.~(\ref{tcDef1})
or (\ref{tcDef2}). 
\label{DynRegimes}}
\end{figure}

For interacting initial conditions $\sigma > 0$ and $\alpha$ not too small,
the characteristic ``s'' shape of the supersoliton appears 
in the regularized sine-Gordon quench. The growth of the supersoliton amplitude is terminated 
after a certain cutoff time $t_\zeta \propto \Delta / v_F (M \zeta)^2$ 
(discussed in more detail below). 
The regularized supersoliton is depicted in Fig.~\ref{SupersolitonReg}.
Interpreting the sine-Gordon quench as the continuum limit of the lattice model 
version, we therefore anticipate the existence of at least three different
dynamical regimes: (1) $0 < t < 1 / (v_F M)$, transient regime, 
(2) $1 / (v_F M) < t < t_\zeta$,``universal''
supersoliton regime, (3) $t > t_\zeta$, 
post-cutoff, non-universal regime. These are sketched in
Fig.~\ref{DynRegimes}.

To obtain an estimate for the cutoff time 
$t_\zeta$,
we analyze
the asymptotic behavior of the density response in Eq.~(\ref{ContEOM}).
We can extract the first correction to Eq.~(\ref{supersolitonSG}) 
in the intermediate time window 
$1/v_F M \ll t \ll t_\zeta$;
the result is 
(c.f.~Appendix \ref{APP--AA})
\begin{align}\label{SupersolitonRegEq}
	\rho(t,x)
	=&
	\frac{Q}{2\sqrt{\pi}\Delta} e^{-(x - t')^2/\Delta^2}
	\nonumber\\
	&-
	\frac{Q}{2 \Delta}
	\frac{
	\Gamma\left(1 - \sigma\right)
	}{
	\Gamma\left(\frac{1 + \sigma}{2}\right)  
	}
	\left[\frac{(M \alpha)^2 t'}{\sqrt{2} \Delta}\right]^{\sigma/2}
	\nonumber\\
	&
	\phantom{-}
	\,
	\times
	\!\!
	\left[
	\begin{aligned}
	&
	F_{\sigma}\left(\frac{x - t'}{\Delta}\right)
	\\&
	\,-
	\left[\frac{(M \zeta)^2 t'}{\sqrt{2} \Delta}\right]^{\frac{1 - \sigma}{2}}
	\frac{
	\Gamma\left(\frac{\sigma - 1}{2}\right)
	}
	{
	2 \Gamma\left(- \sigma\right)
	}
	F_{1}\left(\frac{x - t'}{\Delta}\right)
	\end{aligned}
	\right]
	\nonumber\\
	&+
	\{x \rightarrow -x\},
\end{align}
where the function $F_{\sigma}(z)$ was defined by Eq.~(\ref{FDef}).
The correction grows as $t^{(1 - \sigma)/2}$, but with a sign opposite
to the supersoliton. At intermediate times, the dominant effect is the
suppression of the supersoliton growth. The expansion in 
Eq.~(\ref{SupersolitonRegEq}) is a conserving approximation, because
$F_{\sigma}(z)$ integrates to zero ($\sigma > 0$).

In the limit $\sigma \rightarrow 0$, the third term in Eq.~(\ref{SupersolitonRegEq}) 
vanishes, as expected for the non-interacting quench (which is independent of 
$\alpha$ and $\zeta$). By contrast, ignoring the divergent prefactor we see that the 
second and third terms precisely cancel for $\sigma = 1$. This result obtains because 
the prediction of the unregularized sine-Gordon theory suffers a UV divergence for $\sigma \geq 1$; 
a perturbative expansion about the $\zeta = 0$ limit 
does not exist there.

We define $\mathfrak{f}_1$ as the amplitude ratio of the second and third terms in 
Eq.~(\ref{SupersolitonRegEq}), evaluated on the lightcone ($x = t'$). At argument $z = 0$,
$F_{\sigma}(z)$ equals $2^{\sigma/4} \sqrt{\pi} \,/\, \Gamma[(2 - \sigma)/4]$, a value close 
but not equal to its peak magnitude. For fixed ratio $\mathfrak{f}_1$, we then define the cutoff time 
\begin{equation}\label{tcDef1}
	t_{\zeta}^{(1)} 
	\equiv
	\frac{\Delta}{v_F (M \zeta)^2}
	\left[
	\frac
	{
	2 \mathfrak{f}_1 \,
	\Gamma\left(\frac{1}{4}\right)
	\Gamma\left(- \sigma\right)
	}
	{
	\Gamma\left(\frac{2 - \sigma}{4}\right)
	\Gamma\left(\frac{\sigma - 1}{2}\right)
	}
	\right]^{2/(1 - \sigma)}.
\end{equation}
As an alternative, we compare the integral of the absolute values of the final two terms 
in Eq.~(\ref{SupersolitonRegEq}).
We define
\begin{align}\label{OmegaDef}
	\Omega_{\sigma}
	\equiv
	\int_{-\infty}^{\infty}
	d z
	\,
	\left|
	F_{\sigma}(z)
	\right|.
\end{align}
Eq.~(\ref{OmegaDef}) can be evaluated numerically.
Then, we set $\mathfrak{f}_2$ equal to the ratio of the integrated absolute values
associated with the second and third terms in Eq.~(\ref{SupersolitonRegEq}).
For fixed $\mathfrak{f}_2$, we have
\begin{align}\label{tcDef2}
	t_{\zeta}^{(2)}
	\equiv
	\frac{\Delta}{v_F (M \zeta)^2}
	\left[
	\frac{
	\mathfrak{f}_2 \, 2^{(5 - \sigma)/4} \Omega_{\sigma} \Gamma\left(- \sigma\right)
	}{
	\Omega_1 \Gamma\left(\frac{\sigma - 1}{2}\right)
	}
	\right]^{2/(1 - \sigma)}.
\end{align}

Both $t_\zeta^{(1,2)}$ diverge as $\sigma \rightarrow 0$, as expected for the 
non-interacting quench. 
Eqs.~(\ref{tcDef1}) and (\ref{tcDef2}) are rapidly decreasing functions of 
$\sigma$ that become ill-defined as $\sigma \rightarrow 1$. 
Unfortunately, both definitions are also strongly sensitive to
the value that we assign to arbitrary ratio $\mathfrak{f}_{1,2}$. In particular,
with $\sigma = 0.4$ or $0.7$, for which we present lattice quench data in 
Sec.~\ref{Results}, the values of $t_\zeta^{(1,2)}$ change by several orders
of magnitude as $\mathfrak{f}_{1,2}$ is swept from $0.1$ to $1$. 
The sensitivity reflects the very slow in time (fractional power law) accumulation
of the final term relative to the second in Eq.~(\ref{SupersolitonRegEq}).
Thus, while the definition of a cutoff time with a natural scale $\Delta / v_F(M \zeta)^2$
is conceptually useful, it proves difficult to utilize as a practical tool
in characterizing finite-size numerics.

In the ultimate long-time limit $t \gg t_{\zeta}$, Eq.~(\ref{ContEOM})
has the leading asymptotic behavior
(c.f.~Appendix \ref{APP--AA})
\begin{align}\label{AsymUltimate}
	\rho(t,x)
	\sim&
	\frac{Q}{2\sqrt{\pi} \Delta}
	\left[
	1 
	- 
	\frac{\sqrt{\pi} \Gamma\left(1 + \frac{\sigma}{2}\right)}{\Gamma\left(\frac{1 + \sigma}{2}\right)}
	\left(
	\frac{\alpha}{\zeta}
	\right)^{\sigma}
	\right]
	e^{-(x - t')^2 / \Delta^2}
	\nonumber\\
	&+
	\{x \rightarrow -x\}.
\end{align}
In the regularized continuum quench, the supersoliton eventually gives
way to a pure translation of the initial Gaussian, with a reduced amplitude.
This is completely different from the single particle evolution
resulting from ``relativistic'' confinement $0< M \Delta \ll 1$, 
discussed in Sec.~\ref{WavepacketDyn}. In that case,
the non-dispersive part of the amplitude decays to zero in the
long time limit [Eq.~(\ref{SingPartRelDisp})].
Eq.~(\ref{AsymUltimate}) is not a conserving approximation for any $\sigma$;
the missing density is distributed in a long tail neglected here. 
In fact, for $\zeta = 1$ and $\alpha \geq 0.64$, the amplitude in 
Eq.~(\ref{AsymUltimate}) is \emph{negative} for $0 < \sigma < 1$,
which applies to Fig.~\ref{SupersolitonReg}.
This is the case for the $\sigma = 0.7$ and $\sigma = 1.0$ quenches
discussed in Sec.~\ref{Results}, although our lattice numerics
are limited to system sizes much too small to reach this regime.

\subsubsection{Band curvature lifetime}\label{curvlifetime}

The sine-Gordon theory presented in the previous section accounts
only for lattice effects on the \emph{initial} pre-quench state, by way of the
$\zeta$-regularized correlation function in Eq.~(\ref{BosCorrsReg}).
This is one ingredient in the post-quench evolution
of the lattice density in Eq.~(\ref{dynamics}); the other 
is the set of Green's functions $G^{(1,2,3)}(t,x_i)$ obtained
by Fourier-transforming Eq.~(\ref{GreensFT}). Instead, in Eqs.~(\ref{ContEOM}),
(\ref{SupersolitonRegEq}), and (\ref{AsymUltimate}), we have employed
the continuum $\bar{G}^{(1,2,3)}(t,x)$ defined by Eq.~(\ref{ContGFs}),
which assumes the Lorentz covariant spectrum in Eq.~(\ref{DiracSpec}). 

We find that this regularized correlator + continuum Green's functions 
approximation proves adequate to model most of the lattice quench numerics 
presented in Sec.~\ref{Results}. However, to characterize the dynamics in the limit of very 
long times (in a correspondingly large system), we would need to account 
for the additional effects of band curvature. 
This is of particular importance for the interacting quench, which yields
the ``regularized'' supersoliton in Eq.~(\ref{SupersolitonRegEq}) ($t \ll t_\zeta$) 
or its ultimate fate as the non-dispersing ghost in Eq.~(\ref{AsymUltimate})
($t \gg t_\zeta$). These disturbances propagate at the ``speed of light''
$v_F$, which is replaced by the maximum band velocity $v_{\rm max}(M)$ 
in the lattice model. [$v_{\rm max}(0) = v_F$; see Sec.~\ref{VeloSec} for more 
details.] As a first correction to the continuum dynamics, we consider
the cubic curvature represented by $\mathcal{O}_3$ in Eq.~(\ref{curvature}). 
  
The lifetime $t_3$ is defined as the interval post-quench during which the cubic
curvature can be \emph{ignored}; a crude order-of-magnitude estimate is given by 
\begin{align}\label{t3Def}
	t_3 \sim \frac{\Delta^3}{v_{\rm max}(M)},
\end{align}
where $\Delta$ is the position space width of the initial density inhomogeneity.
Eq.~(\ref{t3Def}) follows from the expansion of the band dispersion in Eq.~(\ref{spec})
about $k_{\rm max}$ such that $v(k_{\rm max}) = v_{\rm max}$ (Sec.~\ref{VeloSec}): 
\[
	E_k 
	= 
	E_{k_{\rm max}} 
	+
	v_{\rm max}\left[\delta k  - \frac{2}{3} a^2 \delta k^3\right] + \ldots
\]
where $\delta k \equiv k - k_{\rm max}$; $a$ is the lattice constant.
For a Gaussian packet of width $\Delta$, the characteristic frequency associated to the
cubic term is $v_{\rm max} a^2/\Delta^3$, giving Eq.~(\ref{t3Def}) with $a = 1$.


\section{Lattice quench results \label{Results}}


In this section, we present numerical results for the XXZ chain quench set up in Sec.~\ref{Q Setup}.
A chain with $N$ (even) sites and periodic boundary conditions is prepared in the ground state
$\gnd$ of $\Hi$, Eq.~(\ref{HiDef}). The interaction strength $\gamma$ is chosen
to reside in the XY range $-1 < \gamma \leq 0$, so that $\gnd$
exhibits gapless power-law correlations for the lattice fermions.
This state is evolved forward in time according to $\Hf$, Eq.~(\ref{HfDef}).
For the Gaussian initial state inhomogeneity induced by $\mu^{(0)}_i$ in Eq.~(\ref{gaussian}), 
we calculate the post-quench dynamics of the density expectation value $\rho(t,x)$, 
Eq.~(\ref{dynamics}). In the generic case of the interacting quench ($\gamma \neq 0$), 
the required initial state correlation function $\mathcal{C}(x_j,x_{j^\prime})$ is computed
numerically using the density matrix renormalization group (DMRG) technique. 
All data shown are for a system of $N = 202$ sites. 

We compare the numerical results for the lattice quench to the regularized continuum
sine-Gordon theory presented in Sec.~\ref{SGRegTheory}. That theory is epitomized
by the continuum approximation to the initial state lattice correlation function
in Eq.~(\ref{BosCorrsReg}) and the density expectation in Eq.~(\ref{ContEOM}). The regularized sine-Gordon 
model contains two length scale parameters $\alpha$ and $\zeta$
that are not defined in the corresponding lattice theory. These parameters
enter via the initial state correlation function $\mathcal{C}(x_j,x_{j'})$ in Eq.~(\ref{BosCorrsReg}). 
The parameter $\alpha$ determines the amplitude of this correlator, while
$\zeta$ acts as an ultraviolet cutoff that renders finite the on-site value of 
$\mathcal{C}(x_j,x_j)$.
For the non-interacting quench ($\sigma=0$), the continuum predictions
are independent of $\zeta$ and $\alpha$.

\subsection{Non-interacting quench \label{NIResults}}

The special case $\gamma = 0$ yields a free Fermi gas
ground state of $\Hi$. Both the initial and final Hamiltonians are trivially diagonalized, 
and we solve for the dynamics exactly. 
For this ``non-interacting'' quench, the initial state correlation function was 
transcribed in Eq.~(\ref{NonIntCorr}), above. 

\begin{figure}[t]
   \includegraphics[scale=0.25]{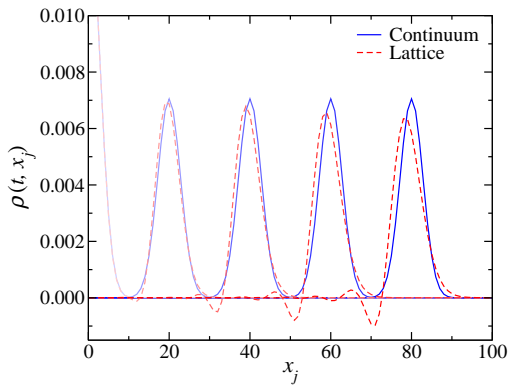}
   \caption{Time slices of a non-interacting quench with a gapless final Hamiltonian
	[$M = 0$ in Eq.~(\ref{HfDef}),
	a 
	``Fermi gas to Fermi gas''
	quench] 
	at times $t=$ 0, 15, 30, 45, and 60;
	fainter (bolder) traces depict earlier (later) times.
	Blue solid lines are the continuum
	prediction [the first term of Eq.~(\ref{SupersolitonRegEq})], 
	and red dashed lines are the result of exact diagonalization of the
	lattice Hamiltonian. The evolution is
	symmetric about $x_j=0$.  
	The relevant quench parameters are $Q=0.10$, $\Delta = 4$. 
	}
   \label{fig:LLtoLL_4}
\end{figure}

\begin{figure}[b]
   \includegraphics[scale=0.25]{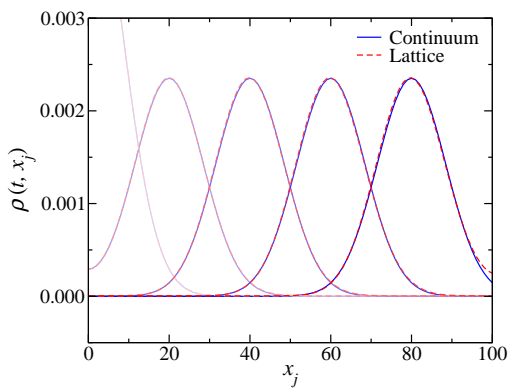}
   \caption{The same as in Fig.~\ref{fig:LLtoLL_4}, but with $\Delta = 12$.
	}
   \label{fig:LLtoLL_12}
\end{figure}

We first investigate the quench into the gapless XX chain, $M = 0$ in $\Hf$.
Since the low-energy field theory description of both the initial and final
states is a free Fermi gas, we refer to this as a ``FG to FG'' quench.
Thus one prepares a density wavepacket at the origin, then simply removes the applied potential
and tracks the resulting dynamics.
Numerical results are depicted in Figs.~\ref{fig:LLtoLL_4} and \ref{fig:LLtoLL_12}
for two different values of $\Delta$.

The continuum prediction is a pure translation of half the initial density profile to
the left and to the right, at the ``speed of light'' $v_F = 2$. The right-moving part
appears as the first term on the right-hand side of Eq.~(\ref{SupersolitonRegEq}).
One can see in
Figs.~\ref{fig:LLtoLL_4} and \ref{fig:LLtoLL_12}
that the agreement between the continuum and the lattice 
quenches is very good, and improves with increasing $\Delta$.  
The slight dispersion seen for $\Delta = 4$ in Fig.~\ref{fig:LLtoLL_4} can
likely be attributed to the deviation of the band spectrum [Eq.~(\ref{spec}) with
$M = 0$] from linearity at wavenumbers $k \sim 1/\Delta$ away from the Fermi wavevector
$k_F = \pi/2$.
As discussed in Sec.~\ref{curvlifetime}, we can associate a lifetime $t_3 \sim \Delta^3/v_F$
to the presence of the cubic non-linearity in the spectrum. Then $t_3(\Delta = 12) \sim 860$,
while $t_3(\Delta = 4) \sim 32$; the latter falls midway in the range of times plotted in Fig.~\ref{fig:LLtoLL_4}. 

We now turn to non-interacting quenches into a {\em gapped} final Hamiltonian.  
Here, we quench from a free Fermi gas into a band insulator.
The periodic potential in Eq.~(\ref{HfDef}) with $M \neq 0$
allows for backscattering umklapp processes, which open up a bandgap with magnitude $4M$.
To compare to the continuum theory, we would like to reach the scaling limit where
all relevant length scales in the problem greatly exceed the lattice spacing, e.g.\ $\Delta \gg a$,
$1/M \gg a$,
while keeping $\Delta \ll N a$.
In addition, we restrict our quench parameters to the ``non-relativistic'' transport regime
$M \Delta > 1$, as explained in Sec.~\ref{WavepacketDyn}, so as to avoid confusing the putative 
supersoliton (in the interacting quench, below) with relativistic propagation induced by excessive 
``squeezing'' of the initial density disturbance relative to the Compton wavelength.
Specifically, for all data presented subsequently
we will fix the product $M \Delta = 3/2$,
and examine
four wavepacket widths $\Delta = $ 4, 6, 12, and 20, yielding
the respective band-gap parameters 
$M = $ 3/8, 1/4, 1/8, and 3/40.

The application of the staggered potential causes adjacent site occupancies to `polarize' opposite
to one another, but this small-scale density effect is not one in which we are interested;
the staggered potential is merely a tool to induce a gap in the spectrum.
We henceforth present results for the {\em relative} particle density, given by
\begin{equation}\label{RelDensity}
	\delta \rho(t,x_j) \equiv \rho(t,x_j)_Q - \rho(t,x_j)_{Q=0}, 
\end{equation}
i.e.\ we subtract the time-dependent density profile originating from a spatially homogeneous ($Q=0$)
initial state.

\begin{figure}[t]
   \includegraphics[scale=0.25]{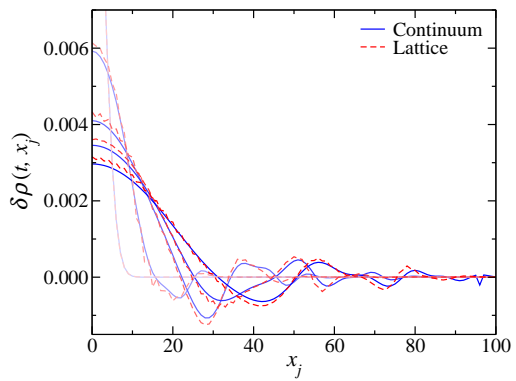}
   \caption{Time slices of a non-interacting quench into a gapped final Hamiltonian
	(a 
	non-interacting Fermi gas
	to band insulator quench) at times $t=$ 0, 15, 30, 45, and 60;
	fainter (bolder) traces depict earlier (later) times.
	Blue solid lines are the continuum
	predictions and red dashed lines are the result of exact diagonalization of the
	lattice Hamiltonian. 
	The continuum data results from a numerical integration of Eq.~(\ref{ContEOM})
	with $\sigma = 0$.
	The evolution is
	symmetric about $x_j=0$.  
	The relevant quench parameters are $Q=0.10$, $\Delta = 4$, $M = 3/8$. 
	}
	\label{fig:nonint_4}
\end{figure}

Figs.~\ref{fig:nonint_12} (in the Introduction) and \ref{fig:nonint_4} 
show the resulting post-quench dynamics for two of the four ($\Delta$, $M$) pairs given above.
The dynamics are strongly dispersive, in stark contrast to the
ultrarelativistic propagation seen in the FG to FG quench.
We see that the initial Gaussian inhomogeneity broadens gradually and does so more slowly for larger
values of the band gap; these non-interacting quench dynamics are grossly similar to the non-relativistic single 
particle wavepacket depicted in Fig.~\ref{DiracPlot}(a).
The behavior is generic 
and we find it to occur for a wide range of non-interacting quench parameters satisfying the non-relativistic 
condition $M \Delta > 1$.

The dispersion arises from a combination of the strong band curvature near the (non-interacting) 
Fermi point and the relatively weak occupancy of the 
conduction band induced by the quench,
as evidenced by the corresponding distribution function plots in Fig.~\ref{fig:nk} with $\sigma = 0$.
We emphasize however that the global momentum distribution in Fig.~\ref{fig:nk} does not encode
information about the inhomogeneity; for this purpose one should consult the Wigner function,
as discussed in Sec.~\ref{WDSec} and Appendix~\ref{APP--DistInhomog}.
For the continuum theory, we find that the ``local'' velocity distribution for
the non-relativistic, non-interacting quench exhibits a strong suppression of velocities $v \gtrsim v_F/M \Delta$
[Eq.~(\ref{VelocDistNI}) in Appendix~\ref{APP--DistInhomog}], due to Pauli-blocking in 
the initial Fermi gas ground state (Sec.~\ref{WDSec}).

The continuum curves in Figs.~\ref{fig:nonint_12} and \ref{fig:nonint_4} obtain from the 
numerical integration of Eq.~(\ref{ContEOM}), with $\sigma = 0$.
The agreement between the lattice and continuum quench dynamics is generally excellent.
For the $\Delta$ values considered, we observe negligible sublattice staggering in the lattice 
$\delta \rho(t,x_j)$; such behavior is a good indicator of 
the near complete separation of the smooth and staggered components of the density. 
This is consistent with the retention of only the smooth component of the initial inhomogeneity
in the regularized sine-Gordon theory of Sec.~\ref{SGRegTheory}.\cite{Ref--Sublat Stag Chem Pot}


\subsection{Interacting quench \label{IResults}}

We turn to the most interesting
case of an {\em interacting} initial state, 
$\gamma \neq 0$ in Eq.~(\ref{HiDef}).
Because the final state is still non-interacting, the dynamics are exactly
given by Eq.~(\ref{dynamics}) above, but the initial state correlation function 
$\mathcal{C}(x_j,x_{j'})$
cannot be obtained via elementary means.
To achieve this task, we employ the density matrix renormalization group 
(DMRG)\cite{whi92,whi93,sch05} due to its ability to treat
relatively large interacting one-dimensional systems.  All calculations were performed
on a chain of size $L=202$ (so that the number of fermions at half-filling, $L/2$, is odd)
with periodic boundary conditions (PBCs).  In standard DMRG, the
relative error introduced with PBCs is significantly larger than that
obtained with open boundary conditions (OBCs).
To achieve the relative error
obtained with $m$ states per block using OBCs, one would need $m^2$
states when using PBCs.  This results in
greatly increased computational times which scale as $m^6$ with PBCs as compared to $m^3$
with OBCs.  Efficient methods to improve
DMRG's ability to handle PBCs are still on-going topics of research
(see e.g.\ Ref.~\onlinecite{pip10} 
and references therein).
In spite of the above considerations,
we found the use of PBCs was necessary to mitigate dynamical boundary effects
appearing during the quench process.  In all calculations presented, we kept up to 200 states
and performed eight sweeps in the DMRG algorithm, yielding a truncation error (discarded weight)
on the order of $10^{-7}$.
We tested the combination of DMRG and exact time evolution for the non-interacting
quench by comparing to the results of exact diagonalization, presented in the previous
section. 

Although our DMRG calculations are
complicated by the use of PBCs and a spatially inhomogeneous
Hamiltonian, one could in principle imagine performing DMRG calculations
for larger systems.  Unfortunately, although the (non-interacting)
dynamics are trivially written down, they suffer from quite poor
polynomial scaling with system size.  Namely, forward and backward
Fourier transforms [each requiring $O(N)$ operations] 
for each of the $N$ sites out to times scaling with the size of the system yields 
a dynamics algorithm which scales as $O(N^4)$.  
Calculating the dynamics for
systems much larger than those considered here 
is currently prohibitive.

\subsubsection{Initial state correlator}

\begin{figure}[t]
   \includegraphics[scale=0.3]{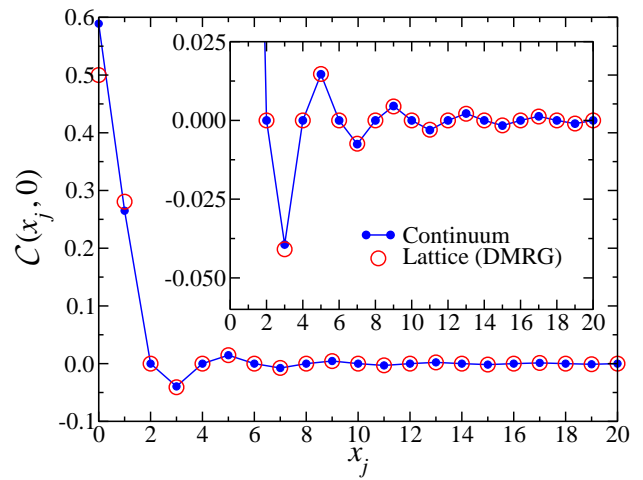}
   \caption{
	Comparison of the 
	initial state
	correlation function $\mathcal{C}(x_j,0)$ 
	defined in 
	Eq.~(\ref{ICCorr}),
	as predicted
	by the regularized ($\alpha=0.75$, $\zeta=1$) LL theory 
	[Eq.~(\ref{BosCorrsReg}), blue dots connected by lines] 
	and as calculated with DMRG (red open circles).
	The inset is a close-up of the same data.
	The relevant parameters are $\sigma$ = 1.0, $Q$ = 0.0.
	}
   \label{fig:corr}
\end{figure}

\begin{figure}[b]
   \includegraphics[scale=0.3]{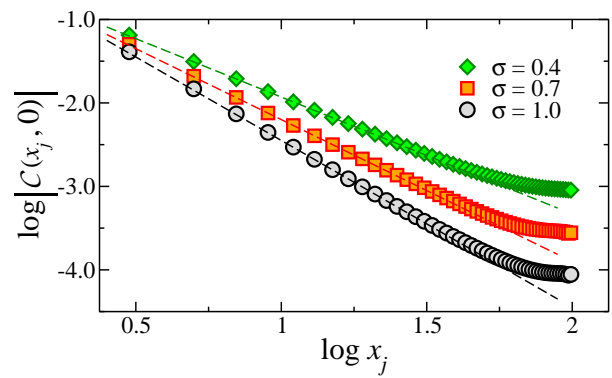}
   \caption{Envelopes of the correlation function $\mathcal{C}(x_j,0)$ calculated by DMRG
	for the dynamical exponents
	$\sigma = 1.0$ (black circles), 0.7 (red squares), and 0.4 (green diamonds)
	-- all for $Q$ = 0.0.  The dashed lines are a guide to
	the eye, exhibiting 
	slopes 
	-2, -1.7, and -1.4 (bottom to top).
	}
   \label{fig:power}
\end{figure}

As a first analysis, 
we consider the initial state correlation function $\mathcal{C}(x_j,0)$ [Eq.~(\ref{ICCorr})]. 
LL theory predicts interaction-dependent power-law
behavior in correlation functions.  
At large separations, Eq.~(\ref{BosCorrsReg}) yields
\begin{equation}\label{LLCorrCrude}
	\mathcal{C}(x_j,0) \sim |x_j|^{-(\sigma+1)},
\end{equation}
where the exponent $\sigma$ 
is taken as the Bethe ansatz result,
Eq.~(\ref{sigma}).
Fig.~\ref{fig:corr} compares the correlation function calculated numerically by DMRG to
the regularized continuum prediction
for $\sigma = 1.0$, with $Q=0$.
[For the cases of non-zero inhomogeneity with small $|Q|$ considered below, 
the $Q$- and $\Delta$-dependencies of $\mathcal{C}(x_j,x_{j'})$ are very minor.]
To fix the continuum result in Eq.~(\ref{BosCorrsReg}),
we make the physically motivated choice $\zeta=a=1$,
associated with nearest-neighbor density-density interactions
on the lattice, while we
adjust the scale-setting prefactor $\alpha$ to best match the
DMRG calculated correlation function
at \emph{large} separations.
This approach yields $\sigma$-dependent values of $\alpha$,
specifically $\alpha =$ 0.75, 0.64, and 0.50 for $\sigma =$ 1.0, 0.7, and 0.4, respectively.
The agreement between the lattice and continuum correlation
functions is seen in Fig.~\ref{fig:corr} to be excellent after such a fitting procedure;
similar agreement is obtained for the other values of $\sigma$.
The regularization parameters obtained in this manner
are also employed in the subsequent continuum calculation of the interacting quench dynamics.

The power-law prediction
emerging from LL theory is seen to be very robust. 
In Fig.~\ref{fig:power}, we plot 
(in log-log scale) the envelope of the DMRG correlation function for the interaction strengths
yielding exponents $\sigma = 1.0,\ 0.7$, and $0.4$.  The 
deviations occuring at the largest separations are
an artifact of the numerics.

\subsubsection{Maximum band velocity \label{VeloSec}}

Below we
compute the quench dynamics originating from an interacting initial state
characterized by the correlation function analyzed above.  We first pause to discuss
a time-rescaling procedure adopted in the following.  
In Sec.~\ref{GlobalDist}, we considered the static post-quench distribution functions $n_{\pm}(k)$
for conduction and valence band fermions 
[excitations of the final band insulating Hamiltonian $\Hf$ or $\Hfc$]
in the lattice and continuum quenches. 
Continuum results for the non-interacting and interacting 
quenches are given by Eqs.~(\ref{DistNI}) and (\ref{DistI}). 
Fig.~\ref{fig:nk} shows lattice quench results for $n_{+}(k)$ obtained from the DMRG initial state 
correlation function associated with the four values of $M$ considered in this section.
Each subplot exhibits
traces for $\sigma = 0$, 0.4, 0.7, and 1.0. 

Fig.~\ref{fig:nk} indicates
that the final state distribution of excited
particles for an interacting quench ($\sigma > 0$) extends 
deep into the conduction band.  
Although the final particle spectrum in Eq.~(\ref{spec}) is quadratic at 
low energies near the band center, the distribution
induced by an interacting quench stretches into the linear ``relativistic''
regime of the spectrum and beyond.
In the ungapped case, the slope $v_F = 2$ for $k$ just above $k_F = \pi/2$;
with $M > 0$,
the maximum group velocity of the band structure
is modified to $v_{\rm max}(M) < 2$.
It is this velocity with which we henceforth rescale time in
the continuum calculations, $t^\prime = v_{\rm max}(M)t$ in Eq.~(\ref{ContEOM}).

For $E_k$ in Eq.~(\ref{spec}), the velocity $v(k;M) \equiv d E_k(M) / dk$
evaluates to
\begin{equation}\label{vmax_k}
	v(k;M) = - \frac{2 \sin(k) \cos(k)}{\sqrt{\cos^2(k) + M^2}}.
\end{equation}
This equation is maximized at a wavevector $k_{\rm max}$ satisfying
\begin{equation}\label{kstar}
	\cos(k_{\rm max}) 
	= \sqrt{\sqrt{M^2+M^4}-M^2}.
\end{equation}
Inserting the solution of Eq.~(\ref{kstar}) into Eq.~(\ref{vmax_k}) yields the maximal
band velocity $v_{\rm max}(M)$.
In Fig.~\ref{fig:nk}, the position of 
$k_{\rm max}$
for each value of $M$ is indicated
by a dashed vertical line. 

The velocity rescaling procedure outlined above 
was {\em not} adopted  in the continuum non-interacting quench data 
exhibited in Figs.~\ref{fig:nonint_12} and \ref{fig:nonint_4}.
For the case $\sigma = 0$, Fig.~\ref{fig:nk} indicates that the 
linear regime is only weakly populated for all but the largest value 
of $M = 3/8$ considered here; see also Eq.~(\ref{DistNI}).
As a consequence, the strongly dispersive dynamics 
in the ``non-relativistic'' transport regime $M \Delta > 1$
are dominated by the low-$k$ bandstructure.
This picture is confirmed by 
the excellent agreement between lattice and continuum results 
in Figs.~\ref{fig:nonint_12} and \ref{fig:nonint_4},
and by the local velocity distribution obtained for the
non-interacting, non-relativistic quench in Eq.~(\ref{VelocDistNI}).
For the interacting quenches considered below, the rescaling
of the velocity is not a systematic incorporation of bandstructure
effects into the continuum Green's functions defined by Eq.~(\ref{ContGFs});
aspects of ultraviolet band curvature beyond the linear regime
have been neglected. Band curvature effects are expected to become important at
post-quench times $t$ later than $t_3$, defined as the cubic
dispersion lifetime via Eq.~(\ref{t3Def}).

\begin{figure}[t]
   \includegraphics[scale=0.25]{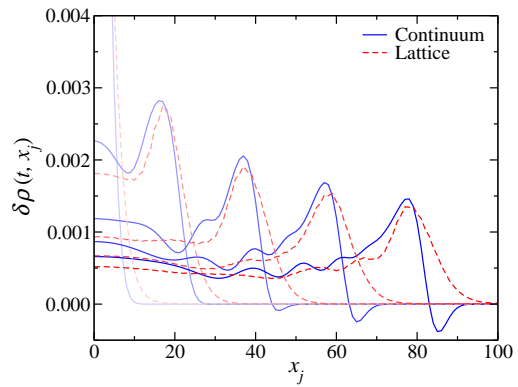}
   \caption{
	Time slices of an interacting quench 
	with $\sigma = 1.0$
	into a gapped final Hamiltonian
	(an interacting Luttinger liquid to band insulator quench) at times $t=$ 0, 15, 30, 45, and 60;
	fainter (bolder) traces depict earlier (later) times.
	Blue solid lines are the continuum
	predictions and red dashed lines are the 
	results of DMRG calculations combined with exact
	time evolution of the
	lattice Hamiltonian. 
	The continuum data obtains from a numerical integration of Eq.~(\ref{ContEOM}),
	with $\alpha = 0.75$ and $\zeta = 1$.
	The evolution is
	symmetric about $x_j=0$.  
	The relevant quench parameters are $Q=0.10$, $\Delta = 4$, $M = 3/8$.
	}
	\label{fig:int_1.0_4}
\end{figure}

\begin{figure}[t]
   \includegraphics[scale=0.25]{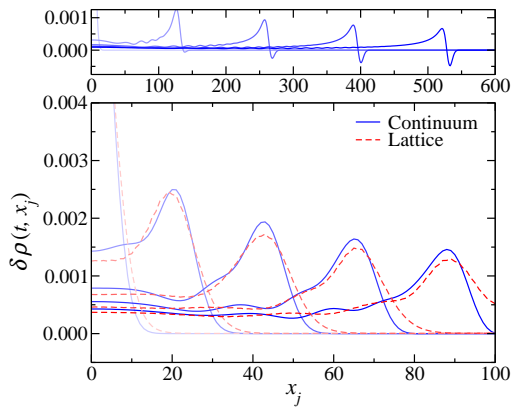}
   \caption{
	Interacting quench with $\sigma = 1.0$ 
	as in Fig.~\ref{fig:int_1.0_4}, but with $\Delta = 6$, $M = 1/4$.
	The continuum result [numerical integration of Eq.~(\ref{ContEOM})]  
	can be obtained for much larger times and system sizes than is currently 
	practical with the interacting numerics (DMRG+dynamics).
	The top panel shows the continuum evolution over a window of length 600.
	}	
   \label{fig:int_1.0_6}
\end{figure}

\begin{figure}[t]
   \includegraphics[scale=0.25]{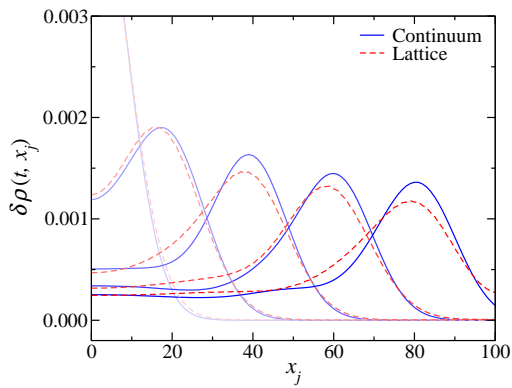}
   \caption{
	Interacting quench with $\sigma = 1.0$ 
	as in Fig.~\ref{fig:int_1.0_4}, but at times $t=$ 0, 12, 24, 36, and 48,
	with $\Delta = 12$, $M = 1/8$.
	}
   \label{fig:int_1.0_12}
\end{figure}

\begin{figure}[t]
   \includegraphics[scale=0.25]{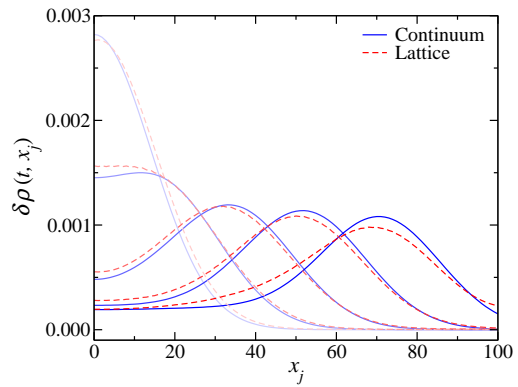}
   \caption{
	Interacting quench with $\sigma = 1.0$ 
	as in Fig.~\ref{fig:int_1.0_4}, but at times $t=$ 0, 10, 20, 30, and 40,
	with $\Delta = 20$, $M = 3/40$.
	}
   \label{fig:int_1.0_20}
\end{figure}

\begin{figure}[t]
   \includegraphics[scale=0.25]{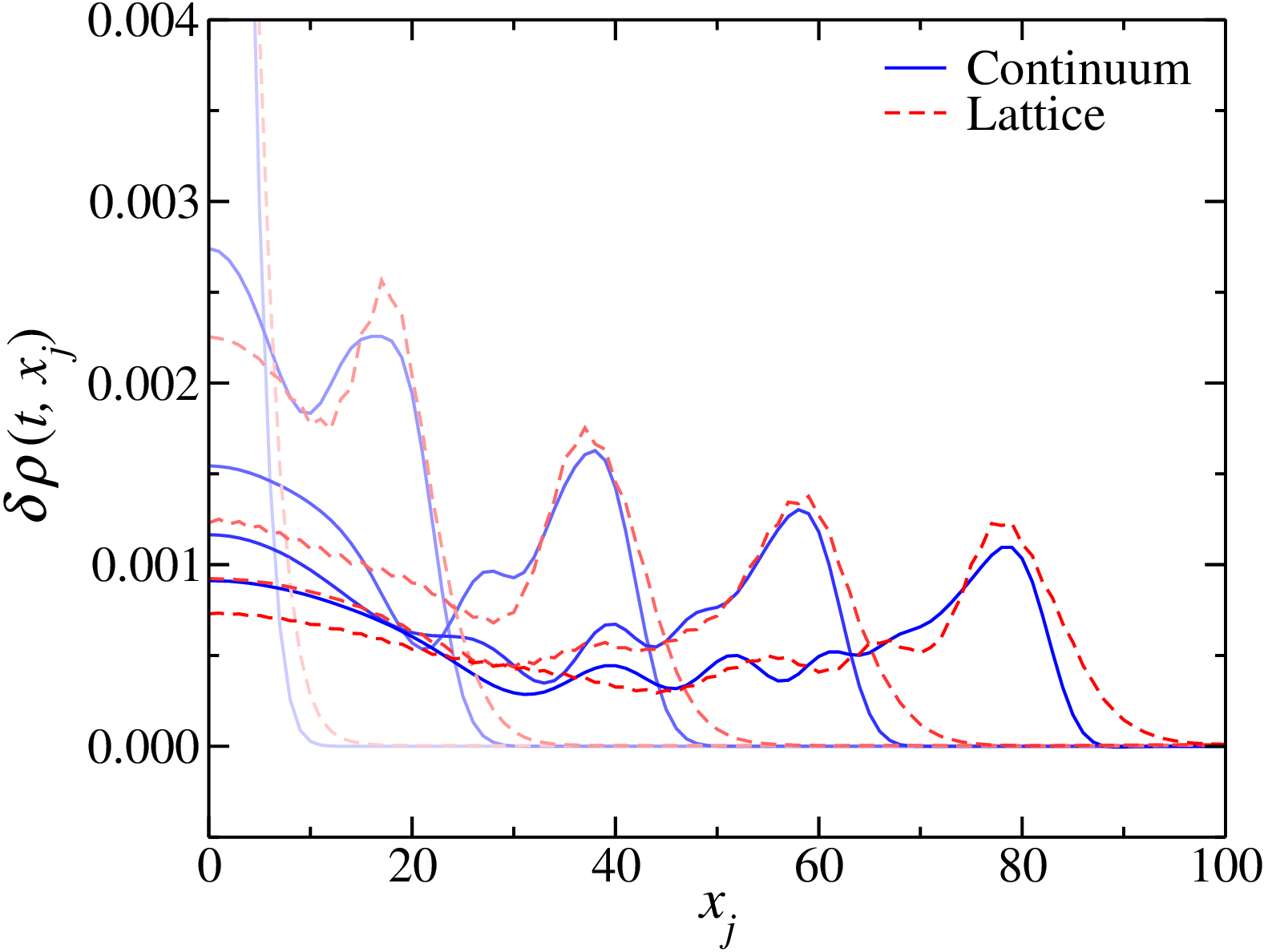}
   \caption{Time slices of an interacting quench 
	with $\sigma = 0.7$ 
	at times $t=$ 0, 15, 30, 45, and 60.
	The continuum data obtains from a numerical integration of Eq.~(\ref{ContEOM}),
	with $\alpha = 0.64$ and $\zeta = 1$.
	The relevant quench parameters are $Q=0.10$, $\Delta = 4$, $M = 3/8$.
	}
	\label{fig:int_0.7_4}
\end{figure}

\begin{figure}[t]
   \includegraphics[scale=0.25]{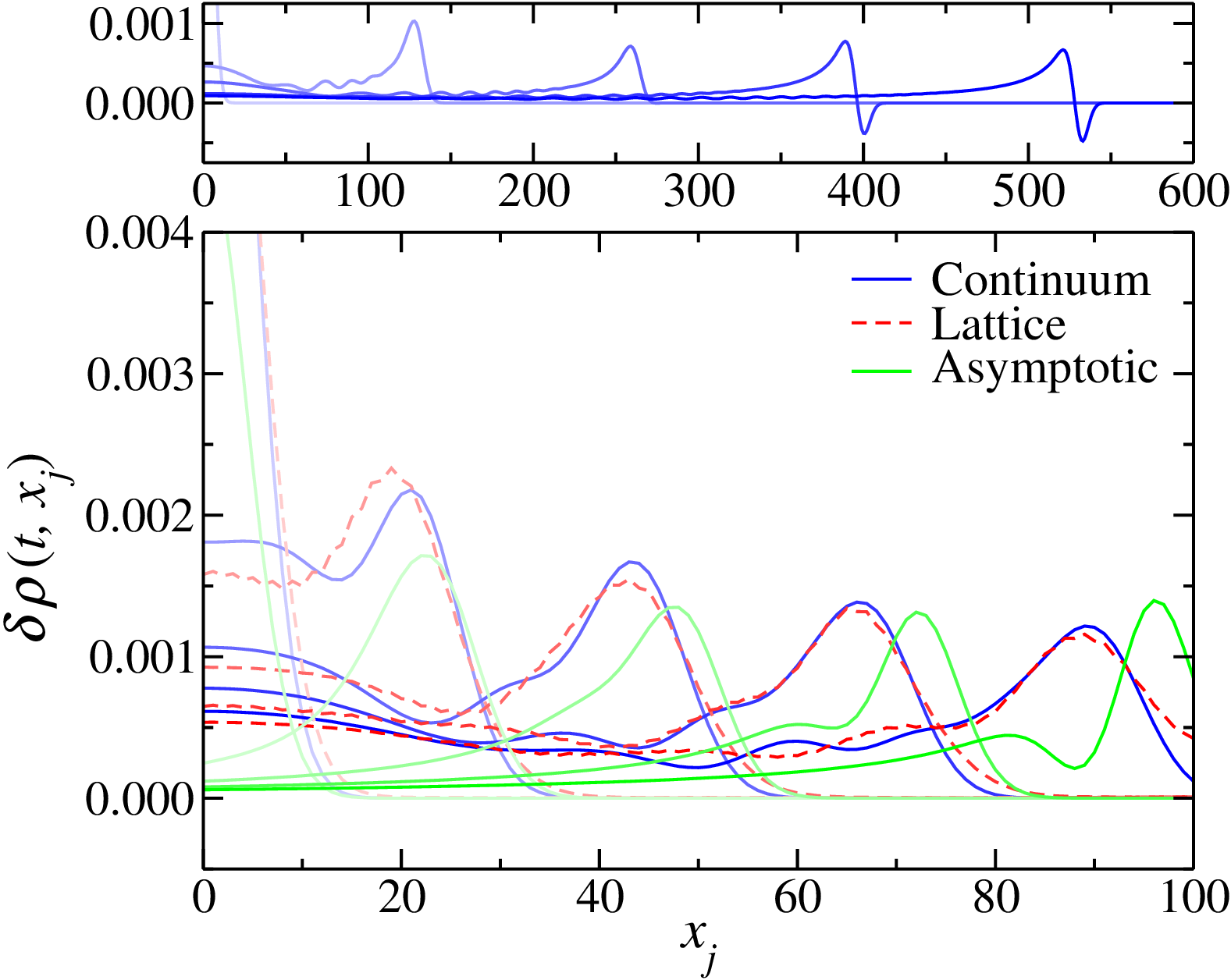}
   \caption{
	Interacting quench with $\sigma = 0.7$ 
	as in Fig.~\ref{fig:int_0.7_4}, but with $\Delta = 6$, $M = 1/4$.	
	The curve marked ``asymptotic'' is the analytical result for the 
	``regularized supersoliton'' in Eq.~(\ref{SupersolitonRegEq}).
	The top panel shows the numerical continuum evolution over a window of length 600.
	}
   \label{fig:int_0.7_6}
\end{figure}

\begin{figure}[t]
   \includegraphics[scale=0.25]{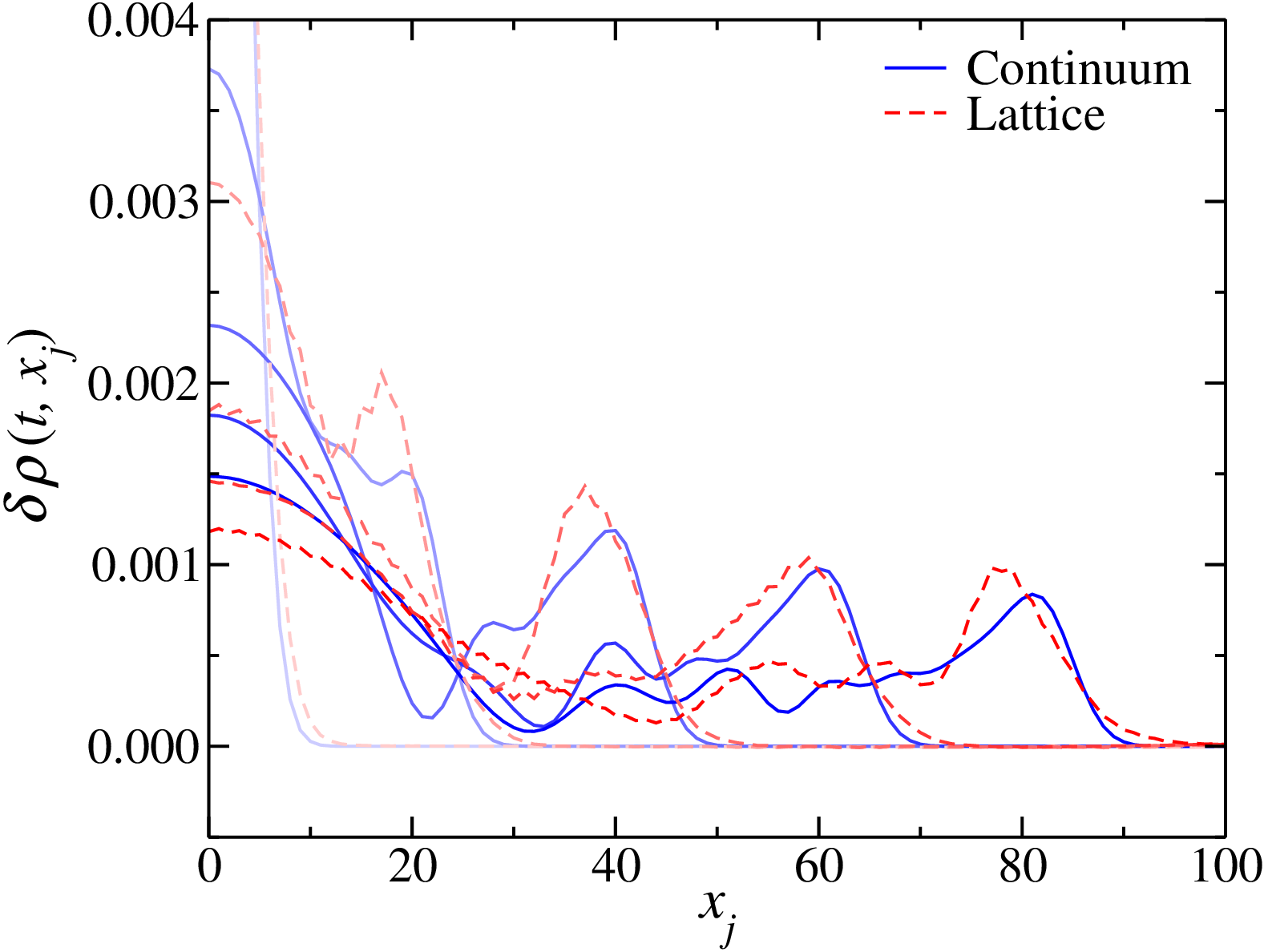}
   \caption{Time slices of an interacting quench 
	with $\sigma = 0.4$ at times $t=$ 0, 15, 30, 45, and 60.
	The continuum data obtains from a numerical integration of Eq.~(\ref{ContEOM}),
	with $\alpha = 0.50$ and $\zeta = 1$.
	The relevant quench parameters are $Q=0.10$, $\Delta = 4$, $M = 3/8$.
	}
	\label{fig:int_0.4_4}
\end{figure}

\begin{figure}[t]
   \includegraphics[scale=0.25]{fig23.eps}
   \caption{
	Interacting quench with $\sigma = 0.4$ 
	as in Fig.~\ref{fig:int_0.4_4}, but with $\Delta = 6$, $M = 1/4$.
	The curve marked ``asymptotic'' is the analytical result for the 
	``regularized supersoliton'' in Eq.~(\ref{SupersolitonRegEq}).
	The top panel shows the numerical continuum evolution over a window of length 600.
	}
   \label{fig:int_0.4_6}
\end{figure}

\begin{figure}[t]
   \includegraphics[scale=0.25]{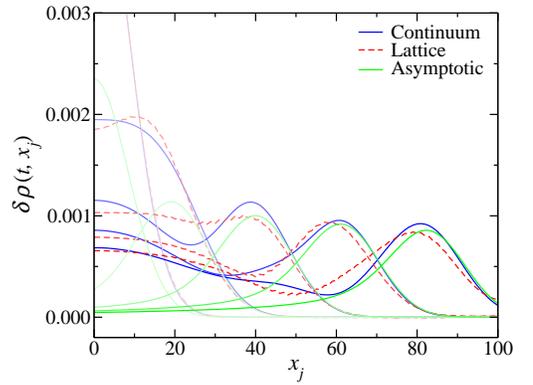}
   \caption{
	Interacting quench with $\sigma = 0.4$ 
	as in Fig.~\ref{fig:int_0.4_4}, but at times $t=$ 0, 12, 24, 36, and 48,
	with $\Delta = 12$, $M = 1/8$.
	The curve marked ``asymptotic'' is the analytical result for the 
	``regularized supersoliton'' in Eq.~(\ref{SupersolitonRegEq}).
	}
   \label{fig:int_0.4_12}
\end{figure}

\subsubsection{Coherent relativistic wave propagation:\\ the ``regularized'' supersoliton}

In Figs.~\ref{fig:int_1.0_4}-\ref{fig:int_0.4_12} we present the interacting quench dynamics associated with
three different values of the exponent $\sigma(\gamma)$ defined by Eq.~(\ref{sigma}), which characterizes the 
initial interacting spin chain described by $\Hi$. The values we choose are 
$\sigma = 1.0$
(Figs.~\ref{fig:int_1.0_4}--\ref{fig:int_1.0_20}), 
$\sigma = 0.7$ 
[Figs.~\ref{fig:int_0.7_4}, \ref{fig:int_0.7_6}, and \ref{fig:int_0.7_12} (in the Introduction)], 
and $\sigma = 0.4$ (Figs.~\ref{fig:int_0.4_4}--\ref{fig:int_0.4_12}); 
these respectively correspond to interaction strengths $\gamma = -0.913$, $-0.872$, and $-0.790$, 
receding from the ferromagnetic transition at $\gamma = -1$ (Fig.~\ref{XXZPhaseDiag}). In the unregularized, 
pure sine-Gordon theory described in Ref.~\onlinecite{supersoliton} and in Sec.~\ref{SG Def}, the quench 
yields the prediction of the supersoliton for $0 < \sigma < 1$ [Eq.~(\ref{supersolitonSG}) and 
Fig.~\ref{supersolitonSGFig}], while $\sigma = 1$ marks the onset of an ultraviolet divergence that must 
be regularized, as in Eq.~(\ref{BosCorrsReg}).

We show data for $\Delta =$ 4, 6, 12, and 20 in Figs.~\ref{fig:int_1.0_4}--\ref{fig:int_1.0_20}, 
respectively; except for $\Delta = 20$, the same values appear
in Figs.~\ref{fig:int_0.7_12} 
and \ref{fig:int_0.7_4}--\ref{fig:int_0.4_12}. We emphasize that all 
quenches have $M = 3 / 2 \Delta$, the same relationship imposed for
the non-interacting case. This constraint puts all of our quenches in the
``non-relativistic'' transport regime, as discussed in Sec.~\ref{WavepacketDyn}.
A single particle wavepacket with $M \Delta = 3/2$ shows only slow
broadening, similar to the non-interacting quench data in Figs.~\ref{fig:nonint_12} and \ref{fig:nonint_4}.

\begin{figure}[t]
   \includegraphics[scale=0.3]{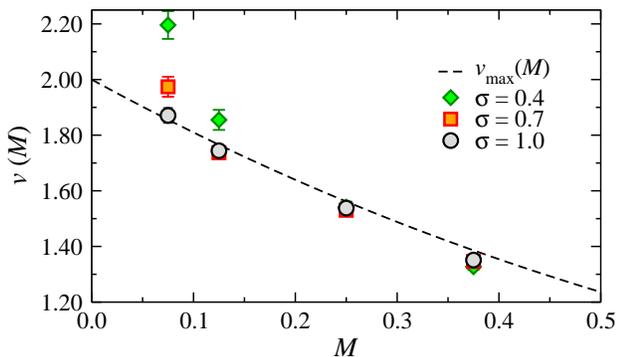}
   \caption{
	The maximum band velocity versus the band gap parameter, $M$,
	as calculated by Eqs.~(\ref{vmax_k}) and (\ref{kstar}) [black dashed line]
	and 
	the propagation velocity of the
	relativistically moving wavepacket in the interacting lattice quench
	for the values $\sigma =$ 0.4 [green diamonds], 0.7 [red squares], and 1.0 [black circles].
	}
   \label{fig:vmax}
\end{figure}

The difference in density dynamics 
for the interacting quenches shown in Figs.~\ref{fig:int_0.7_12} and \ref{fig:int_1.0_4}-\ref{fig:int_0.4_12}
as compared to the non-interacting 
versions in Figs.~\ref{fig:nonint_12} and \ref{fig:nonint_4} 
is remarkable.  For all interacting
parameter sets investigated, we observe a strong separation of dispersive dynamics
localized near the origin 
(the center of the chain and of the initial Gaussian inhomogeneity), 
and well-defined left- and right-moving wavepackets that propagate away 
from the origin
showing minimal dispersion
in their spatial extents.  
[Only the right-moving packet is depicted; the left-mover is an exact mirror image for the
Gaussian initial condition in Eq.~(\ref{gaussian}).]
Furthermore,
we find empirically that 
these wavepackets travel
``relativistically,'' i.e.\ at the maximal band velocity
$v_{\rm max}(M)$
determined above.  By tracking the peak of 
the right-moving wavepacket,
we are able to extract its propagation speed, which we plot in Fig.~\ref{fig:vmax} on top of 
$v_{\rm max}(M)$.  The error bars shown
there originate solely from the linear fit (of peak position vs.\ time).  The deviation seen at low $M$ 
(wide $\Delta$)
for
weak interaction strengths likely originates from the inaccuracy in determining the exact peak location
of such a shallow, wide wavepacket as well as from possible transient
distortion of the wavepacket's shape over its initial
time evolution.  

The continuum data in Figs.~\ref{fig:int_1.0_4}-\ref{fig:int_0.4_12} was obtained by
integrating the regularized sine-Gordon result in Eq.~(\ref{ContEOM}) numerically,
using $\zeta = 1$ and the values of $\alpha$ quoted in the figure captions.
Because the velocity renormalization scheme employed does not represent a fully systematic incorporation
of lattice dispersion details into the continuum Green's functions in Eq.~(\ref{ContGFs}),
we anticipate poorest agreement between lattice and continuum data in the dispersive ``tail'' dynamics occurring near
the origin. By contrast, we find very good agreement for the 
propagating wavepacket's speed and overall 
shape, despite the crude phenomenological regularization of the continuum initial state 
correlator.

The 
largest discrepancies
between continuum and lattice predictions 
occur 
for the smallest $\Delta = 4$,
and are
particularly pronounced in the parameter set ($\sigma=1.0$, $\Delta=4$, $M = 3/8$),
Fig.~\ref{fig:int_1.0_4}. 
A significant deviation for $\Delta = 4$ is also observed in the non-interacting, $M = 0$
quench shown in Fig.~\ref{fig:LLtoLL_4}. Taken together, these results suggest
that band curvature at the ultraviolet scale $k \sim k_{\rm max}(M) + 1/\Delta$ becomes important
in the lattice quench for this case. This behavior is not accounted for in our continuum calculations,
which instead assume the massive Dirac fermion spectrum $\varepsilon_k = v_{\rm max}(M) \sqrt{k^2 + M^2}$ 
for the final state Hamiltonian $\Hfc$, Eq.~(\ref{HfLowNRG}) (here $k$ is measured relative to $k_F$).
For $\Delta = 4$, the cubic curvature lifetime in Eq.~(\ref{t3Def}) $t_3 \sim \Delta^3/v_{\rm max}(M) = 46$, 
within the range of plotted time slices in Fig.~\ref{fig:int_1.0_4}. For $\Delta = 6$, our
estimate for $t_3$ leaps to $140$. 

What can we say about the supersoliton identified in the unregularized, continuum sine-Gordon quench
studied in Ref.~\onlinecite{supersoliton}, reviewed in Sec.~\ref{SG Def}? 
The supersoliton is defined as the asymptotic, long time 
($t \gg 1/v_F M$)
result for the pure sine-Gordon model transcribed in Eq.~(\ref{supersolitonSG}), 
exhibited in Fig.~\ref{supersolitonSGFig}. The supersoliton 
propagates ultrarelativistically
at the ``speed of light'' $v_F$, 
has a particular, 
non-dispersing
``s'' shape and an amplitude that grows in time as $t^{\sigma/2}$.
As articulated in Sec.~\ref{WDSec}, the supersoliton arises due to quasiparticle
\emph{fractionalization}.
Fractionalization of the initial, interacting LL state relative 
to the gas of propagating post-quench fermions induces a power-law excitation of
large momenta in the ``local'' Wigner distribution function $n(k;R)$, 
as exemplified in Eq.~(\ref{WDInt}). By contrast, in the non-interacting quench
momenta $|k| \gg 1/\Delta$ are exponentially suppressed as a consequence of Pauli-blocking (Sec.~\ref{WDSec}).
Through the massive post-quench dispersion, the Wigner function translates
into a ``local'' velocity distribution. In the non-interacting quench with $M \Delta \gg 1$,
only small velocities $v \lesssim v_F/M \Delta$ are significantly excited
[Eq.~(\ref{VelocDistNI}) in Appendix~\ref{APP--DistInhomog}];
in the interacting case, a non-integrable divergence appears at the ``speed of light'' 
$v_F$ [Eq.~(\ref{VelocDistI1})], 
irrespective of $M \Delta$,
signaling the presence of the supersoliton. Although regularization of
the LL correlation functions $\mathcal{C}^{r}_{\phantom{i} s}(x,x')$ 
in Eqs.~(\ref{BosCorrs}) or (\ref{BosCorrsReg}) at short distances ultimately 
cuts off this divergence [Eq.~(\ref{VelocDistI2})], the distinction between the 
interacting and non-interacting quenches survives, 
because the power-law behavior in 
$\mathcal{C}^{r}_{\phantom{i} s}(x,x')$ at \emph{large} distances is enough
to undermine Pauli-blocking, for $\sigma > 0$.

In the XXZ lattice quench studied here,
we do not observe the characteristic ``s'' shape of the supersoliton in any of the 
lattice data. In particular, the density fluctuation $\delta \rho(t,x_j)$ appears strictly 
positive for the interacting quenches, although negative excursions are observed for 
non-interacting quenches, Figs.~\ref{fig:nonint_12} and \ref{fig:nonint_4}. In each 
Fig.~\ref{fig:int_1.0_6}, \ref{fig:int_0.7_6}, and \ref{fig:int_0.4_6} 
($\sigma =$ 1.0, 0.7, and 0.4, respectively, all with $\Delta = 6$), the slim upper panel 
shows the continuum evolution [numerical integration of Eq.~(\ref{ContEOM})] over a window of 
length 600, corresponding to a system size 6 times larger than that used for the lattice quench. 
For $\sigma = 1.0$ and $0.7$, the continuum data 
shows the emergence of a negative peak at times and positions much larger than could be accessed in
the numerical lattice study. The relatively good agreement between the lattice and continuum
results over the 100-site windows in Figs.~\ref{fig:int_1.0_6} and \ref{fig:int_0.7_6} 
suggests the possibility that the ``s''-shape 
can appear in the lattice quench,
but only larger system size studies can resolve this question.\cite{Ref--ContLongTime}

\begin{figure}
   \includegraphics[scale=0.5]{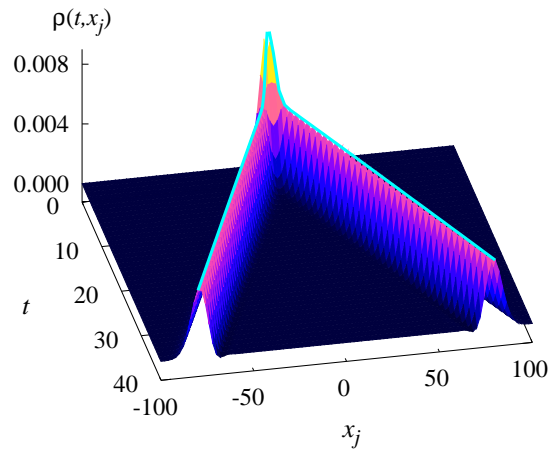}
   \caption{
	Three dimensional view of the 
	Fermi gas to Fermi gas
	lattice quench ($\sigma=0$, $M=0$)
	with $Q=0.1$, $\Delta=6$.
	The cyan line demarks the maximal propagation speed, $v_{\rm max} = v_F = 2$.
	}
   \label{fig:3d_LL_6}
\end{figure}

\begin{figure}
   \includegraphics[scale=0.78]{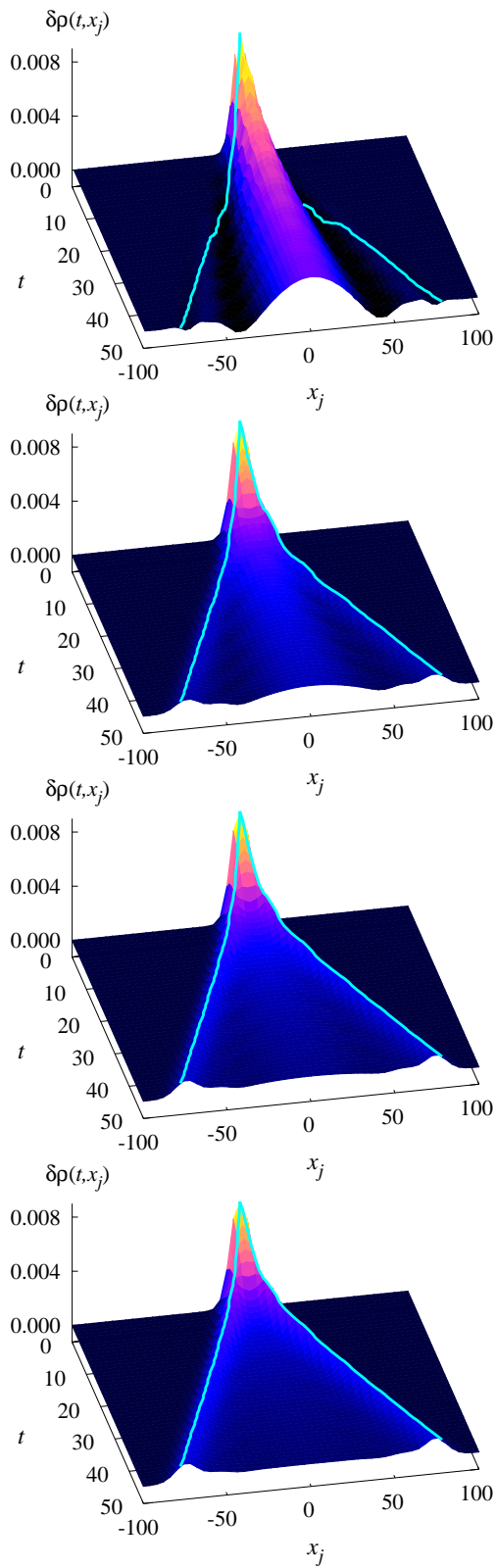}
   \caption{
	Three dimensional view of lattice quenches with $Q=0.1$ and $\Delta=6$,
	into a gapped final Hamiltonian with $M=1/4$
	for four different interaction strengths, $\sigma=0$, 0.4, 0.7, and 1.0 (top to bottom).
	The cyan line demarks the maximal propagation speed, $v_{\rm max}(M=1/4)\approx1.56$.
	}
   \label{fig:3d_gapped_6}
\end{figure}

A more precise way to analyze the interacting lattice quench data is to employ 
Eq.~(\ref{SupersolitonRegEq}). This equation describes a type of ``regularized''
supersoliton that appears at intermediate timescales:
the first two terms are the pure sine-Gordon theory result, while
the third term is the first correction due to a non-zero ultraviolet regularization
parameter $\zeta$. Eq.~(\ref{SupersolitonRegEq}) obtains from 
asymptotic analysis of the ``exact'' regularized sine-Gordon result in 
Eq.~(\ref{ContEOM}), valid for $v_F M \ll t \lesssim t_\zeta$ and $0 < \sigma < 1$;
see also Fig.~\ref{DynRegimes}.
The cutoff time $t_\zeta \propto \Delta / v_F(M \zeta)^2 \propto \Delta^3$
(since $M \Delta = 3/2$ here);
alternative definitions of the $\sigma$-dependent proportionality constant 
are provided in Eqs.~(\ref{tcDef1}) and (\ref{tcDef2}).
In Figs.~\ref{fig:int_0.7_12}, \ref{fig:int_0.7_6}, \ref{fig:int_0.4_6}, and
\ref{fig:int_0.4_12}, we have included time series plots of
Eq.~(\ref{SupersolitonRegEq}) for the quench parameters transcribed in the
captions. For $\Delta = 12$, Figs.~\ref{fig:int_0.7_12} and \ref{fig:int_0.4_12}, 
there is rough agreement between the asymptotic result, the lattice quench, and the 
numerical continuum integration for the latest time steps plotted. 
Eq.~(\ref{SupersolitonRegEq})
fails at earlier times, where transient behavior dominates both the lattice and 
continuum. The asymptotic result 
does not fare as well for $\Delta = 6$,
Figs.~\ref{fig:int_0.7_6} and \ref{fig:int_0.4_6}, although the lattice
data is well-modeled by the numerical integration of Eq.~(\ref{ContEOM}). 
Since the cutoff time $t_\zeta \sim \Delta^3$,
the failure of Eq.~(\ref{SupersolitonRegEq}) for smaller values of $\Delta$ indicates the need 
to retain higher order terms in this expansion; the closest agreement between 
Eq.~(\ref{SupersolitonRegEq}) and
the lattice and continuum data occurs for intermediate time steps in 
Figs.~\ref{fig:int_0.7_6} and \ref{fig:int_0.4_6}.
Since we have slaved $M \Delta = $ const., both $t_\zeta$ due to the
UV regularization of the initial state and $t_3$ due to cubic curvature
in the post-quench Hamiltonian [Eq.~(\ref{t3Def})] scale as $\Delta^3$,
implying that the effects of both types of lattice regularization 
can be simultaneously decreased by increasing $\Delta$ and the system 
size $L$ for fixed $L/\Delta$. 

We do not observe the amplification of the initial inhomogeneity 
predicted for the pure sine-Gordon case\cite{supersoliton}
in the XXZ quenches.
The relativistically propagating wavepacket produced by the interacting quench shows only 
increasing diminishment of its amplitude, for all parameter sets considered. 
This is consistent with the analysis of the regularized sine-Gordon theory
in Sec.~\ref{SGRegTheory}: at $t = t_\zeta$, the prefactor of the second
term in Eq.~(\ref{SupersolitonRegEq}) is proportional to
$(\alpha/\zeta)^{\sigma}$, which is less than or equal to one for
the parameters utilized to model the lattice quench.

In the Introduction, we exhibited in Fig.~\ref{fig:3d_gapped_12} a series of three-dimensional 
number density evolution plots for the lattice quenches with $M \Delta = 3/2$ and $\Delta = 12$, 
with $\sigma \in \{0,0.4,0.7,1.0\}$.
As a benchmark, Fig.~\ref{fig:3d_LL_6} depicts the $\sigma = 0$,  $M = 0$, $\Delta=6$ 
``Fermi gas to Fermi gas'' quench. 
(In this case, the quench consists merely of turning off the initial Gaussian trapping potential 
in a free Fermi gas.)
Fig.~\ref{fig:3d_gapped_6} is the same as Fig.~\ref{fig:3d_gapped_12}, but for the case
$\Delta = 6$. 
The weak undulations seen in the central peak of $\delta \rho(t,x)$ for the non-interacting quenches
(top panels) in Figs.~\ref{fig:3d_gapped_12} and \ref{fig:3d_gapped_6}
occur at the ``Zitterbewegung'' frequency $\omega = 2 M v_F = 4 M$. 
These oscillations appear in the non-relativistic regime for $M \Delta$ not too large,
similar to the single particle wavepacket dynamics discussed in Sec.~\ref{WavepacketDyn};
c.f.\ Eq.~(\ref{DiracNonRelProp}).


\section{Summary and Conclusion}\label{Conc}

\subsection{Summary of results}

In this work, we have performed a systematic study of spatiotemporal density dynamics in a
1D model of lattice
fermions (equivalent to the spin $1/2$ XXZ chain), following a quantum quench. 
The ground state of the XXZ chain in the gapless XY phase, parameterized by the $\hat{S}^z_{i} \hat{S}^{z}_{i+1}$
coupling strength $|\gamma| < 1$, is time-evolved by the non-interacting, band insulator
Hamiltonian obtained by setting $\gamma = 0$, whilst simultaneously turning on a sublattice staggered
magnetic field (chemical potential). As a probe of the quench dynamics, we introduced an additional localized
inhomogeneity into the spin (fermion) density of the initial state, and computed the subsequent evolution
of the density profile expectation value $\rho(t,x_j)$ under the post-quench dynamics generated by 
a translationally-invariant Hamiltonian. 

All quenches studied in this work feature the special property that the dynamics are 
generated by a simple, non-interacting band insulator Hamiltonian, characterized
by a bandgap $4 M$. By contrast, the pre-quench initial condition is the ground state of a 
system of interacting fermions 
possessing a low energy Luttinger liquid description,
except for the special case 
of a free Fermi gas
with $\gamma = 0$, referred to 
as the ``non-interacting'' quench. We used the density matrix renormalization group (DMRG) to 
numerically compute the initial state correlation function required to determine the 
quench evolution in the interacting case, studying chains of 202 sites with periodic boundary 
conditions. 

We identified a qualitative difference in the density dynamics generated by the initial state 
inhomogeneity for the non-interacting versus interacting ($\gamma \neq 0$) quenches. 
For an initial state seeded with a Gaussian density bump of width $\Delta$, in the ``non-relativistic'' 
transport regime ($M \Delta > 1$) we found only dispersive broadening for the non-interacting quench. 
By contrast, an interacting quench with the same value of $M \Delta$ generates coherently 
propagating left- and right-moving density waves, which travel 
``ultrarelativistically'' at the maximum band velocity $v_{\rm max}(M)$ of the post-quench 
spectrum.

We showed that the lattice quench data obtained here could be well-captured by a regularized continuum
sine-Gordon model. The continuum theory is an ultraviolet-modified version of the pure
sine-Gordon quench previously studied in Ref.~\onlinecite{supersoliton}. In that work, 
an ultrarelativistically
propagating density wave dubbed the ``supersoliton''
was identified as the leading asymptotic contribution to 
the exact result
for $\rho(t,x)$, in the case of the interacting quench. The supersoliton exhibits 
a rigid shape and an amplitude that grows in time according to $t^{\sigma/2}$,
where $\sigma > 0$ ($\sigma = 0$) for an initial state possessing (lacking)
interfermion interactions. In the sine-Gordon quench, the supersoliton arises
due to the relative quasiparticle \emph{fractionalization} of the initial
and final (pre- and post-quench) Hamiltonians, quantified by the anomalous
scaling dimension $\sigma/2$ of the post-quench fermions in the initial state.
In this paper, we showed that fractionalization leads to a divergence at $v_F$ in the local 
(Wigner) velocity distribution induced by the density inhomogeneity,
for the interacting quench. 
By contrast, in the non-interacting ($\sigma = 0$), non-relativistic ($M \Delta \gg 1$) 
quench we demonstrated that Pauli-blocking limits the excitation to small
velocities $v \lesssim v_F/M \Delta$.

In the interacting lattice quenches studied here, we did not 
observe amplification of the initial density profile,
nor did we find
the characteristic ``s'' shape of the supersoliton. We 
nevertheless established
that the propagating density waves produced by an interacting lattice quench are well-described
by the regularized sine-Gordon theory. 
For several of the lattice parameter sets studied, 
we demonstrated that the traveling waves of the corresponding continuum theory do
exhibit the characteristic supersoliton shape at length and time scales much larger 
than we can access in the lattice version, owing to computational limitations.
We interpret the waves produced by the interacting quench as
``elementary excitations'' of the non-equilibrium state; in 
the XXZ quench, these waves are ``regularized'' supersolitons.
Using an appropriate lattice definition for $\sigma(\gamma)$, we exhibited
the strong crossover of the post-quench dynamics as a function of the interaction
strength.

\subsection{On field theory methods in quantum quenches}

Beyond the particular dynamical phenomena uncovered 
in this paper, our work provides additional 
support to the idea that standard quantum field theory tools can be useful in studying 
strongly out-of-equilibrium physics
in ``realistic'' microscopic models. This is 
non-trivial, because field theoretic methods are typically employed in condensed matter 
to capture low-energy, long-wavelength equilibrium phenomena such as that observed 
near a quantum critical point. In such established settings, lattice scale details
in the form of irrelevant operators are often safely ignored, and field theory
tools can be used to make robust, sometimes even exact predictions, as a consequence
of \emph{universality}. On the other hand, a sudden quantum quench in a global
parameter of a many-particle system typically injects an extensive quantity of energy;
it is not a priori clear that long-wavelength, continuum methods can provide 
a useful description of the resulting dynamics. Indeed, in the context of a quench,
irrelevant operators encoding lattice-scale details present a serious formal difficulty:
under far-from-equilibrium conditions, renormalizability of the low-energy field theory 
is not necessarily a barrier against their effects. The problem is that long-time
dynamics can become sensitive to ultraviolet details, even if the renormalizable field theory
(i.e., the model obtained by discarding all irrelevant operators) gives an ultraviolet finite
prediction. The difficulty is compounded by the fact that operators irrelevant in the infrared
become relevant in the ultraviolet, rendering perturbative treatments useless for
long-time predictions. 

In the lattice quench studied in this paper, irrelevant operators suppress the amplification effect 
seen in the pure sine-Gordon model, an integrable field theory in 1+1 dimensions. 
A systematic improvement of the pure sine-Gordon theory in order to describe a 
particular ``parent'' microscopic model would require a non-perturbative resummation
of irrelevant operator effects, a difficult task. Nevertheless, we have demonstrated
that a phenomenological regularization of the sine-Gordon theory (equivalent to the
``resummation'' of a \emph{particular} irrelevant operator, characterizing the finite range $\zeta$ 
of the nearest-neighbor density interactions on the lattice) gives good agreement with the 
XXZ chain
dynamics, at least for the system sizes considered here.

\subsection{Open questions and extensions}\label{Extensions}

A key question is the survivability of the ultrarelativistic density packet
dynamics for longer times and larger system sizes. 
We identified two lifetimes $t_\zeta$ and $t_3$ that characterize the
temporal duration, post-quench, over which lattice effects on the dynamics 
in the initial and final Hamiltonians can be safely ignored.
For fixed $M \Delta > 1$ (non-relativistic quench), both $t_{\zeta,3} \propto \Delta^3$,
which implies that the lattice effects can be systematically reduced by
working with larger system sizes $L$ and wavepacket widths $\Delta$, such
that $L/\Delta$ is held constant.

Finally, the effects of interparticle interactions in the dynamical evolution 
pose a particularly interesting question; would the presence of a non-trivial 
S-matrix for the massive, post-quench spectrum of quasiparticle excitations tend 
to encourage or retard the formation and/or decay
of the ``regularized'' supersoliton? The answer likely
hinges upon the presence or absence of integrability for the post-quench Hamiltonian.
In particular, it would be interesting to study the density dynamics of an
XXZ chain quench from the XY phase to the gapped, Mott-insulating Ising AFM 
that occurs for $\gamma > 1$. 

The considerable flexibility afforded to tune control parameters in ultracold atom 
experiments, coupled with the excellent decoupling of these systems from the environment
has brought quench physics in (near) integrable models within observational reach.\cite{Weiss06,Gritsev10}
Despite the powerful methods developed to solve equilibrium properties, so far 
only limited analytical progress on non-equilibrium dynamics in integrable models has been 
made,\cite{Gritsev07,Polkovnikov07,HastingsLevitov08,Polkovnikov09,Fioretto10,Mossel10,Gritsev10} 
with the exception of systems that possess an underlying description in terms of free 
particles.\cite{IucciCazalilla10,Uhrig09,BarouchMcCoyDresden70,Polkovnikov05,Cincio07,Rossini09}
Numerical work using the time-dependent density matrix renormalization group (t-DMRG) 
by Manmana et.\ al.\ in Refs.~\onlinecite{Manmana07,Manmana09} 
on XXZ chain quenches between and within the XY and Ising phases
has revealed the ``light-cone effect'' predicted by Cardy and Calabrese\cite{CardyCalabrese06},
as well as evidence for topological defect formation\cite{KibbleZurek,Polkovnikov09} upon quenching into 
the gapped Ising phase. These studies were limited to relatively small system sizes (50 sites).
A variant of t-DMRG was used in Ref.~\onlinecite{Barmettler09}
to investigate the decay of N\'eel order in XXZ quenches, while a hybrid Bethe ansatz/numerics
approach was used in Ref.~\onlinecite{MosselCaux10} to determine the evolution
of a ferromagnetic domain wall state.
Given the complexity of the pure analytical approaches, it seems likely that a
numerical (or hybrid) scheme has the best chance of addressing the effects of interactions
on the post-quench dynamics articulated in this paper.

We emphasize that even in equilibrium, the effects of integrability and irrelevant operators
on correlation functions at non-zero temperature $T > 0$ remain
subjects of some controversy.\cite{DBT1,DBT2,DBT3,DBT4,DBT5,DBTRev,DamleSachdev98,Fujimoto99,Altshuler06,Langer09}
For massive 1D systems, Sachdev and Damle\cite{DamleSachdev98} gave well-reasoned arguments
that transport at $T>0$ should be diffusive.
This is also the naive expectation for a 1+1-D theory, in the absence of other special 
properties. However, Bethe ansatz results on integrable models appear to support the 
possibility of a non-zero Drude weight at non-zero temperature, indicative of 
ballistic transport.\cite{Fujimoto99,Altshuler06} One might expect that the 
incorporation of an integrability breaking perturbation (such as an irrelevant operator) 
introduces an additional timescale, beyond which the space-time retarded Green's function for 
the appropriate observable (e.g.\ a spin-spin correlation function) would transition 
from ballistic to diffusive behavior.

Lancaster and Mitra\cite{LancasterMitra10} have investigated a quench deep into
the Mott insulating breather regime of the sine-Gordon model,\cite{BosonizationRev3,Rajaraman} 
starting from a LL with an inhomogeneous ``domain wall'' density profile. In this case, 
the post-quench spectrum can be approximated by massive free bosons; because there is no 
fractionalization, the supersoliton does not occur. 
Quenches of an inhomogeneous LL with a domain wall density profile into the breather regime 
of sine-Gordon, incorporating interactions,
were further investigated in Ref.~\onlinecite{LancasterGullMitra10}, using the semiclassical
truncated Wigner approximation (TWA).\cite{Polkovnikov10} In this case, the authors uncovered
a persistent current, which could signal the preservation of ballistic post-quench transport. 
Previous work\cite{Polkovnikov10} has shown that the TWA provides a good approximation 
for quenches into the breather-dominated regime studied in Ref.~\onlinecite{LancasterGullMitra10}.

With respect to
the phenomena discussed in the present paper,
the TWA is known to fail\cite{Polkovnikov10} 
in the ``quantum'' (breatherless) regime of the sine-Gordon model,\cite{BosonizationRev3,Rajaraman} 
where fermionic solitons and antisolitons compose the spectrum. 
The supersoliton has been found at the special Luther-Emery point separating
the semiclassical and quantum regimes, where there are no breathers and the
fermions do not interact.\cite{supersoliton}
In the Ising phase 
of the XXZ chain, there are also no breathers, and the spectrum consists solely of interacting,
fermionic spinons. 
Moreover, the post-quench dynamics of massive, interacting fermions may differ 
between the continuum sine-Gordon and lattice XXZ models.

\begin{acknowledgments}
DMRG calculations were performed with a modified version of the ALPS library.\cite{alb07}
T.C.B. and D.R.R. would like to thank Garnet Chan, Emanuel Gull, and Steven White for
helpful discussions regarding the numerics.
M.S.F. thanks Natan Andrei and Deepak Iyer for helpful discussions on integrable
models.
T.C.B. was supported in part by a DOE Office of Science Graduate Fellowship.
M.S.F. and E.A.Y. acknowledge support by the National Science Foundation under Grant
No.~DMR-0547769, and by the David and Lucile Packard Foundation.
D.R.R. was supported by the National Science Foundation under Grant No.~CHE-0719089.
\end{acknowledgments}


\appendix


\section{Asymptotic analysis\label{APP--AA}}

In this appendix, we sketch the method used
to obtain the long-time asymptotic results of Eqs.~(\ref{supersolitonSG}), 
(\ref{SupersolitonRegEq}), and (\ref{AsymUltimate}) in the text. All three derive from the 
exact expression for the ``regularized'' sine-Gordon quench, Eq.~(\ref{ContEOM}).

All component integrals in Eq.~(\ref{ContEOM}) feature oscillatory
Bessel function kernels; these enter through the Green's functions $\bar{G}^{(1,2,3)}(t,y)$,
Eq.~(\ref{ContGFs}). Defining $\gamma \equiv M t'$, Eq.~(\ref{ContEOM}) 
can be expressed as 
\begin{align}\label{ContEOMAsym}
	\rho(t,x) 
	=&
	\frac{\rho_0(x - t')}{2}
	-
	\frac{c_N \alpha^\sigma \gamma}{2 t'^{\sigma+1}}
	\left[
	I_{1}(t',x)
	+
	I_{2}(t',x)
	\right]
	\nonumber\\
	&+
	\{x \rightarrow -x\}.
\end{align}
For simplicity, we consider here only the linear response to the
initial inhomogeneity $\rho_0(x)$, for the unregularized case
with $\zeta = 0$. Then the integrals $I_{1,2}$ are
\begin{align}
	I_1
	=&
	\int_{-1}^{1}
	\frac{\sqrt{\frac{1+z}{1-z}} \, J_1\ts{\left(\gamma \sqrt{1 - z^2}\right)} \, d z}{(1 - z)^{1 + \sigma}}
	\int_{x - t'}^{x - z t'} d y\, \rho_0(y),
	\label{I1Def}
\end{align}
\begin{align}
	I_2
	=
	\frac{\gamma}{2}
	&
	\int_{-1}^{1}
	d Z
	\int_{0}^{2(1 - |Z|)}
	\frac{d z_d}{z_d^{1 + \sigma}}
	\int_{x - t' Z + t' \frac{z_d}{2}}^{x - t' Z - t' \frac{z_d}{2}} d y\, \rho_0(y)
	\nonumber\\
	&\times
	\left[
	\begin{aligned}
	&
	\ts{\sqrt{\frac{(1+ Z)^2 - z_d^2/4}{(1 - Z)^2 - z_d^2/4}}}
	\\
	&\;\;\;\times
	J_1\sss{\left[\gamma \sqrt{1 - \left(Z + \frac{z_d}{2}\right)^2}\right]}
	J_1\sss{\left[\gamma \sqrt{1 - \left(Z - \frac{z_d}{2}\right)^2}\right]}
	\\&
	+
	J_0\sss{\left[\gamma \sqrt{1 - \left(Z + \frac{z_d}{2}\right)^2}\right]}
	J_0\sss{\left[\gamma \sqrt{1 - \left(Z - \frac{z_d}{2}\right)^2}\right]}
	\end{aligned}
	\right].
	\label{I2Def}
\end{align}

The basic method is to slice up the domain of each integral into pieces 
belonging to one of two varieties: type (i) regions throughout which one can employ 
the large argument asymptotic series for the Bessel functions, and type (ii) crossover 
domains where one cannot. For type (i) regions, the Bessel functions are replaced
by cosines; in the absence of a point of stationary phase or some other obstruction,
these integrals can be systematically evaluated by repeated integration-by-parts.
Successive integrations bring inverse powers of $\gamma = M \tp$ from the cosine
argument, which tend to suppress the contribution of the remainder in the long
time limit. To ensure the
convergence of the series obtained for a type (i) region, it is necessary to carefully
consider the specification of its boundary. 

Type (ii) regions, as well as points of stationary phase appearing in type (i) domains
must be isolated and evaluated by expanding the rest of the integrand 
in the local neighborhood. 
A useful trick to extract the long-time, leading asymptotic contributions
is to let each region boundary vary with $\gamma$ according 
to a power law. For example, the dominant contribution
to $I_1$ in Eq.~(\ref{I1Def}) in the limit $\gamma \rightarrow \infty$ comes
from the narrow type (ii) region $1 - \delta z_0 \leq z \leq 1$,
where $0 < \delta z_0 \ll 1$.
We let $\delta z_0 \equiv \gamma^{-\psi}$, with $\psi > 0$. 
Then we perform iterated integration-by-parts upon the neighboring type (i) region with 
$-1 + \delta z'_0 \leq z < 1 - \gamma^{-\psi}$ (assuming $\delta z'_0 > 0$). 
To ensure that this series converges and produces a subleading contribution, one leverages 
an additional constraint (an upper bound) upon the exponent $\psi$;
for Eq.~(\ref{I1Def}), $0 < \psi < 2$ does the job.  
Knowing the allowed range of $\psi$ in turn determines
the character of the type (ii) $1 - \gamma^{-\psi} \leq z \leq 1$ integration.
In this way, we isolate and evaluate the leading contributions to
$I_{1,2}$ in the long time limit, obtaining the asymptotic behavior of Eq.~(\ref{ContEOMAsym}).

Execution of the above-described program is straight-forward, but tedious; details 
are omitted here. In the remainder of this appendix, we indicate
the results by identifying the key elements leading to 
the unregularized supersoliton formula, Eq.~(\ref{supersolitonSG}). 

As discussed above, $I_1$ is dominated by the contribution near $z = 1$.
Expansion of the rest of the integrand gives
\begin{align}
	I_{1} 
	&\sim
	t'
	\gamma^{2 \sigma - 1}
	\,\rho_0(x - t')\,
	2^{1 + \sigma}
	\int_{0}^{2 \gamma} 
	\frac{d y \, J_{1}(y)}{y^{2 \sigma}}
	\nonumber\\
	&\sim 
	t'
	\gamma^{2 \sigma - 1}
	\,\rho_0(x - t')\,
	\frac{2^{1 - \sigma} \Gamma(1 - \sigma)}{\Gamma(1+\sigma)}.
	\label{I1Eval}
\end{align}	
The $I_2$ integration is dominated by the region with $1 - \delta Z_0 \leq Z \leq 1$,
where $0 < \delta Z_0 \ll 1$. 
For the initial density profile, we assume the Gaussian bump in Eqs.~(\ref{rho0Def}) and (\ref{GaussianSG}).
Making the change of variables $Z \equiv 1 - u/2 \gamma^2$ and $ z_d \equiv u r/\gamma^2$, 
one finds 
\begin{align}\label{I2Def2}
	I_2 
	\sim
	-t'
	\gamma^{2 \sigma - 1}
	\rho_0(x - t')
	\left[
	\bar{I}_{2,a}
	+
	\bar{I}_{2,b}
	\right],
\end{align}
where
\bsub
\begin{align}
	\bar{I}_{2,a}
	=&
	\int_{0}^{2 \gamma^2} 
	\frac{d u}{u^{\sigma}}
	\mathcal{K}(u),
	\\
	\bar{I}_{2,b}
	=&
	\int_{0}^{2 \gamma^2} 
	\frac{d u}{u^{\sigma}}
	\mathcal{K}(u)
	\nonumber\\
	&\phantom{\int_{0}^{0}} 
	\times
	\left\{
	\exp\left[
	- \frac{(x -\tp)}{M \Delta^2} \frac{u}{\gamma}  - \frac{1}{(2 M \Delta)^2} \frac{u^2}{\gamma^2} 
	\right] 
	-1
	\right\},
	\label{I2bDef}
\end{align}
\esub
with
\begin{align}
	\mathcal{K}(u)
	&=
	\int_{0}^{1}
	\frac{d r}{r^{\sigma}}
		\frac{
		1
		}{
		\sqrt{1 - r^2} 
		}
		J_{1}\left[\sqrt{u(1 - r)}\right]
		J_{1}\left[\sqrt{u(1 + r)}\right].
\end{align}
To leading order,
\begin{align}
	\bar{I}_{2,a}
	&\sim
	2^{1 - \sigma}
	\frac{\Gamma(1 - \sigma)}{\Gamma(1 + \sigma)}.
	\label{I2a}
\end{align}
The dominant contribution to the kernel $\mathcal{K}(u)$ 
\emph{relevant to the evaluation of $\bar{I}_{2b}$}
obtains from the large-$u$ behavior of the non-oscillatory term
\begin{align}
	\mathcal{K}(u)
	\sim&
	\frac{1}{\pi \sqrt{u}}
	\int_{0}^{1} 
	\frac{d r\,
	\cos\left[
	\sqrt{u}
	\left(
	\sqrt{1 + r} - \sqrt{1 - r}
	\right)
	\right]
	}{r^\sigma (1-r^2)^{3/4}}
	\nonumber\\
	\sim&
	\frac{u^{\frac{\sigma - 2}{2}}}{\pi}
	\int_{0}^{\infty} 
	\frac{d x}{x^\sigma} 
	\cos\left(x \right) 
	\nonumber\\
	\sim&
	\frac{u^{\frac{\sigma - 2}{2}}}{\pi}
	\Gamma(1 - \sigma) 
	\sin\left(\frac{\pi \sigma}{2}\right).
\end{align}
Assuming the Gaussian bump in Eqs.~(\ref{rho0Def}) and (\ref{GaussianSG}),
Eq.~(\ref{I2bDef}) then evaluates to
\begin{align}
	\bar{I}_{2,b}
	\sim
	\gamma^{-\sigma/2} \,
	&
	\frac{
	2 \Gamma(- \sigma)
	}{
	\Gamma\left(\frac{\sigma}{2}\right)
	\left(\sqrt{2} M \Delta \right)^{\sigma/2}}
	\,
	\exp\left[(x - t')^2/2 \Delta^2\right]
	\nonumber\\
	&\times
	D_{\sigma/2}\left[\sqrt{2} \left(\frac{x - \tp}{\Delta}\right)\right].
	\label{I2b}
\end{align}
In this equation, $D_{\nu}(z)$ denotes the parabolic cylinder function.
Combining Eqs.~(\ref{I2Def2}), (\ref{I2a}), and (\ref{I2b}) yields
\begin{align}\label{I2Eval}
	I_1 
	+ 
	I_2 
	\sim
	-
	\frac{
	2 Q \left(\frac{M^2 t'^3}{\sqrt{2} \Delta}\right)^{\sigma/2} \Gamma(- \sigma)
	}{
	M \sqrt{\pi} \Delta \Gamma\left(\frac{\sigma}{2}\right)
	}
	F_{\sigma}\left(\frac{x - \tp}{\Delta}\right),
\end{align}
where $F_{\sigma}(z)$ was defined by Eq.~(\ref{FDef}).
We obtain Eq.~(\ref{supersolitonSG}) from Eqs.~(\ref{ContEOMAsym}) and (\ref{I2Eval}),
using Ref.~\onlinecite{Ref--cN}.


\section{Fractionalization in the sine-Gordon model\label{APP--FracSec}}

In this appendix, we demonstrate that the natural quasiparticle degrees of freedom
in the continuum Luttinger liquid Hamiltonian defined by Eqs.~(\ref{HiLowNRG}) and (\ref{HiLowNRGB}) 
are \emph{fractionalized} with respect to the $\psi$ fermions that appear in the continuum
insulator Hamiltonian $\Hfc$, Eqs.~(\ref{HfLowNRG}) and (\ref{HfLowNRGB}). 

We begin by defining canonically rescaled boson variables
\begin{equation}
	\Phi \equiv \sqrt{K} \phi, \quad \Theta \equiv \theta/\sqrt{K},
\end{equation}
so that Eq.~(\ref{HiLowNRGB}) can be written as
\begin{align}\label{HiLowNRGBRS}
	\Hic
	&=
	\int 
	d x
	\left[
	\frac{u}{2}
	\left( 
	\frac{d \Phi}{d x}
	\right)^2
	+
	\frac{u}{2}
	\left( 
	\frac{d \Theta}{d x}
	\right)^2
	- \frac{\sqrt{K} \mu^{(0)}(x)}{\sqrt{\pi}} 
	\frac{d \Theta}{d x}
	\right]
	\nonumber\\
	&= 
	\int
	d x
	\left[
	-
	u 
	\,
	\chi^\dagger 
	\left(
	i\,
	\hat{\Sigma}^3
	\frac{d}{d x}
	\right)
	\chi
	-
	\sqrt{K}
	\mu^{(0)}(x)
	: \chi^\dagger \chi :
	\right].	
\end{align}	
On the second line of this equation, we have refermionized to obtain an 
expression in terms of some new, effectively non-interacting Dirac spinor $\chi$.\cite{BosonizationRev2} 
The field $\chi$ carries scaling dimension $1/2$ in the Luttinger liquid with Luttinger parameter $K$, 
and creates or annihilates the ``natural'' propagating quasiparticle degrees of freedom in that phase.
The $\chi$ particles propagate at the sound velocity $u$, rather than the bare Fermi velocity $v_F$.

Comparing Eqs.~(\ref{HiLowNRG}) and (\ref{HiLowNRGBRS}), we see that the chemical potential
$\mu^{(0)}(x)$ has been rescaled by a factor of $\sqrt{K}$ in the $\chi$ language.
This indicates that the $\chi$ fermion carries a fraction $\sqrt{K}$ of the
conserved $\psi$ fermion number charge. We can see this explicitly by considering
the bosonic expressions for the components of $\psi$ and $\chi$; in terms
of the original boson variables $\phi$ and $\theta$ in Eq.~(\ref{HiLowNRGB}),
these read
\begin{align}
	\psi(x)
	&\equiv
	\begin{bmatrix}
	\psi_1 \\
	\psi_2
	\end{bmatrix}
	\nonumber\\
	&=
	\frac{1}{\sqrt{2 \pi \alpha}}
	\begin{bmatrix}
	\exp\left\{i \sqrt{\pi} \left[ \phi(x) + \theta(x) \right] \right\}\\
	\exp\left\{i \sqrt{\pi} \left[ \phi(x) - \theta(x) \right] \right\}
	\end{bmatrix},
	\\
	\chi(x)
	&\equiv
	\begin{bmatrix}
	\chi_1 \\
	\chi_2
	\end{bmatrix}
	\nonumber\\
	&=
	\frac{1}{\sqrt{2 \pi \alpha}}
	\begin{bmatrix}
	\exp\left\{i \sqrt{\pi} \left[\sqrt{K} \phi(x) + \frac{1}{\sqrt{K}}\theta(x) \right] \right\}\\
	\exp\left\{i \sqrt{\pi} \left[\sqrt{K} \phi(x) - \frac{1}{\sqrt{K}}\theta(x) \right] \right\}
	\end{bmatrix}.\label{chiB}
\end{align}
Number charge conservation is associated with the $U(1)$ transformation
\[	
	\phi \rightarrow \phi + \frac{1}{\sqrt{\pi}}
	\Xi
	,\quad \theta \rightarrow \theta.
\]
so that
\[
	\psi \rightarrow 
	e^{i \Xi}
	\psi,\quad \chi \rightarrow 
	e^{i \sqrt{K} \Xi} 
	\chi.
\]

Finally, we note that $\chi$ is non-local when expressed in terms of $\psi$ (and vice-versa) for
any $K \neq 1$, since the right-hand side of Eq.~(\ref{chiB}) must then involve a ``string''
in the argument of the exponential, i.e.\ an integral of the $\psi$ current components $\{J^0,J^{1}\}$
from minus infinity to the argument $x$; see Eq.~(\ref{Currents}).


\section{Wigner functions for the post-quench quasiparticles\label{APP--DistInhomog}}

In this appendix, we define the Wigner functions for the particle $a_k$ and hole $b_k$ operators
of the massive, post-quench Hamiltonian $\Hfc$ [Eq.~(\ref{HfLowNRGPH})], in the (regularized) 
continuum sine-Gordon quench. We then transcribe results for the local velocity ``distributions''
induced by the inhomogeneous $\rho_0(x)$ [Eq.~(\ref{rho0Def})] for the non-interacting
and interacting quenches.
To simplify notation we set the Fermi velocity
\begin{equation}\label{vFone}
	v_F \equiv 1 
\end{equation}
in what follows.

The particle and hole Wigner distribution functions at time $t = 0$ (immediately post-quench)
are defined by
\bsub\label{WignerDist}
\begin{align}
	\np(k;R)
	\equiv&
	\int 
	d x_d\,
	e^{-i k x_d}
	\gndmuc 
	a^\dagger\left(R - {\textstyle{\frac{x_d}{2}}}\right) 
	a\left(R + {\textstyle{\frac{x_d}{2}}}\right) 
	\gndmu,
	\\
	\nm(k;R)
	\equiv&
	\int 
	d x_d\,
	e^{-i k x_d}
	\gndmuc 
	b^\dagger\left(R - {\textstyle{\frac{x_d}{2}}}\right) 
	b\left(R + {\textstyle{\frac{x_d}{2}}}\right) 
	\gndmu,
\end{align}
\esub
where $\gndmu$ denotes the ground state of $\Hic$, Eq.~(\ref{HiLowNRG}). 
Both the real space density profile $\rho_{0,\pm}(R)$ (at time $t = 0$) and 
the global distribution function $n_{\pm}(k)$ can be extracted from Eq.~(\ref{WignerDist}):
\bsub
\begin{align}\label{density}
	\rho_{0,+}(R)
	=&
	\int
	\frac{d k}{2\pi}\,
	\np(k;R)
	\nonumber\\
	=&
	\gndc 
	a^\dagger\left(R\right) 
	a\left(R\right) 
	\gnd,
\end{align}
\begin{align}\label{dist}
	n_{+}(k)
	=&
	\int_{-\epsilon}^{\epsilon} \frac{d Q}{2 \pi}
	\int d R \, 
	\exp\left(-i Q R \right)
	\np(k;R)
	\nonumber\\
	=&
	\int_{-\epsilon}^{\epsilon} \frac{d Q}{2 \pi}
	\gndc 
	a^\dagger\left(k - {\textstyle{\frac{Q}{2}}}\right) 
	a\left(k + {\textstyle{\frac{Q}{2}}}\right) 
	\gnd.
\end{align}
\esub
For a translationally invariant system, the ``point-split'' integration in Eq.~(\ref{dist})
picks up the delta function contribution at $Q = 0$; we are to take $\epsilon \rightarrow 0$ 
at the end of the calculation.

From Eq.~(\ref{Psitoab}), $a$ and $b$ are related to the right- ($\psi_1$) and left-movers ($\psi_2$)
via
\bsub
\begin{align}
	a(k) 
	=& 
	\beta(k) \psi_1(k)
	-
	i
	\beta(-k) \psi_2(k),
	\\
	b(k) 
	=& 
	\beta(k) \psi_1^\dagger(-k)
	-
	i
	\beta(-k) \psi_2^\dagger(-k),
\end{align}
\esub
where
\begin{equation}\label{beta}
	\beta(k) \equiv \sqrt{\frac{1}{2}\left[1 + \frac{k}{\varepsilon(k)}\right]},
\end{equation}
and $\varepsilon(k) = \sqrt{k^2 + M^2}$.

We define $\delta\npm(k;R)$ as the linear response to
$\rho_0(x)$, subtracting the homogeneous (global) distribution.
Using the correlation functions in Eq.~(\ref{BosCorrs}) and
incorporating the ultraviolet regularization $\zeta$ as in 
Eqs.~(\ref{BosCorrsReg}) and (\ref{ContEOM}), we obtain
\begin{align}
	\delta\np&(k;R)
	=
	-\delta\nm(k;R)
	\nonumber\\
	=&
	c_N \alpha^\sigma
	\begin{aligned}[t]
	\int&
	\frac{d q}{2\pi}
	\frac{\tilde{\rho_0}(q)}{q}
	\,e^{i q R}
	B(k;q)
	\\&
	\times
	\left[
	\begin{aligned}
	&
	\sgn\left(k + \frac{q}{2}\right)
	\mathcal{G}_{\sigma}\left(\left|k + \frac{q}{2}\right|;\zeta\right)
	\\&
	-
	\sgn\left(k - \frac{q}{2}\right)
	\mathcal{G}_{\sigma}\left(\left|k - \frac{q}{2}\right|;\zeta\right)
	\end{aligned}
	\right],
	\end{aligned}
	\label{WDpm}
\end{align}
where
\begin{align}
	B(k;q)
	=&
	\beta\left(k - {\textstyle{\frac{q}{2}}}\right)\beta\left(k + {\textstyle{\frac{q}{2}}}\right) 
	+
	\beta\left({\textstyle{\frac{q}{2}}} - k\right)\beta\left(-{\textstyle{\frac{q}{2}}} - k\right). 
	\nonumber
\end{align}
The kernel $\mathcal{G}_\sigma(|p|;\zeta)$ is defined by Eq.~(\ref{GDef}).
Eq.~(\ref{WDpm}) is identical to the Wigner distribution for the right-mover
$\psi_1$ in Eq.~(\ref{WD}), except for the $M$-dependent ``structure factor'' $B(k;q)$.

We consider first the non-interacting quench ($\sigma = 0$), wherein 
$\mathcal{G}_\sigma(|p|;\zeta) = \pi/2$. We assume the Gaussian density 
profile $\rho_0(x)$ in Eqs.~(\ref{rho0Def}) and (\ref{GaussianSG}).
As discussed below Eq.~(\ref{WDNI}), for the non-relativistic ($M \Delta \gg 1$), 
non-interacting quench, Pauli-blocking slaves the $k$-dependence of $\delta\np(k;R)$ to that
of the initial density profile $\tilde{\rho}_0(q = 2 k)$, suppressing
the contribution of momenta $|k| \gtrsim 1/\Delta$. 
Using the dispersion in Eq.~(\ref{DiracSpec}) to convert momentum to velocity,
we obtain the local velocity ``distribution'' at $R = 0$ (the center of the density bump),
\begin{equation}\label{VelocDistNI}
	\begin{aligned}
	&\delta\np(v;R = 0)
	\sim
	\frac{Q M}{2 v^2 (M \Delta)^2}
	\frac{
	\exp\left(
	-  
	\frac{v^2 (M \Delta)^2}{1 - v^2}
	\right)
	}{
	\sqrt{1 - v^2}
	}.
	\end{aligned}
\end{equation}
This equation applies when $v \gg 1/M \Delta$, for the non-relativistic regime
($M \Delta \gg 1$) of the non-interacting quench ($\sigma = 0$).
The exponential strongly suppresses velocities $v \gg 1/M \Delta$.

For the 
interacting case,
we are interested in
a ``soft quench'' (Sec.~\ref{SoftQuenchSec}), defined as  
the
regime where $1/\Delta \ll M \ll 1/\zeta$, i.e.\ a non-relativistic
initial condition, and an effective Compton wavelength much larger
than the ultraviolet scale $\zeta$, which is of order the lattice spacing.
As in Eq.~(\ref{WDInt}) the position and momentum dependencies factorize.
Converting to velocity, we obtain
\bsub
\begin{align}\label{VelocDistI1}
	\frac{
	\delta\np(v;R)
	}{
	\rho_0(R)}
	\sim&
	\frac{c_1(\sigma) \, v^{\sigma - 1}}{(1 - v^2)^{1 + \sigma/2}}
	\nonumber\\
	&\times
	\left[
	1
	+
	2^{\sigma -1} \left(\frac{v M \zeta}{\sqrt{1 - v^2}}\right)^{1 - \sigma} 
	\frac{\Gamma\left(\frac{\sigma - 1}{2}\right)}{\Gamma\left(\frac{1 - \sigma}{2}\right)}
	\right],
\end{align}
valid for $(1/M \Delta) \ll v \lesssim 1 - (M \zeta)^2/2$,
and
\begin{align}\label{VelocDistI2}
	\frac{
	\delta\np(v;R)
	}{
	\rho_0(R)
	}
	\sim&
	\frac{c_2(\sigma) \, v^{\sigma/2 - 1}}{(1 - v^2)^{1 + \sigma/4}}
	\exp\left(-\frac{v M \zeta}{\sqrt{1 - v^2}}\right),  
\end{align}
\esub
valid for $1 - (M \zeta)^2/2 \lesssim v \leq 1$.
The prefactors in these equations are given by
\begin{align}
	c_1(\sigma)
	=&
	\frac{(M \alpha)^\sigma \, \sigma \pi \Gamma\left(\frac{1 - \sigma}{2}\right)}{2^{\sigma + 1}\,\Gamma\left(\frac{1 + \sigma}{2}\right)},
	\nonumber\\
	c_2(\sigma)
	=&
	\left(\frac{M \alpha^2}{\zeta}\right)^{\sigma/2}
	\frac{\sigma \pi^{3/2}}{2^{1 + \sigma/2} \, \Gamma\left(\frac{1 + \sigma}{2}\right)}.
	\nonumber
\end{align}
Eqs.~(\ref{VelocDistI1}) and (\ref{VelocDistI2}) apply to the interacting
quench with $0 < \sigma < 1$. For $M \zeta \ll 1$, i.e.\ a Compton wavelength
much larger than the lattice spacing, Eq.~(\ref{VelocDistI1}) exhibits a 
strong non-integrable singularity approaching $v = 1$. For any $\zeta > 0$,
this divergence is ultimately cut off, as in Eq.~(\ref{VelocDistI2}). 
The exponential velocity suppression in the latter equation 
is weaker than that in Eq.~(\ref{VelocDistNI}), and originates in the 
ultraviolet behavior of the regularized Luttinger liquid correlation function, 
rather than the initial density profile.



\begin{thebibliography}{99}
\bibitem{BosonizationRev2}
	For a review, see e.g.
	T. Giamarchi,
	\textit{Quantum Physics in One Dimension}
	(Oxford University Press, Oxford, 2004).
\bibitem{BosonizationRev3}
	For a review, see e.g.
	A. O. Gogolin, A. A. Nersesyan, and A. M. Tsvelik,
	\textit{Bosonization and Strongly Correlated Systems}
	(Cambridge University Press, Cambridge, 1998).
\bibitem{Cazalilla04}
	M. A. Cazalilla, 
	J. Phys. B {\bf 37}, S1 (2004).
\bibitem{FisherGlazman96}
	For a review, see 
	M. P. A. Fisher and L. I. Glazman, in
	\textit{Mesoscopic Electron Transport},
	edited by Sohn, Kouwenhoven, and Sch\"on
	(Kluwer, The Netherlands, 1997).
\bibitem{BosonizationRev1}
	For a review, see e.g.
	R. Shankar, Acta Phys. Pol. {\bf 26}, 1835 (1995).
\bibitem{LLDCTransport}
	D. L. Maslov and M. Stone, 
	Phys. Rev. B {\bf 52}, 5539(R) (1995);
	I. Safi and H. J. Schulz,
	{\textit{ibid.}} {\bf 52}, 17040(R) (1995);
	V. V. Ponomarenko,
	{\textit{ibid.}} {\bf 52}, 8666(R) (1995).
\bibitem{BlochDalibardZwerger08}
	I. Bloch, J. Dalibard, and W. Zwerger,
	Rev. Mod. Phys. {\bf 80}, 885 (2008).
\bibitem{Bloch02}
	M. Greiner, O. Mandel, T. W. H\"ansch, and I. Bloch,
	Nature {\bf 419}, 51 (2002).
\bibitem{Weiss06}
	T. Kinoshita, T. Wenger, and D. S. Weiss,
	Nature {\bf 440}, 900 (2006).
\bibitem{Stamper-Kurn06}
	L. E. Sadler, J. M. Higbie, S. R. Leslie, M. Vengalatorre, and D. M. Stamper-Kurn,
	Nature {\bf 443}, 312 (2006).
\bibitem{Weiler08}
	C. N. Weiler, T. W. Neely, D. R. Scherer, A. S. Bradley, M. J. Davis,
	and B. P. Anderson,
	Nature {\bf 455}, 948 (2008).
\bibitem{Ott04}
	H. Ott, E. de Mirandes, F. Ferlaino, G. Roati, G. Modugno, and M. Inguscio,
	Phys. Rev. Lett. {\bf 92}, 160601 (2004);
	L. Pezz\`e, L. Pitaevskii, A. Smerzi, S. Stringari, G. Modugno, E. de Mirandes, 
	F. Ferlaino, H. Ott, G. Roati, M. Inguscio,
	{\textit{ibid.}} {\bf 93}, 120401 (2004);
	N. Strohmaier, Y. Takasu, K. G\"unter, R. J\"ordens, M. K\"ohl, H. Moritz,
	and T. Esslinger,
	{\textit{ibid.}} {\bf 99}, 220601 (2007).
\bibitem{Schneider10}
	U. Schneider, L. Hackerm\"uller, J. P. Ronzheimer, S. Will, S. Braun,
	T. Best, I. Bloch, E. Demler, S. Mandt, D. Rasch, A. Rosch,
	arXiv:1005.3545
\bibitem{Sommer11}
	A. Sommer, M. Ku, G. Roati, and M. W. Zwierlein,
	arXiv:1101.0780.
\bibitem{Rigol07}
	M. Rigol, V. Dunjko, V. Yurovsky, and M. Olshanii, 
	Phys. Rev. Lett. {\bf 98}, 050405 (2007);
	M. Rigol, V. Dunjko, and M. Olshanii, 
	Nature (London) {\bf 452}, 854 (2008).
\bibitem{Kollath07}
	C. Kollath, A. M. L\"auchli, and E. Altman,
	Phys. Rev. Lett. {\bf 98}, 180601 (2007).
\bibitem{Manmana07}
	S. R. Manmana, S. Wessel, R. M. Noack, and A. Muramatsu,
	Phys. Rev. Lett. {\bf 98}, 210405 (2007).
\bibitem{MoeckelKehrein08}
	M. Moeckel and S. Kehrein, 
	Phys. Rev. Lett. {\bf 100}, 175702 (2008);
	Ann. Phys. (N.Y.) {\bf 324}, 2146 (2009);
	M. Eckstein, M. Kollar, and P. Werner, 
	Phys. Rev. Lett. {\bf 103}, 056403 (2009).	
\bibitem{Barmettler09}
	P. Barmettler, M. Punk, V. Gritsev, E. Demler, and E. Altman,
	Phys. Rev. Lett. {\bf 102}, 130603 (2009);
	New J. Phys. {\bf 12}, 055017 (2010).
\bibitem{Rigol09}
	M. Rigol, 
	Phys. Rev. Lett. {\bf 103}, 100403 (2009).
\bibitem{SabioKehrein10}
	J. Sabio and S. Kehrein, 
	New J. Phys. {\bf 12}, 055008 (2010).
\bibitem{KibbleZurek}
	T. W. B. Kibble, J. Phys. A {\bf 9}, 1387 (1976);
	W. H. Zurek Nature (London) {\bf 317}, 505 (1985).
\bibitem{Polkovnikov05}
	W. H. Zurek, U. Dorner, and P. Zoller,
	Phys. Rev. Lett. {\bf 95}, 105701 (2005);
	A. Polkovnikov,
	Phys. Rev. B {\bf 72}, 161201(R) (2005);
	S. Deng, G. Ortiz, and L. Viola,
	{\textit{ibid.}} {\bf 80}, 241109(R) (2009). 
\bibitem{Cincio07}
	L. Cincio, J. Dziarmaga, M. M. Rams, and W. H. Zurek,
	Phys. Rev. A {\bf 75}, 052321 (2007).
\bibitem{PolkovnikovGritsev08}
	A. Polkovnikov and V. Gritsev,
	Nature Physics {\bf 4}, 477 (2008).
\bibitem{Polkovnikov09}
	C. De Grandi, V. Gritsev, and A. Polkovnikov,
	Phys. Rev. B {\bf 81}, 012303 (2010);
	{\bf 81}, 224301 (2010).
\bibitem{BarouchMcCoyDresden70}
	E. Barouch, B. McCoy, and M. Dresden,
	Phys. Rev. A {\bf 2}, 1075 (1970).
\bibitem{CardyCalabrese06}
	P. Calabrese and J. Cardy,
	Phys. Rev. Lett. {\bf 96}, 136801 (2006);
	J. Stat. Mech. P06008 (2007).
\bibitem{IucciCazalilla10}
	M. A. Cazalilla, 
	Phys. Rev. Lett. {\bf 97}, 156403 (2006);
	A. Iucci and M. A. Cazalilla,
	Phys. Rev. A {\bf 80}, 063619 (2009);
	New J. Phys. {\bf 12}, 055019 (2010).
\bibitem{Gritsev07}
	V. Gritsev, A. Polkovnikov, and E. Demler,
	Phys. Rev. B {\bf 75}, 174511 (2007).
\bibitem{Polkovnikov07}
	V. Gritsev, E. Demler, M. Lukin, and A. Polkovnikov,
	Phys. Rev. Lett. {\bf 99}, 200404 (2007).
\bibitem{HastingsLevitov08}
	M. B. Hastings, L. S. Levitov,
	arXiv:0806.4283.
\bibitem{Uhrig09}
	G. S. Uhrig,
	Phys. Rev. A {\bf 80}, 061602(R) (2009);
	B. D\'ora, M. Haque, and G. Zar\'and,
	arXiv:1011.6655. 
\bibitem{Manmana09}
	S. R. Manmana, S. Wessel, R. M. Noack, and A. Muramatsu,
	Phys. Rev. B {\bf 79}, 155104 (2009).
\bibitem{Rossini09}
	D. Rossini, A. Silva, G. Mussardo, G. Santoro,
	Phys. Rev. Lett. {\bf 102}, 127204 (2009);
	D. Rossini, S. Suzuki, G. Mussardo, G. E. Santoro, A. Silva,
	Phys. Rev. B {\bf 82}, 144302 (2010).
\bibitem{Mathey10}
	L. Mathey and A. Polkovnikov,
	Phys. Rev. A {\bf 81}, 033605 (2010).
\bibitem{MosselCaux10}
	J. Mossel and J.-S. Caux,
	New J. Phys. {\bf 12}, 055028 (2010).
\bibitem{LancasterMitra10}
	J. Lancaster and A. Mitra,
	Phys. Rev. E {\bf 81}, 061134 (2010).
\bibitem{LancasterGullMitra10}
	J. Lancaster, E. Gull, and A. Mitra,
	Phys. Rev. B {\bf 82}, 235124 (2010).
\bibitem{Schmitteckert04}
	P. Schmitteckert,
	Phys. Rev. B {\bf 70}, 121302 (2004).
\bibitem{Kollath05}
	C. Kollath, U. Schollw\"ock, J. von Delft, and W. Zwerger,
	Phys. Rev. A {\bf 71}, 053606 (2005).
\bibitem{Langer09}
	S. Langer, F. Heidrich-Meisner, J. Gemmer, I. P. McCulloch, and U. Schollw\"ock,
	Phys. Rev. B {\bf 79}, 214409 (2009).
\bibitem{supersoliton}
	M. S. Foster, E. A. Yuzbashyan, and B. L. Altshuler,
	Phys. Rev. Lett. {\bf 105},
	135701 (2010).
\bibitem{Cai11}
	Z. Cai, L. Wang, X. C. Xie, U. Schollw\"ock, X. R. Wang, M. Di Ventra, and Y. Wang,
	Phys. Rev. B {\bf 83}, 155119 (2011).
\bibitem{Heidrich-Meisner09}
	F. Heidrich-Meisner, S. R. Manmana, M. Rigol, A. Muramatsu, A. E. Feiguin,
	and E. Dagotto, 
	Phys. Rev. A {\bf 80}, 041603(R) (2009).
\bibitem{Kajala11}
	J. Kajala, F. Massel, and P. T\"orm\"a,
	arXiv:1101.6025v1.
\bibitem{Heidrich-Meisner08}
	F. Heidrich-Meisner, M. Rigol, A. Muramatsu, A. E. Feiguin, and E. Dagotto,
	Phys. Rev. A {\bf 78}, 013620 (2008).
\bibitem{LutherEmery74}
	A. Luther and V. J. Emery,
	Phys. Rev. Lett. {\bf 33}, 589 (1974).
\bibitem{Rajaraman}
	For a review, see e.g.
	R. Rajaraman,
	\textit{Solitons and Instantons}
	(North-Holland, Amsterdam, 1982).
\bibitem{Sutherland}
	For a review, see e.g. 
	B. Sutherland,
	\textit{Beautiful Models}
	(World Scientific, Singapore, 2004).
\bibitem{Ref--Pauli matrix}
	We employ the standard basis for all Pauli matrices.
\bibitem{Ref--Phi Plus Minus}
	The functions $\phi_{\{0,+,-\}}$ in Eq.~(\ref{DiracNonRelProp}) are given by
	\begin{align}
		\phi_0 
		&= 
		\frac{t x^2}{M \Delta^4 \delta^2(t)},
		\nonumber
		\\
		\phi_{\pm} 
		&= 
		\frac{t\left[(M x)^2 - 1\right]}{M^3 \Delta^4 \delta^2(t)} \pm \frac{2 x}{M \Delta^2 \delta^2(t)}.
		\nonumber
	\end{align}
\bibitem{Ref--Dirac Parity TRI}
	We employ the following definitions of the time-reversal $\mathcal{T}$ and parity 
	$\mathcal{P}$ transformations in the low energy Dirac theory outlined in Eqs.~(\ref{HfLowNRG})--(\ref{DiracDef}):
	\begin{align}
		(\mathcal{T}):&\quad
		\Psi(x) \rightarrow \hat{\sigma}^1 \Psi^{*}(x),\; i \rightarrow - i,
		\nonumber
		\\
		(\mathcal{P}):&\quad
		\Psi(x) \rightarrow \hat{\sigma}^2 \Psi(-x).
		\nonumber
	\end{align}
	The time-reversal transformation $\mathcal{T}$ is antiunitary and squares
	to one (spinless/spin-polarized fermions).
	These conventions are consistent with appropriate ``microscopic'' definitions 
	for the lattice model in Eq.~(\ref{HfDef}).
\bibitem{Ref--Sublat Stag Chem Pot}
	Technically, the lattice potential $\mu_{i}^{(0)}$ in Eq.~(\ref{HiDef})
	appears in the continuum as
	\[
	\mu_{i}^{(0)} \sim \mu^{(0)}(x_i) + (-1)^{x_i} \mu^{(0)}_{s}(x_i).
	\]
	In this equation, $\mu^{(0)}(x)$ [$\mu^{(0)}_s(x)$] gives the slowly-varying
	envelope for the smooth (sublattice-staggered) component of $\mu_{i}^{(0)}$.
	We have neglected the sublattice-staggered component in Eq.~(\ref{HiLowNRG}), 
	because the initial Gaussian ``bump'' assumed in Eq.~(\ref{gaussian}) gives
	a negligible contribution to $\mu^{(0)}_s(x)$ for $\Delta$ larger than
	a couple of lattice spacings.
\bibitem{QFT}
	See, e.g.,
	J. Zinn-Justin,
	\textit{Quantum Field Theory and Critical Phenomena}, $4^{\mathrm{th}}$
	ed. (Clarendon Press, Oxford, 2002).
\bibitem{Ref--cN}
	The normalization constant $c_N$ is determined by enforcing the fermionic sum rule
	(canonical anticommutation relations) on the correlation function 
	$\mathcal{C}^{i}_{\phantom{i} j}$ in Eq.~(\ref{BosCorrs}). The result is
	\[
		c_N = \sqrt{\pi}\frac{\Gamma\left(1 + \frac{\sigma}{2}\right)}{\Gamma\left(\frac{1 + \sigma}{2}\right)}.
	\]
\bibitem{Ref--SigmaLE}
	We note that Eq.~(\ref{SigmaDefSG}) is slightly different from a corresponding expression 
	in Ref.~\onlinecite{supersoliton}. 
	In the language of that paper, the fermion $\psi$ appearing in Eqs.~(\ref{HfLowNRG}) and
	(\ref{HiLowNRG}) denotes the LE point quantum soliton; by applying the bosonization 
	transformation directly to $\Hic$ as expressed in terms of $\psi$, the LE point is
	effectively shifted from $K = 1/4$ to $K = 1$.
\bibitem{Ref--CFT}
	Vertex operators are primary fields in the free boson conformal field
	theory; a conventional normalization scheme sets the coefficient
	of the 2-point correlator in Eq.~(\ref{BosCorrs}) equal to one.\cite{CFT}
\bibitem{Ref--Causality}
	While the exact asymptotic expression in Eq.~(\ref{supersolitonSG}) conserves
	the particle number, it is not strictly causal. The function $F_\sigma(z)$ 
	in Eq.~(\ref{FDef}) exhibits a power-law tail $\propto |z|^{-1 - \sigma/2}$ for
	$z \rightarrow -\infty$, inducing a finite (i.e., not exponentially suppressed)
	density disturbance at arbitrarily large $|x|$ in Eq.~(\ref{supersolitonSG}) 
	for any $t' \gtrsim 1/M$. This is an artifact of the asymptotic analysis, not the
	exact bosonization result, because neglected terms in Eq.~(\ref{supersolitonSG})
	(which feature amplitudes that decay in time)
	cancel these tails, shifting the acausal contribution inside the lightcone.
	The causal response of the exact result can be seen from the numerical 
	integration depicted in Fig.~\ref{supersolitonSGFig}.
\bibitem{Ref--PrimaryField}
	Although the natural quasiparticle degrees of freedom in the interacting
	Luttinger liquid $\Hic$ [Eq.~(\ref{HiLowNRG}) with $\gamma \neq 0$]
	are ``fractionalized'' with respect to the post-quench fermion
	$\psi$ [Eq.~(\ref{HfLowNRG})], the latter remains an eigenoperator of
	the renormalization group.\cite{CFT} 
\bibitem{LP}
	E. M. Lifshitz and L. P. Pitaevskii,
	\textit{Physical Kinetics}
	(Pergamon, London, 1981).
\bibitem{Gutzwiller}
	M. C. Gutzwiller, 
	\textit{Chaos in Classical and Quantum Mechanics}
	(Springer-Verlag, New York, 1990).
\bibitem{Cirac10}
	J. I. Cirac, P. Maraner, and J. K. Pachos,
	Phys. Rev. Lett. {\bf 105}, 190403 (2010).
\bibitem{RG}
	See, e.g., 
	N. Goldenfeld,
	\textit{Lectures on Phase Transitions and the Renormalization Group}
	(Perseus Books, Reading Mass., 1992);
	J. Cardy,
	\textit{Scaling and Renormalization in Statistical Physics}
	(Cambridge University Press, Cambridge, 1996).
\bibitem{DBT1}	
	S. Sachdev, T. Senthil, and R. Shankar,
	Phys. Rev. B {\bf 50}, 258 (1994);
	S. Sachdev, 
	{\textit{ibid.}} {\bf 50}, 13006 (1994).
\bibitem{DBT2}	
	H. Castella, X. Zotos, and P. Prelov\v{s}ek,
	Phys. Rev. Lett. {\bf 74}, 972 (1995);
	X. Zotos, F. Naef, and P. Prelov\v{s}ek,
	Phys. Rev. B {\bf 55}, 11029 (1997).
\bibitem{DBT3}	
	A. Rosch and N. Andrei, 
	Phys. Rev. Lett. {\bf 85}, 1092 (2000);
	K. Saito, 
	Phys. Rev. B {\bf 67}, 064410 (2003);	
	S. Fujimoto and N. Kawakami, 
	Phys. Rev. Lett. {\bf 90}, 197202 (2003).
\bibitem{DBT4}	
	X. Zotos and  P. Prelov\v{s}ek,
	Phys. Rev. B {\bf 53}, 983 (1996);
	B. N. Narozhny, A. J. Millis, and N. Andrei,
	{\textit{ibid.}} {\bf 58}, R2921 (1998);
	J. V. Alvarez and C. Gros,
	Phys. Rev. Lett. {\bf 88}, 077203 (2002);
	F. Heidrich-Meisner, A. Honecker, D. C. Cabra, and W. Brenig,
	Phys. Rev. B {\bf 68}, 134436 (2003).
\bibitem{DBT5}	
	J. Benz, T. Fukui, A. Kl\"umper, and C. Scheeren,
	J. Phys. Soc. Jpn. Suppl. {\bf 74}, 181 (2005);
	J. Sirker, R. G. Pereira, and I. Affleck,
	Phys. Rev. Lett. {\bf 103}, 216602 (2009);
	T. Prosen, arXiv:1103.1350.
\bibitem{DBTRev}
	For a recent overview, see e.g.\	
	J. Sirker, R. G. Pereira, and I. Affleck,
	Phys. Rev. B {\bf 83}, 035115 (2011).
\bibitem{DamleSachdev98}
	K. Damle and S. Sachdev,
	Phys. Rev. B {\bf 57}, 8307 (1998);
	Phys. Rev. Lett. {\bf 95}, 187201 (2005).
\bibitem{Fujimoto99}
	S. Fujimoto, 
	J. Phys. Soc. Jpn. {\bf 68}, 2810 (1999);
	R. M. Konik,
	Phys. Rev. B {\bf 68}, 104435 (2003).
\bibitem{Altshuler06}
	B. L. Altshuler, R. M. Konik, A. M. Tsvelik,
	Nucl. Phys. B {\bf 739}, 311 (2006).
\bibitem{Ref--RQFT}
	A renormalizable quantum field theory is defined\cite{QFT} by the condition 
	that the coupling strengths associated with all irrelevant operators are pinned to zero.
	This notion becomes important when exploring theories asymptotically
	free in the ultraviolet, such as QCD.
\bibitem{whi92}
	S. R. White,
	Phys. Rev. Lett. {\bf 69},
	2863 (1992).
\bibitem{whi93}
	S. R. White,
	Phys. Rev. B {\bf 48},
	10345 (1993).
\bibitem{sch05}
	U. Schollw{\" o}ck,
	Rev. Mod. Phys. {\bf 77},
	259 (2005).
\bibitem{pip10}
	P. Pippan, S. R. White, and H. G. Evertz,
	Phys. Rev. B {\bf 81}, 081103(R) (2010).
\bibitem{Ref--ContLongTime}
	The ultimate long-time behavior of the regularized sine-Gordon result
	in Eq.~(\ref{ContEOM}) gives the pure Gaussian translation in Eq.~(\ref{AsymUltimate}).
	For the values of $\alpha$ and $\zeta$ employed above, the amplitude of
	the Gaussian is negative, compensated by a long positive density tail 
	neglected in Eq.~(\ref{AsymUltimate}). 
\bibitem{Gritsev10}
	V. Gritsev, T. Rostunov, and E. Demler,
	J. Stat. Mech. P05012 (2010).
\bibitem{Fioretto10}
	D. Fioretto and G. Mussardo,
	New J. Phys. {\bf 12}, 055015 (2010).
\bibitem{Mossel10}
	J. Mossel, G. Palacios, and J.-S. Caux,
	J. Stat. Mech. LO9001 (2010).
\bibitem{Polkovnikov10}
	For a review, see 
	A. Polkovnikov,
	Ann. Phys. (N.Y.) {\bf 325}, 1790 (2010).
\bibitem{alb07}
	A. F. Albuquerque, 
	Journal of Magnetism and Magnetic Materials {\bf 310},
	1187 (2007).
\bibitem{CFT}
	See, e.g., 
	P. Di Francesco, P. Mathieu, and D. S\'en\'echal, 
	\textit{Conformal Field Theory}
	(Springer-Verlag, New York, 1996).
\end{thebibliography}
\end{document}